\documentclass[twoside,twocolumn,9pt]{article}
\usepackage{extsizes}
\usepackage[super,sort&compress,comma]{natbib} 
\usepackage[version=3]{mhchem}
\usepackage[left=1.5cm, right=1.5cm, top=1.785cm, bottom=2.0cm]{geometry}
\usepackage{balance}
\usepackage{widetext}
\usepackage{times,mathptmx}
\usepackage{sectsty}
\usepackage{graphicx} 
\usepackage{lastpage}
\usepackage[format=plain,justification=raggedright,singlelinecheck=false,font={stretch=1.125,small,sf},labelfont=bf,labelsep=space]{caption}
\usepackage{float}
\usepackage{fancyhdr}
\usepackage{fnpos}
\usepackage[english]{babel}
\usepackage{array}
\usepackage{droidsans}
\usepackage{charter}
\usepackage[T1]{fontenc}
\usepackage[usenames,dvipsnames]{xcolor}
\usepackage{setspace}
\usepackage[compact]{titlesec}
\usepackage{stackengine}
\usepackage{url}

\usepackage{epstopdf}
\usepackage[version=3]{mhchem}
\usepackage[flushleft]{threeparttable}
\usepackage{booktabs,caption,fixltx2e}
\usepackage{todonotes}
\definecolor{cream}{RGB}{222,217,201}

\begin{document}

\pagestyle{fancy}
\thispagestyle{plain}
\fancypagestyle{plain}{

}

\makeFNbottom
\makeatletter
\renewcommand\LARGE{\@setfontsize\LARGE{15pt}{17}}
\renewcommand\Large{\@setfontsize\Large{12pt}{14}}
\renewcommand\large{\@setfontsize\large{10pt}{12}}
\renewcommand\footnotesize{\@setfontsize\footnotesize{7pt}{10}}
\makeatother

\renewcommand{\thefootnote}{\fnsymbol{footnote}}
\renewcommand\footnoterule{\vspace*{1pt}%
\color{cream}\hrule width 3.5in height 0.4pt \color{black}\vspace*{5pt}} 
\setcounter{secnumdepth}{5}

\makeatletter 
\renewcommand\@biblabel[1]{#1}            
\renewcommand\@makefntext[1]%
{\noindent\makebox[0pt][r]{\@thefnmark\,}#1}
\makeatother 
\renewcommand{\figurename}{\small{Fig.}~}
\sectionfont{\sffamily\Large}
\subsectionfont{\normalsize}
\subsubsectionfont{\bf}
\setstretch{1.125} 
\setlength{\skip\footins}{0.8cm}
\setlength{\footnotesep}{0.25cm}
\setlength{\jot}{10pt}
\titlespacing*{\section}{0pt}{4pt}{4pt}
\titlespacing*{\subsection}{0pt}{15pt}{1pt}

\fancyfoot{}
\fancyfoot[RO]{\footnotesize{\sffamily{1--\pageref{LastPage} ~\textbar  \hspace{2pt}\thepage}}}
\fancyfoot[LE]{\footnotesize{\sffamily{\thepage~\textbar\hspace{3.45cm} 1--\pageref{LastPage}}}}
\fancyhead{}
\renewcommand{\headrulewidth}{0pt} 
\renewcommand{\footrulewidth}{0pt}
\setlength{\arrayrulewidth}{1pt}
\setlength{\columnsep}{6.5mm}
\setlength\bibsep{1pt}

\makeatletter 
\newlength{\figrulesep} 
\setlength{\figrulesep}{0.5\textfloatsep} 

\newcommand{\topfigrule}{\vspace*{-1pt}%
\noindent{\color{cream}\rule[-\figrulesep]{\columnwidth}{1.5pt}} }

\newcommand{\botfigrule}{\vspace*{-2pt}%
\noindent{\color{cream}\rule[\figrulesep]{\columnwidth}{1.5pt}} }

\newcommand{\dblfigrule}{\vspace*{-1pt}%
\noindent{\color{cream}\rule[-\figrulesep]{\textwidth}{1.5pt}} }

\makeatother

\twocolumn[
  \begin{@twocolumnfalse}

\noindent\LARGE{\textbf{Histological coherent Raman imaging: a prognostic review}} 
\vspace{0.3cm}

\noindent\large{Marcus T. Cicerone\textit{$^{a\ast}$} and Charles H. Camp, Jr.\textit{$^{a}$}}

\vspace{0.2cm}

\noindent\normalsize{Histopathology plays a central role in diagnosis of many diseases including solid cancers. Efforts are underway to transform this subjective art form to an objective and quantitative science. Coherent Raman imaging (CRI), a label-free imaging modality with sub-cellular spatial resolution and molecule-specific contrast possesses characteristics which could support the qualitative-to-quantitative transition of histopathology. In this work we briefly survey major themes related to modernization of histopathology, review applications of CRI to histopathology and, finally, discuss potential roles for CRI in the transformation of histopathology that is already underway.} 

 \end{@twocolumnfalse} \vspace{0.6cm}

  ]

\renewcommand*\rmdefault{bch}\normalfont\upshape
\rmfamily
\section*{}
\vspace{-1cm}


\footnotetext{\textit{$^{a}$~Address, 100 Bureau Drive, Gaithersburg, MD, USA. Fax: +1.301.975.4977 ; Tel: +1.301.975.8104; E-mail: cicerone@nist.gov}}



\footnotetext{$\ast$ Official contribution of the National Institute of Standards and Technology; not subject to copyright in the United States.}



\section{Introduction}
Histopathology has its roots in sixteenth century biology, \cite{Milestones2009} and  currently plays a key role in diagnosis of many cancers and other diseases. Nowadays, disease assignments from histology are generally based on the spatial arrangement of cell and tissue features that are highlighted using stains and sample preparation methods that were developed near the beginning of the 20$^{th}$ century.\cite{Titford:JHistotech:2009}  The current standard is subjective, consensus-based tissue analysis. This practice has held for many decades, however, it is known to suffer from several important sources of variability,\cite{Hunninghake2001,VanRhijn2010,AllsbrookJr2001,Costantini2003,Longacre2005} including differences in staining protocols and tissue analysis approaches among individuals and institutions. Further, while there are general agreements regarding sets of features that indicate a particular diagnosis,\cite{StanfordHistoCrit} knowledge of and proficiency in identifying these varies significantly with level and emphasis of training among pathologists. Together, these factors lead to some degree of diagnostic imprecision. Given the central role of histology in influencing diagnosis and subsequent treatment regimens, and it is not surprising that efforts are underway to increase objectivity, repeatability, and precision of histopathological analysis. 

Over the past several decades, a number of approaches have been pursued to increase the amount of quantitative information extracted from histology samples, and thus support improved analytical performance of histopathology approaches. Immunohistochemistry\cite{Coons1941} (IHC), introduced in the early 1940s, enhances image information content through antibody-based contrast agents for diagnostically valuable targets, such as cell surface markers and structural proteins. This approach can yield information on the spatial distribution of these diagnostic targets and has proven very useful in some cases,\cite{Fletcher2002} but also has important limitations.\cite{Werner2000,Sabah2003,Kirkegaard2006} Other efforts have focused on quantifying image information and presenting it to physicians in the form of computer aided diagnostic (CAD) tools.\cite{} A requirement for widespread use of CAD methods is that clinical tissue images be digitized. Although histopathology analysis is generally performed directly through the microscope, with no digital image ever acquired, digital recording of whole slide images (WSI) is becoming increasingly common, particularly for training purposes.\cite{Weinstein:HumPath:2009}

Vibrational spectroscopies such as Raman scattering and infrared absorption (IR) offer significant potential to increase the information content of  histology images. Like IHC, these modalities provide chemically specific and diagnostically relevant information, but without the need of labeling. IR spectroscopy was applied as early as 1952 for qualitative characterization of normal and neoplastic tissue.\cite{Woernley:Cancer:1952} Several IR and Raman histopathological studies in the 1990's through the 2000's utilized multivariate analysis methods to provide probabilities of disease within regions of interest.\cite{Kendall:TheJournalOfPathology:2003,Bhargava:ApplSpectrosc:2004} Some studies also provided relative abundances of major tissue components having diagnostic value.\cite{Haka2005} While studies such as these showed impressive sensitivity and specificity, they often did not provide image information that could be easily integrated into a diagnostic decision. IR imaging was not of sufficient spatial resolution, and Raman spectral acquisition is sufficiently slow as to preclude imaging without significant under sampling. These have been important limitations for a field where image interpretation is central. 

The recent advent of coherent Raman microscopies\cite{Zumbusch1999} have facilitated rapid label-free imaging, and bright infrared (IR) sources now available allow for IR imaging with higher spatial resolution.\cite{Nasse:NatMeth:2011,Yeh:AnalChem:2015} When spatially resolved, vibrational spectroscopy can provide familiar and diagnostically important information such as cell arrangement, phenotype, cellular and subcellular morphologies, structural proteins, and intracellular lipids. Because they require little or no sample preparation, these modalities can provide this data with minimized latency between excision and availability of actionable information. Further, the intrinsically digital image format and potential for rich chemical contrast makes vibrational histopathology a strong candidate for application of CAD.

Here we review the application of coherent Raman methods to histopathology, and look ahead through the lens of their potential for meeting the needs of histopathology. In order to better evaluate this potential, we first provide brief overviews of histopathology workflow, diagnostic criteria \& uncertainty, and technology-related efforts underway to improve histopathology, including application of infrared and spontaneous Raman spectroscopy.

\section{Histopathology} 

\subsection{Histopathology Workflow - A Brief Outline}

Histopathology, the microscopic analysis of tissues for the purpose of identifying disease, may be performed on tissue samples pre-operatively, intra-operatively, or post-operatively. Pre-operative analysis is performed to help establish a detailed diagnosis and treatment regimen. Intra-operative and post-operative histopathology is performed to ensure complete resection of diseased tissue by confirming absence of disease in margins of extracted or \textit{in situ} tissue. In cases where the benefit outweighs the risk, it is common practice to remove grossly healthy tissue surrounding the tumor to decrease the likelihood of positive margins. When excising a margin of nominally healthy tissue is contraindicated, such as for brain, the surgeon takes care not to remove more tissue than is justified. In such cases, intra-operative histopathology studies are sometimes performed on flash-frozen resected tissue to ensure that the entire tumor has been removed. The tissues are analyzed immediately, but an intra-operative histology analysis typically takes 20 to 30 minutes, and so prolongs the surgery. 

Pre- and post-operative histopathology studies are typically done in a batch mode, and usually take a day or two to wend their way through the process, requiring approximately 13 hours of actual processing time. In this process, excised tissue is first grossly examined (by eye) for diagnostic information, and prepared for microscopy. Tissue samples are typically fixed and embedded within paraffin wax, and some of the tissue is sliced in 10 to 50 $\nu$m thick sections. These formalin-fixed paraffin-embedded (FFPE) sections are then stained (typically with hematoxylin and eosin [H\&E]) and mounted for inspection. In a large hospital, a pathologist may inspect 500 histology slides in a day. Diseased regions are frequently small, so the clinician may spend a significant amount of time combing through normal tissue. In order to search more efficiently, pathologists will often identify landmark species such as blood cells and vessels, lymphocytes and ducts, which are frequently found in vicinity of cancerous lesions. Preliminary microscopic analysis and diagnosis are often carried out by a single pathologist. In difficult cases, a panel of experts may review slides from a case, and additional staining may be requested to identify diagnostically important species that H\&E staining does not highlight. Ultimately, analysis and diagnosis are generally reviewed and signed-out by an attending physician.

\subsection{Diagnostic Criteria}
The primary diagnostic criteria for most cancer types are related to intercellular organization.\cite{StanfordHistoCrit} There are a number of additional criteria that are specific for various cancer types. These include cellular features such as number density of mitotic events, infiltration of immune cells, degree of cell differentiation within and surrounding the tumor, the degree of necrosis, abnormal nuclear shape or intracellular lipid levels, the appearance of over-developed mitochondria, and chromatin clumping.\cite{StanfordHistoCrit,ProtAtlas,Fridman2012,Schubert2006} Tissue structural features are also important for some cancer types, such as prostatic adenocarcinoma.\cite{Humphrey:JClinPath:2007} These features may include acinar organization, mineralized tissue, mucin or polysaccharide  deposits, the presence  or absence of structural proteins, such as laminin, elastin, or collagen. In the course of qualitative tissue classification, the pathologist must evaluate whether observed features fall in or out of the normal range. These judgments are subjective and thus susceptible to variability which is in a sense uncontrolled, because it cannot be quantified on a case-by-case basis.

\subsection{Diagnostic Uncertainty}\label{SecDU}
While it is not currently feasible to assign diagnostic probabilities on a case-by-case basis, most studies address the issue of diagnostic uncertainty through analyzing outcomes in groups of patients. These studies can be classified into several types, and, while the results of these studies vary significantly, a few general trends emerge.

One type of study involves mandatory or voluntary review of a wide range of tissues by "peer"  physicians (i.e., not necessarily subspecialty experts). In this class, Kronz et al.\cite{Kronz1999} and Raab et al.\cite{Raab2005} report on inter-institution or institution-wide studies where large numbers of cases (6171 and 6162 respectively) covering many tumor types were reviewed. Kronz et al. found 1.5\% discordant diagnosis resulting in a major modification in therapy or prognosis was observed.\cite{Kronz1999} The majority of these cases involved a change between benign and malignant or a major change in tumor classification, and changes involving only a modification of tumor grade or stage were not included. That study found significant variability in diagnostic discordance with type of tissue. Serosal surfaces and the female reproductive tract tissues had 9.5\% and 5.1\% diagnostic discordance respectively, the highest found in the study. Raab et al\cite{Raab2005} estimated an error rate 6.7\% from self-reported discrepancies upon second review at 72 institutions. In that study, only 1.1\% of discordant diagnosis resulted in a major modification in therapy or prognosis.

Addressing these types of studies, Ho et al.\cite{Ho2006} argue that statistics based on second review likely reflect an underrepresentation of errors since the reviewer often has knowledge of the original diagnosis and the sign-out pathologist. They argue that such \textit{a priori} knowledge has led to biased review in similar situations.\cite{Renshaw2001} 

In another class of study, previously diagnosed cases for a single tissue type are analyzed by a panel of sub-specialists.  Higher discordant diagnosis rates are typically found in these types of studies.  Lurkin et al.\cite{Lurkin:BmcCancer:2010} report on a review of all sarcoma cases (366) in the Rhone-Alpes region of France over a 12 month period. They found that 19\% of cases resulted in change of type or invalidation of diagnosis, and 27\% of cases resulted in change of grade or subtype of diagnosed cancer. Bruner et al.\cite{Bruner1997} found similar results for 500 neuropathology cases submitted to a specialist review committee. They found that 9\% of reviews resulted in immediate significance for therapy or intervention, and 19\% resulted in a change in type or grade of glioma.  Similarly, specialist review of 602 prostate adenocarcinoma cases\cite{Nguyen2004} led to a change in the Gleason score by at least 1 point in 44\% of cases, and patients' risk category was increased in 11\% of cases. Likewise, of 340 patients presenting for second opinions regarding breast cancer, 80\% resulted in some diagnostic change, with 8\% of reviews leading altered surgical therapy.\cite{Staradub2002} In another study, 131 bladder carcinoma cases underwent secondary review, with 18\% exhibiting significant discrepancies.\cite{Coblentz2001} In a gynecological oncology\cite{Selman1999} study of 295 referred patients, 5\%, resulted in diagnostic changes that had major therapeutic or prognostic implications.

Arriving at definitive numbers for diagnostic uncertainty is outside the scope of this review. However, it does seem clear from the studies cited above that the uncertainty is significant for some cancer types. This widely recognized fact was reflected in a recent survey of pathologists,\cite{BestDoctors:2013} which showed that physicians on average expect 10\% diagnostic uncertainty from histopathology for cancer, and that lack of sub-specialty expertise is seen as the most important factor contributing to misdiagnoses. Based on these facts, one may be inclined to favor the sub-specialist review literature, and the slightly higher discordant diagnosis numbers found therein. The overall picture is that difficulty in diagnosis varies significantly with tissue and tumor type, and many human factors figure in precision and accuracy of diagnostics.\cite{Hollensead:JSurgOncol:2004}\cite{BestDoctors:2013}

\subsection{Impact of Diagnostic Uncertainty}
Table \ref{tbl:RSCa} shows published figures for cancer rates, inter-observer diagnostic variability among pathologists, and cost associated with initial treatment. It appears that approximately 200,000 people each year in the U.S. are either incorrectly informed that they have cancer, and subsequently undergo invasive treatments, or are incorrectly told that they do not have cancer, and miss the opportunity for potentially life-saving treatments. One might assume that half of misdiagnosed cases are false positive and half are false negative. Under this naive assumption, one estimates the cost of unnecessary procedures is \$2.5 B annually in the U.S. Of course, the cost of  missed treatment is harder to estimate, but misdiagnoses are uniformly the largest reason for medical lawsuits, which cost \$55.6 B annually, or 2.4\% of all health spending.\cite{Mello:HealthAff:2010}

\begin{table*}
	\small
	\caption{\ Major cancer types and diagnostic uncertainty}
	\begin{threeparttable}
	\begin{tabular*}{\textwidth}{@{\extracolsep{\fill}}lccccc}
		\toprule
		Type & \stackanchor{New}{Cases\cite{CancerFacts:2016}}& \stackanchor{Diagnostic}{Uncertainty} & \stackanchor{Initial}{Cost\cite{CancerCost2012}} & \stackanchor{Unnecessary}{Treatment Cost$^\dagger$} & \stackanchor{Raman}{Uncertainty} \\
		\midrule
		Breast          & 250 k & 20\%\cite{Longacre2005} &\$23 k &\$575 M&5\%\cite{Haka2005}\\
		Lung            & 225 k & 10\%\cite{Hunninghake2001} &\$61 k &\$680 M&10\%\cite{Huang2003}\\
		Prostate        & 180 k & 30\%\cite{AllsbrookJr2001} &\$20 k &\$540 M&5\%\cite{Stone2004}\\
		Colon \& Rectal & 135 k & 10\%\cite{Costantini2003} &\$52 k &\$350 M&5\%\cite{Stone2004}\\
		Bladder         & 77 k  & 40\%\cite{VanRhijn2010} &\$21 k &\$320 M&7\%\cite{BakkerSchut2006}\\
		Melanoma        & 76 k  & 25\%\cite{Gniadecka2004} &\$5.5 k&\$52 M &5\%\cite{Gniadecka2004}\\
		\bottomrule
	\end{tabular*}
	\begin{tablenotes}
	\small
	\item $\dagger$These values are estimates assuming diagnostic uncertainty is evenly distributed between precision and accuracy.
	\end{tablenotes}
	\end{threeparttable}
	\label{tbl:RSCa}
\end{table*}

The magnitude and gravity of this problem is not lost on physicians and technologists, and the field of histopathology has been slowly evolving to provide better informed and precise diagnoses. Several innovations have been in process of adoption over the past decades. 

\subsection{Immunohistochemistry}
Immunohistochemistry (IHC), first introduced in the 1940s, typically uses fluorescently tagged antibodies to provide image contrast specific to proteins of diagnostic value. Most IHC tags are targeted to functional proteins, such as cell surface markers;\cite{Uhlen2005} however, soluble\cite{Welsh:PNAS:2003} and structural\cite{StanfordHistoCrit,Honda:JCB:1998,Mauri:PASEB:2005,Alfonso:Proteo:2005} proteins are also targeted. The IHC approach facilitates mechanism-linked disease detection, and has enjoyed some important successes. For example, a protein (KIT), found to be mechanistically related to gastrointestinal stromal tumors (GISTs), is easily visualized through IHC labeling, and has found widespread use in diagnosing GISTs.\cite{Fletcher2002} While it now seems that KIT labeling is also prone to false positives in some cases,\cite{Sabah2003} the use of IHC for KIT detection has nevertheless simplified detection of GISTs, which was notoriously controversial with regard to classification, line(s) of differentiation, and prognostication.\cite{Fletcher2002} In spite of successes, IHC in general has some drawbacks including variability of labeling,\cite{Werner2000} and incompatibility with some stains. Compatibility issues include background fluorescence from a stain, interfering with IHC detection, and stains chemically modifying the affinity antibodies used in IHC. H\&E has both of these issues. Additionally, while IHC images can provide specific chemo-spatial information, they are also interpreted subjectively which can lead to variability in diagnostic outcomes.\cite{Kirkegaard2006} 

Improved diagnostic precision has been achieved by multiplexing classical and IHC stains, where the various contrast agents are imaged together.\cite{Levenson:Cytometry:2006}  Alternatively, cyclic IHC labeling\cite{Schubert2006,Zrazhevskiy:NatProt:2013,Gerdes2013,Riordan2015} of samples facilitates registered imaging of many proteins. Such approaches can lead to improved specificity for cancer characterization, but can add considerable complexity and, in the case of serial IHC, require up to a day for each labeling cycle.\cite{Riordan2015}  
 
\subsection{Whole slide imaging}

Pathology slides are typically viewed and analyzed directly through a light microscope for diagnostic purposes. Image digitization is not a common practice for histopathology, but that is now beginning to change. Whole-slide imaging\cite{Schubert1994} (WSI), which entails digitization and storage of entire tissue slides, was introduced in the mid 1990s. Image acquisition speeds were initially too slow to be of practical use, but current WSI instruments have pixel acquisition rates on the order of 3 MHz, allowing them to generate bright-field images of a 1 cm$^2$ tissue sample at 500 nm resolution in about 2 minutes; roughly the time a pathologist might spend on a tissue sample of similar size. WSI is now used widely in teaching environments,\cite{Weinstein:HumPath:2009} and has been fully adopted in some hospital systems.\cite{Pantanowitz2015} WSI is also useful for quality assurance, consultation, and telemedicine applications.\cite{Pantanowitz2015,Ho2006} 

\begin{figure}
	\includegraphics[width=9cm]{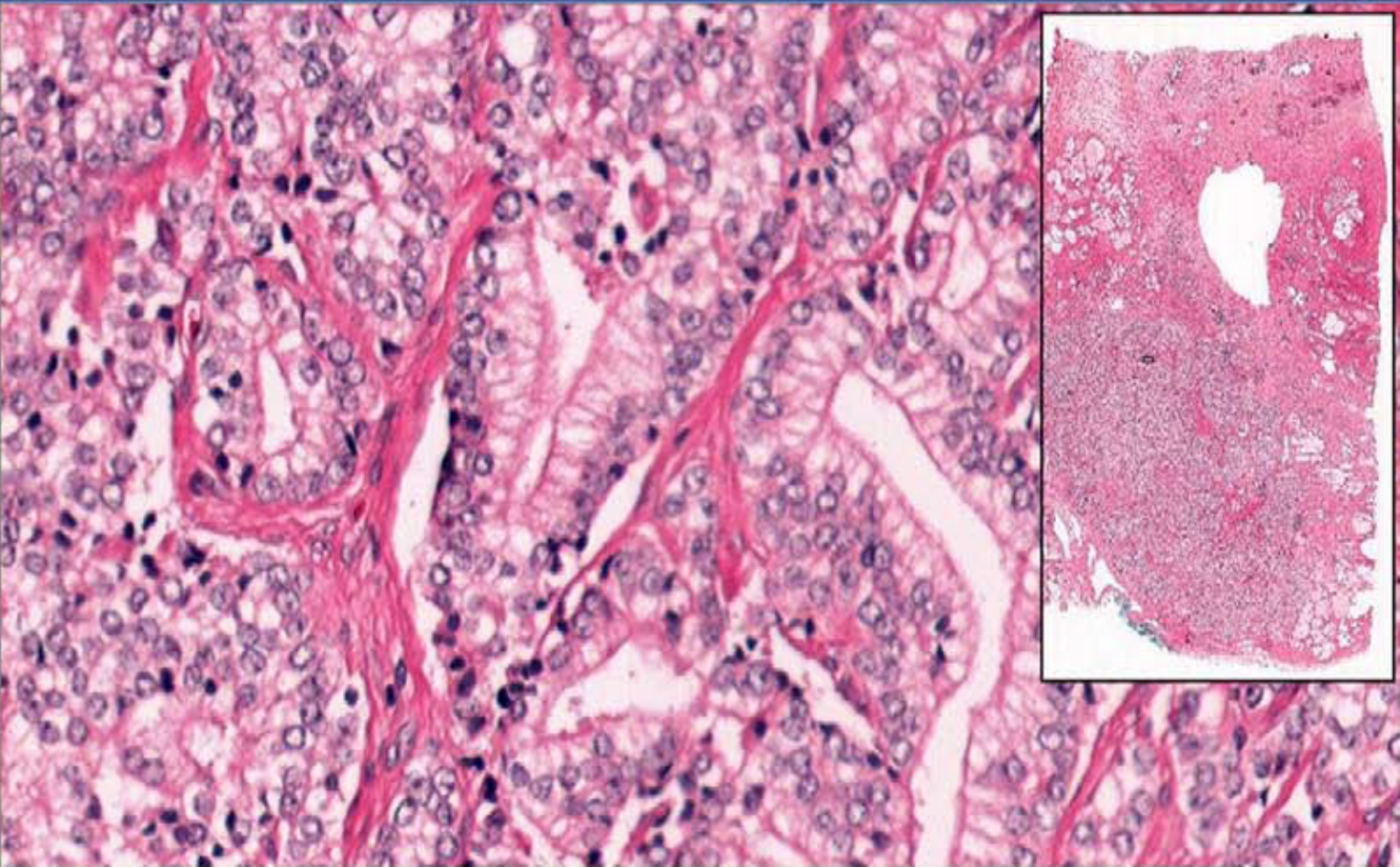}
	\caption{WSI image of prostate adenocarcinoma tissue from set of Genitourinary Block training in the preclinical
		curriculum at the University of Arizona College of Medicine, in Tucson, AZ.\cite{Weinstein:HumPath:2009}}
	\label{fgr:WSI_mag}
\end{figure}

Figure 1 shows an example of a prostate adenocarcinoma display from a pedagogical WSI application, allowing simultaneous view of an image thumbnail, and a detailed sub-image section.\cite{Weinstein:HumPath:2009} This sort of data presentation has obvious pedagogical and practical benefits, but large, high-resolution images such as this may occupy $\approx$1 Gb of disk space. Thus, the work of a single pathologist may occupy 0.5 Tb / day, and storage space requirements are often cited as a potential barrier to widespread acceptance of WSI.\cite{Pantanowitz:JPatholInform:2011} Another barrier is slightly reduced image quality compared to direct microscopic observation.\cite{Pantanowitz2015} There are further concerns about variation in sample preparation and staining practices, and that image capture and coloration can vary over an important range among WSI instruments. Consideration of these issues is currently underway at the U.S. Food and Drug Administration.\cite{WSI:FDA:2016}

\subsection{Computer Aided Diagnosis (CAD)}
CAD approaches support improved diagnostic precision by providing image-derived, diagnostically relevant information to pathologist. Efforts towards CAD began in the late 1990s with detection of malignant masses in radiological images of breast tissue,\cite{Mendez1998} and CAD is now routinely used in radiology for identification of lesions. Further, it seems that the prevalence and sophistication of CAD is likely to grow. In 2015, IBM acquired Merge Healthcare, for its 30 billion radiological and histological images, with the intent of using the image data to train Watson, their physician assistant software.\cite{Orcutt2015} 

While radiological CAD is now well-integrated into the diagnostic process, histological CAD is not used routinely. This dichotomy may be due in part to the fact that the former simply provides evidence for the presence of tumors, while the latter must address the more complex questions of tumor type and grade, and deals with images of higher complexity. Numerous algorithms have been proposed for diagnosis of specific cancers,\cite{He:SGAMA:2010,Mousavi:JPathInform:2015,Barker:MedicalImageAnalysis:2016} but there has not been significant clinical adoption. Histological CAD, however, is beginning to emerge from the research-only phase, as a number of companies are becoming involved in developing these tools.\cite{Bengtsson:CytA:2017} 

Histological CAD systems may exploit a large range of image features to derive clinically significant information,\cite{He:SGAMA:2010} including object size and shape, intensity, color, texture, and inter-object organization. Approaches for extracting these inputs from images have been reviewed recently.\cite{Gurcan:IEEE:2009,Irshad:IEEEReview:2014} The image processing associate with CAD will generally include some or all of the following steps: 1) Preprocessing, such as color and intensity normalization and de-noising. 2) Feature identification using thresholding or segmentation approaches. 3) Quantification of relevant feature characteristics, such as shape, size, spatial arrangement, texture, or color. 4) Reduction of data dimensionality using methods such as principle component analysis. 5) Classification based on the reduced-dimension dataset. Steps 2 through 5 (and sometimes step 1) are collectively referred to as "machine learning". We briefly discuss these steps below.

\subsubsection{Preprocessing}
Image analysis and quantification ultimately depend on spatial variation in color and intensity. It is therefore crucial that variations in tissue properties of interest give rise to the same range of image color and intensity variations within a sample, or between samples. However, in practice, it is often difficult to achieve sufficient image uniformity. Many factors such as inconsistency in staining or storage conditions, illumination brightness and uniformity or sample tilt can induce significant variation in response of the captured image to actual changes in tissue properties. 

Use of blank images\cite{Marty:BioTechnique:2007} or reference materials can reduce inter-image variability, but this is often insufficient. In order to minimize these variations, preprocessing steps such as color and intensity normalization are often employed. Color convolution\cite{Ruifrok:AQCH:2001} quantifies staining densities of a small number of contrast agents based on the fact that light absorptivity of individual stains is linear in log space. This approach continues to be developed, and can be used to map image color and intensities from many samples onto common range,\cite{Macenko:IEEE:2009} assuming each image contains similar ranges of features. However, color and intensity normalization approaches can lead to significant artifacts when based on assumptions that the proportion of pixels or range of stain intensities from each image in a set should be identical.\cite{Khan:IEEE:2014} In cases where a given stain vector may not be appropriate for all samples, individual vectors can also be normalized to a single image.\cite{Khan:IEEE:2014} 

\subsubsection{Image Segmentation}
Segmentation is the process of grouping like pixels into super-pixels, and generating boundaries around regions or objects with similar properties in the image. There are many segmentation approaches. The simplest of these is thresholding, where pixels in an image are converted to 0 or 1 depending on whether they meet a threshold criterion. The threshold may be set by consideration of pixel values in the overall image\cite{Otsu:Auto:1975} or of local regions if the image is expected to have trivial nonuniformities. \cite{Chang:IEEE:2000} Watershed\cite{Roerdink:FundaInform:2000} is another common segmentation approach where the boundaries of enclosed object shapes are defined by local maxima in pixel values. K-means clustering is another approach which finds, for some number (k) of clusters, the cluster location and boundaries that minimize a merit function, designed, for example, to group pixels of similar intensity. Fuzzy clustering\cite{Bezdek:CompGeo:1984} and Gaussian mixture models\cite{x} allow for the possibility of overlapping objects, assigning each pixel to one or more contiguous objects that are defined in such a way to minimize variation of a particular property within the boundaries of the object. The optimal configurations of these clusters are often found using expectation maximization approaches.\cite{Moon:IEEE:1996}

Another approach uses active contours, which are deformable splines that encompass image segments. They are constructed to minimize some energy function that may have contributions dictating the stiffness of the spline, the placement of the spline on the image (e.g., lowest energy at a minimum image gradient location), and a term that encapsulates \textit{a priori} knowledge of image segment shape or location.\cite{Irshad:IEEEReview:2014} These splines can be parameterized (SNAKES)\cite{Kass:IJCV:1988} or unparameterized (level set methods).\cite{Suri:IEEE:2002} Each approach has strengths and weaknesses, with SNAKES being more robust to noise, and level set methods having greater flexibility to accommodate variable topology. In cases where salient aspects of the image segment shape is known, Bayesian analysis of template comparison\cite{Naik:IEEE:2008} has been successful, and found robust to image variation. Each of these segmenting approaches may have trouble with overlapping features of interest, and active contours are prone to inclusion of unwanted background objects.\cite{Gurcan:IEEE:2009} Also, Bayesian methods are, in  general, computationally expensive.

Of course, pixel-level and active contour approaches are not mutually exclusive. It is possible to leverage the strengths of each in a single algorithmic approach.\cite{Alilou:CMIAG:2013} 

\subsubsection{Feature Selection}\label{Sec:FS}
Image segmentation prepares one to select features that may be used for quantification or in predictive model construction. This step frequently reduces complexity of the data to be considered and may be performed with or without user supervision. Supervised methods require the user to specify properties of features to be found, whereas unsupervised methods discover these properties autonomously from the data. The latter class includes principle component analysis (PCA),\cite{Jolliffe:PCA:2002} which finds a minimum set of orthogonal, linear combinations of the original data dimensions that retain the majority of signal variation. Exploratory factor analysis (EFA)\cite{Joliffe:StatMeth:1992} is a similar approach, but differs in that it allows for a component of random error in apparent associations between features and observed variables. Independent component analysis (ICA)\cite{Hyvarinen:ICA:2001} is also similar to PCA, except that ICA attempts to unmix signals by using the non-Gaussian character of individual signal elements to separate them into components that are as statistically independent from one another as possible.

Linear discriminant analysis (LDA) is a commonly used supervised feature selection approach that assumes independent variables are normally distributed. It is similar to PCA, except that the feature types are specified by the user. A comparison of PCA and LDA can be found in Martinez et al.,\cite{Martinez:IEEE:2001} who prefer PCA when the training data set is small, other factors being equal.

Similar to methods mentioned above, manifold learning (ML) approaches project an M dimensional data set to N dimensions, where, N < M, and preferably, N $\ll$ M. Both supervised and unsupervised ML approaches exist, and these differ from those above in that there is no assumption of linearity in the relationship between the features of interest and the original data.\cite{Gurcan:IEEE:2009} 

\subsubsection{Feature Quantification}
Once diagnostically important image features are identified, their properties may be quantified. Spatial arrangement of cells is almost always important, and the Feldman group\cite{Doyle:ISBI:2007, Naik:IEEE:2008} provide an example of how this can be characterized. They start with Voroni tesselation, in which the image is partitioned into convex polygons such that each polygon contains exactly one nucleus and all points within the polygon are closer to the center of their nucleus than to that of any other. From this starting point, they use Delaunay triangulation to draw lines between nuclei that shared a Voroni boundary, and then choose a set of lines that form a continuous structure of minimum lenght - a minimum spanning tree.  From characterization of the tree and a dozen other similar metrics, were able to discriminate Gleason 3, 4, and normal regions for prostate,\cite{Doyle:ISBI:2007} and evaluate Bloom-Richards criteria for breast cancer.\cite{Naik:IEEE:2008} Automated analysis and classification compared favorably to a manual approach, yielding better results in every categorization except cancer vs non-cancer in breast tissue.

\subsubsection{Classification}
Once image features have been quantified, the associated metrics may be used to classify tissues regarding their disease state. One approach is to use decision trees, which are a series of questions, the answers to which lead directly to a classification. These approaches have a number of attractions; they are similar to the approach a human might use to classify a tissue and are transparent. Another is that they have low systematic bias.\cite{Breiman:MachineLearning:2001} A potential drawback is that they tend to have high variance, but this can be reduced using approaches such as random forests,\cite{Breiman:MachineLearning:2001,Valkonen:CytA:2017} where many decision trees (the forest) are grown on varying subsets of the data, and decision nodes of the final tree are based on statistical sampling of the nodes in the forest.

Deep learning is another route to classification. These approaches are an evolution of neural networks in which the data vectors are iteratively transformed using nonlinear functions in such a way as to emphasize image aspects having predictive power. Using these approaches, segmentation, feature selection, and classification are all done at once, and all without user intervention. However, the entire process is opaque to the user in that there is no way to know what features the model is using for classification. Thus, there is thus no way to know when the model is likely to fail. In fact, only to the extent that the data set on which the deep learning model was trained is comprehensive, can the characterization be reliable. By way of apparent counterexample, however, Cirean et al. were able to use a relatively small sampling of mitotic events to create a highly effective deep learning model for detecting mitoses in H\&E images.\cite{Cirean:MICCAI:2013} Nonetheless, building a comprehensive data set for deep-learning image recognition is apparently the intent of IBM in acquiring a company with access to 30 billion radiological and histology images.\cite{Orcutt2015} 

\subsubsection{Roles for CAD in Histopathology}\label{Sec:CAD_Roles}
CAD assists the pathologist in one or more of the image analysis steps leading to diagnosis. The assistance could range from simply providing quantitative image metrics, to contextual reminders of disease mimics to apparent probabilities of various diagnoses.

Even the most basic level of CAD assistance - that of providing image feature identification, such as of nuclei or regions of potential metastases\cite{Valkonen:CytA:2017} may be quite valuable. In many cases, a pathologist may search for regions of interest (ROIs) that cover only a small fraction of the sample. Identifying such regions up front, or even identifying landmark species that typically signify such regions could provide a significant time savings.\cite{Riordan2015} 

Feature quantification might be considered the next level of sophistication where CAD approaches could be of high value. We find evidence for this by considering the work of Fuchs and Buhmann,\cite{Fuchs:CMIG:2011} who found 10\% disagreement on location and size of nuclei in a renal cell carcinoma sample between two sub-specialists, and inter- and intrareader disagreement of 42\% and 21\% respectively for counts of normal and atypical nuclei among five pathologists who were not sub-specialists. This level of variability is comparable to the overall diagnostic uncertainty discussed in Section \ref{SecDU}, and suggests that correct identification of relevant features may play a significant role in diagnostic uncertainty for this and other cancers. Image analysis approaches discussed in the previous sections can provide  quantitative characterization of key features, such as cell density\cite{Mousavi:JPathInform:2015} or mitotic event counts\cite{Cirean:MICCAI:2013} and nuclear shape metrics\cite{Doyle:ISBI:2007, Naik:IEEE:2008} at precision levels similar to that of sub-specialist analysis in the Fuchs and Buhmann work.\cite{Fuchs:CMIG:2011} 

While CAD work is largely done using H\&E,\cite{Cirean:MICCAI:2013} improved approaches to image contrast would likely lead to increased reliability of CAD outputs.\cite{Bengtsson:CytA:2017} Naturally, an expanded contrast pallet serves to put image-derived metrics on a more solid footing.\cite{Riordan2015} Even the simplest pixel-level thresholding methods can be quite reliable when pixel intensity values are related to specific chemical or functional information. For example, pixel-level analysis of vibrational spectra resulted in accuracies of 94\%-100\% for classification of ten disparate histologic classes,\cite{Fernandez:NatBio:2005} 98\% for positive prostate nuclei from IHC images with contrast for androgen receptor protein,\cite{Singh:AnalQuantCyto:2004} and 95\% for micro-metastasis from cytokeratin-stained lymph node sections.\cite{Weaver:ModernPath:2003}

\section{Spectroscopic Histopathology}
Vibrational spectroscopy is a rich source of enhanced spatio-chemical information that has been applied to histopathology, albeit primarily in a research mode. From such studies it appears that much of the information sought through staining, and perhaps more, is available without labeling, from techniques such as Raman or IR spectroscopy that can measure molecular vibrations intrinsic to the species of interest in the specimen.

It is common to describe molecular vibrations in terms of oscillatory motions along vectors ($Q$), called "normal modes." These modes constitute collective motion of molecular substituents that preserves the overall symmetry of the molecule during motion. The modes have discrete allowed states with energies approximately equal to $(\nu+1/2)\hbar\omega_m$, where $\nu$ are integer values, and $h\omega_m$ is the ground vibration frequency. The ground state vibrational frequencies and energies vary depending on the strengths of bonds and masses of atoms involved in the normal mode motion. Transitions with $\Delta \nu=\pm 1$ are most often measured, and those with energies in the range (500 to 3100) $cm^{-1}$ are typically of interest in biological systems. Signatures in the higher energy range, (2800 to 3100) $cm^{-1}$ arise from transitions between states of modes involving symmetric or asymmetric stretching of C-H bonds. The range (1800 to 2800) $cm^{-1}$ is referred to as the quiescent region because transitions of natural biological molecules typically do not appear in this range. The greatest variety of vibrational transitions in biological molecules occur in the fingerprint range (500 to 1800) $cm^{-1}$. 

Transitions between these states are usually detected through absorption, emission, or inelastic scattering of light. The probability for photon absorption or emission is significant when $\partial\mu/\partial Q\ne 0$, where $\mu$ is a mode's permanent dipole moment. By contrast, inelastic scattering is important when $\partial \alpha/\partial Q \ne 0$, where $\alpha$ is the molecular polarizability. Infrared and Raman spectroscopies are respectively based on absorption and inelastic scattering, and are often complimentary since many modes for which $\partial\mu/\partial Q = 0$ will have $\partial \alpha/\partial Q \ne 0$, and \textit{vice versa}. 

\begin{figure}
	\includegraphics[width=9cm]{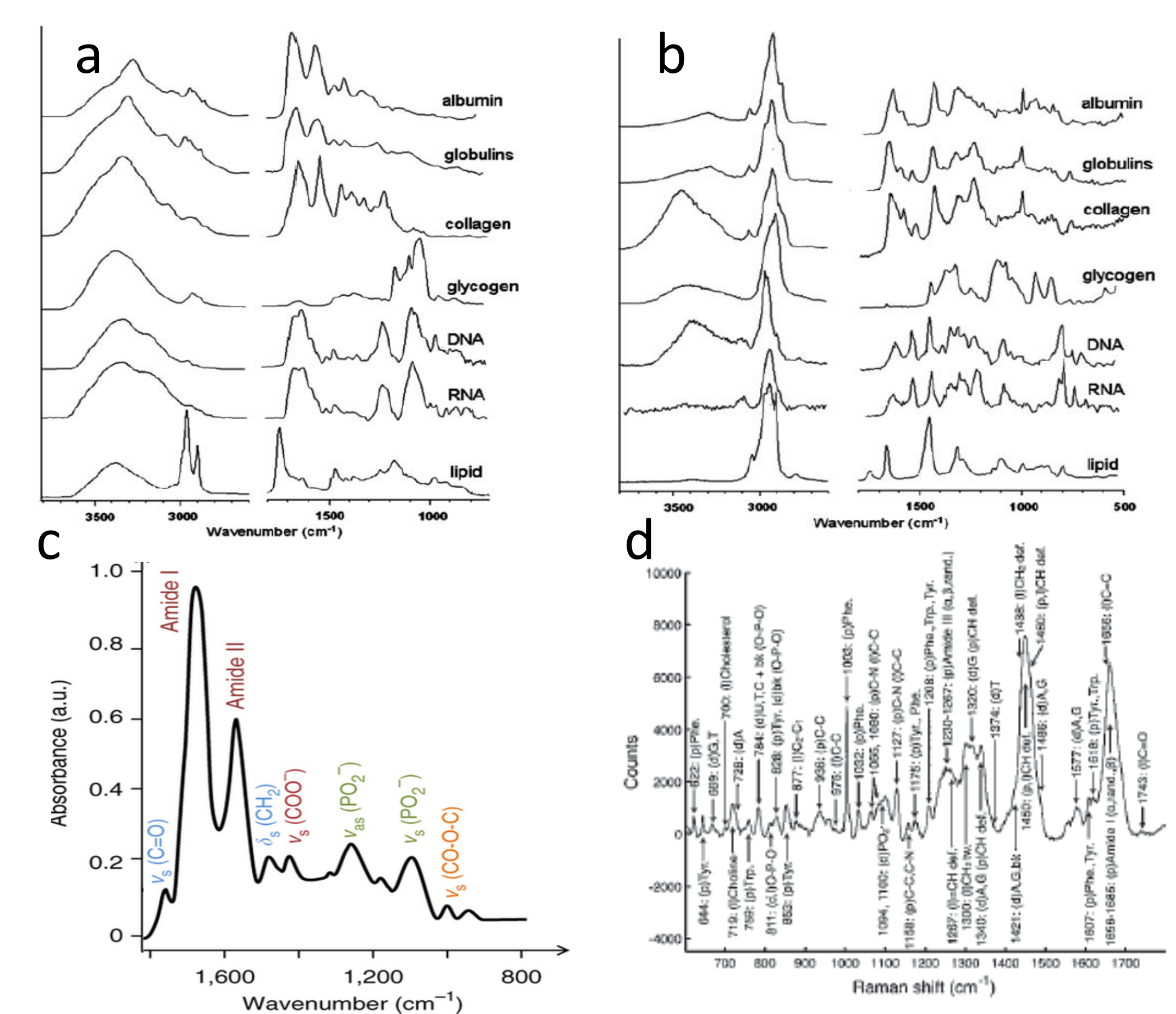}
	\caption{a: IR spectra of major cell components.\cite{Diem:JPB:2013} b: Raman spectra of major cell components.\cite{Diem:JPB:2013} c: IR fingerprint spectrum of whole cells\cite{Baker:NatProt:2014}. d: Raman fingerprint spectrum of whole cells\cite{Matthews:PMB:2011}}
	\label{fgr:IR-Raman_spec}
\end{figure}

Figure \ref{fgr:IR-Raman_spec}, panels a and b show IR and Raman spectra for major classes of tissue constituents. Since these have distinct signatures in both spectroscopies, they can be identified by either method. On the other hand, IR absorption lines are slightly broader. This is of consequence in the fingerprint region, where spectral lines are quite congested. Panels c and d of Fig \ref{fgr:IR-Raman_spec} show fingerprint regions of IR and Raman spectra from cells, illustrating that the width of the individual IR peaks makes it difficult to separate them into more than just a few distinct contributions without peak fitting. On the other hand, Raman spectra from similar systems contain a large number of easily resolvable peaks, providing a much higher level of chemical specificity. The differences can be quantified in terms of information content, or entropy, $S_i=ln(\Omega_i)$, where $\Omega_i$ is the maximum number of distinct spectral states one can discriminate. Here, $\Omega_i=SNR^M$ where M is the number of spectral peaks and SNR is the signal to noise ratio in each peak. We can estimate this value by noting approximately 15 peaks in the IR spectrum, 5 peaks in the Raman CH region, and 45 peaks in the Raman fingerprint region. Assuming SNR=15 for IR peaks and for CH stretch Raman peaks, and SNR=3 for Raman fingerprint, we obtain $\Omega_i=$ $10^{17}$, $10^{6}$, and $10^{21}$ for IR, Raman CH, and Raman fingerprint spectra respectively ($10^{27}$  for the entire Raman spectrum). Partly for this reason, Raman is found to have better specificity for classifying cancer.\cite{Krafft:JRS:2006}

Given their capacity for encoding chemical information, it is not surprising that these spectroscopies have been applied to many general aspects of disease diagnoses.\cite{Austin:Analyst:2016,Eberhardt:ERMD:2015} They have also been used for guidance in tissue sampling,\cite{Stevens:ChemSocRev:2016} augmented bases for stratified diagnosis,\cite{Montgomery2001} and intraoperative guidance during tissue-conserving resection.\cite{Jermyn:OSA:2014,Hollon2016} Of particular present interest is the considerable body of literature showing that vibrational spectroscopy provides diagnostically significant information for cancer diagnosis in conjunction with histopathology. There have been many excellent reviews on this topic in recent years,\cite{Manoharan1996,Petibois:TrendBiotech:2006,Meyer:JBO:2011,Diem:JPB:2013,Pence:ChemSocRev:2016,Austin:Analyst:2016,Shipp:AOP:2017}  so here we will only summarize relevant aspects. 

Histology-related studies using vibrational spectroscopy began to appear in the 1960s\cite{Fukuda:Histo:1966} for IR, and in the 1980s for Raman,\cite{DelMul:HumPath:1984} and to date, there have been thousands of publications. However, it appears that only in the past decade or so that IR and Raman technology have been sufficiently developed to produce meaningful studies.\cite{Diem:JPB:2013} Recent work provides rather convincing evidence that vibrational spectroscopy can detect diagnostically important changes in tissues, routinely yielding objective sensitivity and specificity figures generally in the 85\% to 95\% range for detection and classification of cancers, even when spectra on which discrimination is based are usually taken at just a few spots from normal and diseased portions of tissue. 

In one example, Pence et al.\cite{Pence:ChemSocRev:2016} discuss 26 large studies of epithelial cancers (> 50 patients), among which, the average sensitivity and specificity obtained using Raman spectroscopy were $89\pm7\%$ and $88\pm8\%$ respectively. Such precision generally exceeds that possible by subjective morphological analysis. Similar examples are listed in the final column of Table \ref{tbl:RSCa}, indicating a reduction in diagnostic uncertainty by more than a factor of two in most cases. Table \ref{tbl:RSCa} also provides prevalence, histology-based diagnostic uncertainty and initial treatment costs for the six most common cancers in the US. Assuming naively that diagnostic uncertainty is evenly distributed between precision and accuracy, we estimate that using Raman spectroscopy to supplement histology could save \textgreater\$2B/ yr in unneeded treatments, and reduce the burden on patients and families associated with missed opportunities for earlier treatment. 

\begin{table*}	
	\caption{Major cancer markers detected through vibrational spectroscopy}
	\begin{threeparttable}
		\small
		\begin{tabular}{p{3.5cm}p{4cm}p{9.8cm}}
			\toprule
			Marker & \stackanchor{Relevant}{Cancers}& \stackanchor{Spectral}{Lines $[cm^{-1}]$}\\
			\midrule
			
			&&\\
			\textbf{Metabolic Markers}&&\\
			DNA (relative abundance)& breast\cite{Haka2005} larynx\cite{Stone:Lrng:2000}&668, 678, 728, 750, 785$\nu_{\ce{O-P-O}}$ 825 $\nu_{\ce{O-P-O}}$ 1093$\nu_{\ce{PO^-_2}}$, 1336, 1488, 1580\\
			
			lipid (structure and relative abundance)& breast\cite{Haka2005} colon\cite{Stone2004,Beljebbar:CRO:2009} bladder\cite{BakkerSchut2006} skin\cite{Gniadecka2004,Gniadecka:PCPB:1997}& 1309$\tau_{\ce{CH2}}$ 1445$\delta_{\ce{CH2}}$ 1654$\nu_{\ce{C-C}}$ 2850$\nu_{\ce{CH2},sym}$ 2880$\nu_{\ce{CH2},asym}$ 2920$\nu_{\ce{CH3},sym}$ 2960$\nu_{\ce{CH3},asym}$\\
			
			cholesterol ester& breast\cite{Haka2005} brain\cite{Koljenovic:LI:2002} & 430, 702, 1302, 1442, 1740$\nu_{\ce{C=O}}$\\

			glycogen&bladder\cite{BakkerSchut2006} prostate\cite{Stone2004} brain\cite{Koljenovic:LI:2002} & 472, 481, 846, 932\\
			
			creatinine&breast\cite{Rehman:JRS:2007}&600$\nu_{\ce{N-CH3}}$ 678$\nu_{\ce{C=O}}$ 685$\tau_{\ce{C-S}}$ 692$\delta_{\ce{CO2}}$ 840$\delta_{\ce{N-CH2}}$ 903$\nu_{\ce{C-C-N}}$\\
			
			&&\\
			\textbf{Structural Molecules, etc.}&&\\
			protein structure & skin\cite{Gniadecka2004} melanoma\cite{Gniadecka2004}&1660$\nu_{\ce{C-O}}$ (Amide I) 2850$\nu_{\ce{CH2},sym}$ 2880$\nu_{\ce{CH2},asym}$ 2920$\nu_{\ce{CH3},sym}$ 2960$\nu_{\ce{CH3},asym}$\\

			keratin& epithelial cancers\cite{Chen:SciRep:2016} & 623$\nu_{\ce{CS}}$, 643$\nu_{\ce{CS}}$, 850$\delta_{\ce{CCH aromatic}}$, 885$\rho_{\ce{CH_2}}$ 933, 1002$\nu_{\ce{C-C}}$ (ring breathing), 1031 (phenylalanine), 1200 ~ 1350 (Amide III) 1450$\tau_{\ce{CH}}$ 1650$\nu_{\ce{C=O}}$ (Amide I), 2940$\nu_{\ce{CH3}}$\\ 		
			
            collagen & breast\cite{Haka2005} prostate\cite{Stone2004} colon\cite{Stone2004} lung\cite{Huang2003,Kaminaka:JRS:2001} skin\cite{Nijssen:JBO:2007} &855, 938, 1004$\nu_{\ce{C-C}}$ 1260$\nu_{\ce{CN}},\delta_{\ce{NH}}$ 1314$\omega_{\ce{CH3CH2}}$ 1445$\delta_{\ce{CH2}}$ 1660$\nu_{\ce{C-O}}$ 2940 \\
			 
			elastin&lung\cite{Huang2003}& 1442$\delta_{\ce{CH2}}$ 1660 $\nu_{\ce{C-O}}$\\
			
			carotenoids & bladder\cite{BakkerSchut2006} lung\cite{Huang2003} brain\cite{Koljenovic:LI:2002} &1159, 1523\\
			
			polysaccharides &brain\cite{Mizuno:JRS:1994}&856\\
			
			&&\\
			\textbf{Others}&&\\
			phospholipids & lung\cite{Huang2003} brain\cite{Koljenovic:LI:2002}&719 1442$\delta_{\ce{CH2}}$ \\
			
			\ce{CaC2O4}, \ce{Ca5(PO4)3} & breast\cite{Haka2005} brain\cite{Mizuno:JRS:1994} &912$\nu_{\ce{C-C}}$ 960$\nu_{\ce{P=O}}$ 1477$\nu_{\ce{C-O}}$\\
			
			phosphatidylcholine & adenocarcenoma\cite{Diem:JPB:2013}& 719, 1666$\nu_{\ce{C-O}}$\\

			trypotphan&lung\cite{Huang2003} prostate\cite{Stone2004}&650$\tau_{\ce{C-C}}$ 1260$\nu_{\ce{CN}},\delta_{\ce{NH}}$\\
			
			proline & prostate\cite{Stone2004}& 939$\nu_{\ce{C-C}}$\\
			
			tyrosine&prostate\cite{Stone2004}& 1176$\nu_{\ce{C-H}}$ 1217$\nu_{\ce{C-C6H5}}$\\

			\bottomrule
		\end{tabular}
		\begin{tablenotes}
			\small
			\item $\nu$-stretch, $\delta$-scissoring, $\rho$-rocking, $\omega$-wag, $\tau$-twist
		\end{tablenotes}
	\end{threeparttable}
	\label{tbl:VibMarkers}
\end{table*} 

\subsection{Vibrationally Detected Cancer Markers}
The impressive diagnostic value of vibrational spectroscopy accrues from its inherent chemical specificity. It can reveal information that is otherwise available only by methods such as IHC, polymerase chain reaction or gene arrays,\cite{Diem:JPB:2013} leading to more precise tissue segmentation,\cite{Fernandez:NatBio:2005} and providing new prognostic information.\cite{Kwak2015} 

Table \ref{tbl:VibMarkers} gives a partial list of cancer biomarkers that have been used in Raman studies to classify neoplastic and cancerous tissues, and lists spectral peaks used to identify those markers. Most markers have several peaks, facilitating increased robustness in detection, to the extent that noise in separate peaks is uncorrelated, but even correlated noise can be suppressed if peak ratios can be used. The fact that most of the peaks used in these Raman studies are found in the fingerprint region is due to the high spectral density and low peak widths found there, and underscores the importance of this spectral range for chemical specificity. 

Many of the markers listed in Table \ref{tbl:VibMarkers} are commonly visualized in pathology through IHC or classical stains, but, of course those methods require labeling. Not only is vibrational contrast label-free, it can also be semi-quantitative when peak amplitude ratios are used. Accordingly, peak pairs that provide information on protein-to-DNA or protein-to-lipid ratios, or on degree of unsaturation in fatty acids frequently yield highly reliable metrics of disease type and grade.\cite{Manoharan:PCPB:1998,Mizuno:JRS:1994,Stone:Lrng:2000,Huang2003,Liu:JPPB:1992,Utzinger:ApplSpec:2001,You:SciRep:2016}

Macromolecular structure and material phase information are also available from vibrational spectroscopy, and can be useful in cancer detection and diagnosis. For example, degree of calcification is a known marker for many cancers. \cite{StanfordHistoCrit} Using classical stains, it is difficult to distinguish between types of calcification, but such discrimination turns out to be important. Haka et al.\cite{Haka2005} showed that Raman spectroscopy can be used to easily distinguish \ce{CaC2O4} from \ce{Ca5(PO4)3}, and that their relative abundances correlate with malignancies in breast cancer. Raman spectroscopy also contains structural and hydrogen bonding information for lipid, protein, and nucleic acids, which has diagnostic value for many cancers.\cite{Rehman:JRS:2007,Haka2005,Gniadecka2004,Gniadecka:PCPB:1997}

Spectral changes between healthy and diseased tissue can be striking,\cite{Huang2003,Kast:Biopolymers:2008} or subtle,\cite{Diem:JPB:2013} but generally appear in the context of highly complex spectra from the tissue. Thus, while it is sometimes possible to determine identity of diagnostically relevant species though differences between spectra from normal and diseased tissue,\cite{Koljenovic:LI:2002} this is sometimes not possible even when the primary marker species is known.\cite{Chen:SciRep:2016} For this reason researchers often derive diagnostic information from tissue spectra using spectral pattern recognition approaches.

\subsubsection{Machine Learning on Spectral Features}
Many of the machine learning methods used for image feature selection (see Section \ref{Sec:FS}) are also used for spectral feature selection and classification.  Unsupervised feature selection methods such as principle component analysis (PCA, linear transform), singular value decomposition (SVD, linear transform), and kernel PCA (KPCA, nonlinear transform) generate orthonormal basis vectors containing correlated spectral features.\cite{Miljkovic:Anal:2010,Vajna2011} These basis functions may contain positive and negative spectral elements, so their interpretation is not always clear. However, they are useful for denoising or dimensionality reduction, in which only a subset of basis vectors are retained and spectral information of interest is reconstituted through linear combinations of the retained vectors.\cite{Butler2016} Other unsupervised feature selection methods, such as nonnegative matrix factorization\cite{Vajna2011,Masia2013} (NMF; also `positive matrix factorization'), positivity constrained multivariate curve resolution\cite{Andrew1998,Patel2011,Vajna2011,Zhang2013} (MCR) and vertex component analysis\cite{Miljkovic:Anal:2010,Alfonso-Garcia2017} (VCA), are designed to retrieve physically meaningful, positive-valued spectra from which abundance maps can be directly generated. MCR and NMF techniques return components of spectra that carry discriminating information, and may or may not return realistic spectra of molecular constituents, depending on their implementation. VCA assumes that there are `pure pixels' for each molecular entity (i.e., the unmixed spectrum of each pure chemical component is represented at least once among all the spectral pixels). Methods that are explicitly designed to return spectra corresponding to real molecules are termed `endmember extraction' or `spectral unmixing'.  

These unsupervised feature extraction methods frequently serve as the input to unsupervised and supervised classification methods. Unsupervised methods, such as K-means clustering\cite{Miljkovic:Anal:2010,Alfonso-Garcia2017} (KMC), divisive correlation cluster analysis (DCCA), and agglomerative hierarchical cluster analysis (AHCA), link together related spectra based on spectral proximity metrics.\cite{Miljkovic:Anal:2010} Supervised methods, such as linear discriminant analysis\cite{Austin:Analyst:2016} (LDA), partial least squares\cite{Berger1999,Bergholt:Gastroenterology:2014} (PLS), support vector machines\cite{Widjaja2008, Rosch2005} (SVM), and random forest classifiers,\cite{Teh2009, Kallenbach-Thieltges2013, Alfonso-Garcia2017} rely on training data to generate a classification model. These models may be conservative or relaxed according to clinical need. For applications such as tissue-conserving resection,\cite{Jermyn:OSA:2014} one may choose a conservative model to minimize false positives. Alternatively, if the aim is to reduce the burden on pathologists, one could design a model that may include false positives, but automatically classifies the most clear-cut cases, leaving the rest for inspection.\cite{Old:Analyst:2016} Training these models may be arduous, requiring many datasets and extensive validation; but post-training classification may be rapid.\cite{Bergholt:Gastroenterology:2014,Alfonso-Garcia2017} There are, additionally, `semi-supervised', methods that aim to incorporate known and unknown information into classification problems, such as hybrid linear analysis\cite{Berger1998} (HLA) semi-supervised PCA-linear discriminant analysis\cite{Lloyd2014} (PCA-LDA).

Deep learning\cite{LeCun2015} (DL) and artificial neural networks\cite{Gniadecka1997,Goodacre1998} (ANN) constitute an important family of methods that have received significant attention recently. These methods rely on layers of interconnected decision units or `neurons'; typically 1-2 layers for ANN approaches, and potentially many layers for DL approaches. Whether employed as supervised or unsupervised learning, the weights and the interconnection strengths of the neurons are iteratively modified to optimize a merit function. One of the primary strengths of ANNs and DL is that nonlinear and extremely complex interconnections can be created within the network that may not be possible with traditional linear or nonlinear methods such as those mentioned above. A significant challenge to ANNs and DL, as with all supervised techniques, is the requirement that the training set be comprehensive in that it is representative of all future data. Another challenge is that these approaches are `black box' in that it is not currently clear how to understand what data features are found to be important, and how they are used in the evolved ANN/DL neural connection architecture. Interpretation of ANNs and DL is an active area of research.\cite{Manescu2017}

\subsubsection{Spatio-Spectral Analysis}
The diagnostically relevant information contained in vibrational spectroscopy from tissue samples is useful for histology to the extent that it can be spatially resolved on relevant length scales . Raman and IR spectroscopies are both conducive to spatially resolved acquisition, and both have been used in an imaging modality with tissues.

\begin{figure}[h]
	\centering
	\includegraphics[width=9cm]{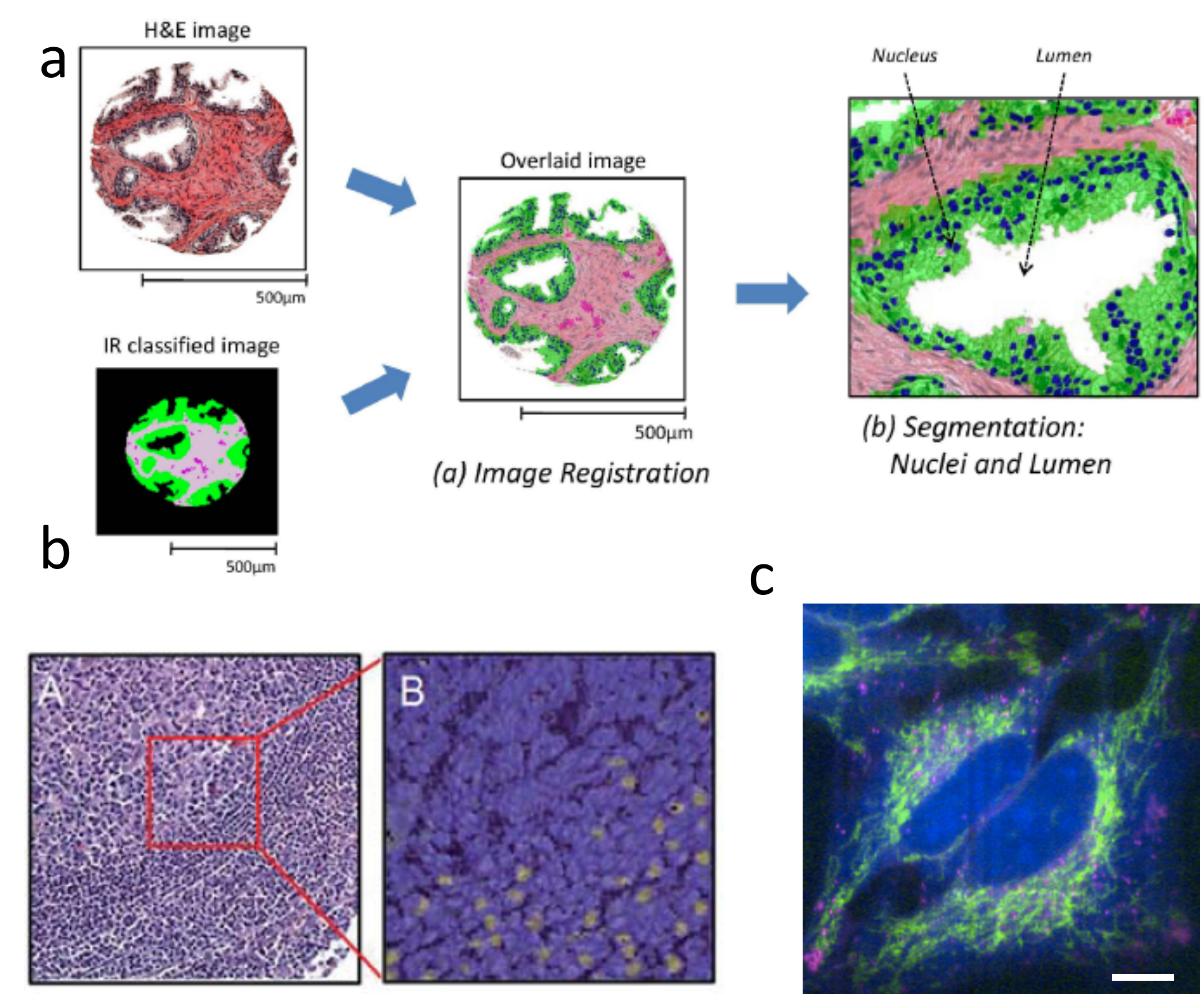}
	\caption{IR and Raman histopathology: (a) Infrared absorption spectra contain sufficient information for classification of tissue, and provide value through indicating regions of interest.\cite{Kwak:BmcCancer:2011} (b) A: H\&E stained section of lymph node germinal and mantle zone, and B: a sub-image with contrast from Raman spectroscopy showing activated lymphocytes in yellow.\cite{Diem:JPB:2013} (c) Spatially multiplexed Raman spectroscopy with 30 ms effective spectral acquisition time using  500 mW of 532 nm light. The cytochrome-c distribution is indicated in yellow.\cite{Okada:PNAS:2012}}
	\label{fgr:RamanImaging}
\end{figure}

Thanks to recent commercial availability of quantum cascade lasers, approaches\cite{Nasse:NatMeth:2011} for IR imaging at spatial resolution as high as $1\,\mu m$ are now feasible, making it possible to generate gross tissue maps based on IR absorption. Since the chemical specificity of IR spectroscopy is sufficient to easily discriminate among tissue regions such as epithelium and stroma,\cite{Pilling:Anal:2017} and between benign and malignant tissue,\cite{Fernandez:NatBio:2005} IR maps can be merged with H\&E images to indicate tissue regions of interest,\cite{Kwak2015} as shown schematically in Figure \ref{fgr:RamanImaging} a.  Morphological features extracted directly from such maps have been used to classify tissue samples as cancer or non-cancer, with high accuracy.\cite{Kwak:BmcCancer:2011} This automated ability could provide considerable value, as mentioned in Section \ref{Sec:CAD_Roles}.

From a spatial resolution perspective, Raman imaging has an important advantage over IR in that it uses visible or near infrared wavelengths, so can reach spatial resolution 2 to 3X higher than with IR, making cellular and sub-cellular imaging possible. The combination of high resolution and rich fingerprint spectra allows for significant information in addition to  providing classification of tissue regions. For example, Raman imaging can provide label-free identification of cell type,\cite{Chan:AnalChem:2009} distribution\cite{Matthaus:ApplSpec:2006,Miljkovic:Anal:2010} and content of organelles.\cite{Okada:PNAS:2012}  
Important functional information is also available, such as DNA to protein ratio of nuclei,\cite{Schulze:Anal:2013} the cell cycle state,\cite{Matthaus:ApplSpec:2006, Boydston:VibSpec:2005,Boydston:BioSpec:1999} and whether immune cells such as leukocytes are activated.\cite{Chan:AnalChem:2008,Diem:JPB:2013} Examples of cellular function and organelle distribution maps generated with Raman spectroscopy are shown in Fig.\ref{fgr:RamanImaging} b and c.

Although IR and spontaneous Raman imaging can provide useful diagnostic information, both have practical drawbacks. For example, conventional glass slides can be problematic for both methods as they emit fluorescent light when excited with visible and near IR light, and absorb IR light at wavelengths of 3.5 $\mu m$ and longer, which constitute the fingerprint region for IR. Consequently, both approaches often require special substrates - quartz or \ce{CaF} substrates for Raman studies, and \ce{CaF} or \ce{BaF_2} substrates are used in IR. On the other hand, glass slides are transmissive to IR wavelengths in the range $\lambda$= (2.8 to 3.5)$\mu m$, allowing only the CH stretch spectral region to be used.\cite{Pilling:Anal:2017} 
Another consideration is that the intrinsically low Raman scattering levels can be easily masked by fluorescence; thus, in addition to use of specialty substrates, it is often necessary to photobleach samples before acquiring Raman spectra.

The benefits of spectral image acquisition should be considered in context of required imaging time. These considerations differ significantly for IR and Raman spectral imaging. IR absorption cross sections for fingerprint and CH stretch modes are typically on the order of $10^{-18}cm^2$, so most of the light will be absorbed for a high density of absorber with (5 to 10) $\mu m$ path length. Having such a large effect on transmitted light, signal can be acquired very quickly. Yeh et al.\cite{Yeh:AnalChem:2015} describe work wherein 128 X 128 pixel wide field IR absorption images were acquired serially on a fixed region of the tissue while the laser wavelength was scanned. In total, 282 frames were acquired, yielding IR absorption images over a 1128 $cm^{-1}$ range in the fingerprint region, with an effective pixel rate of 1.6 MHz, and spectral acquisition rate of 3 kHz. In cases where fewer spectral points are required, some imaging speed increase can be obtained by using only several discrete frequencies. For example, Tiwari et al.\cite{Tiwari:AnalChem:2016} found that they could reliably identify diseased cardiovascular tissue regions with on-the-order of 10 discrete spectral points.

Allowing for imaging of $\sim$10 discrete contrast frequencies and a 5 $\mu m$ pixel size, tissue could be mapped at 0.5 $cm^2/min$, similar to WSI instruments currently in use. Further, if the discrete frequencies are exclusively in the \ce{CH} stretch region, glass slides could be used, and there would be minimal impact on the histology work flow, as IR imaging can be performed on H\&E stained slides.\cite{Pilling:Anal:2017}

In contrast to IR absorption, the Raman signal is quite small, and imaging with this contrast mechanism is proportionally slow. The differential scattering cross-section of Raman-scattered light collected of over solid angle ($\Omega$) is given by $d\sigma/d\Omega\propto|\partial \alpha/\partial Q|^2$, and is on the order of $10^{-30} cm^2\,sr^{-1}$ for most modes. Accordingly, spectral acquisition rates typically range from 5 to 0.01 Hz.\cite{Popp:AGIE:2016,Chen:SciRep:2016} Using spontaneous Raman scattering, one cannot excite only selected vibrational frequencies, so there is no benefit to discrete spectral imaging as in IR. On the other hand, some increase in acquisition speed can be achieved through spatial multiplexing. Figure \ref{fgr:RamanImaging} c displays a spontaneous Raman image acquired at a 90 Hz effective spectral acquisition rate\cite{Okada:PNAS:2012} through line-focused (rather than spot-focused) excitation light. Because practical detector arrays are presently limited to 2 dimensions, and one of those is devoted to spectral variations, spontaneous Raman signal is limited to 1 dimensional multiplexing such as line excitation. Of course, spatially multiplexed signal generation requires light sources with proportionally scaled power. 

Given the much slower acquisition speeds for spontaneous Raman scattering, it is difficult to find an obvious application in the histopathology work flow. For example, pixel sizes of 27 $\mu m$ and a 90 Hz spectral acquisition rate achieved through spatial multiplexing would provide appropriate sample throughput, but with such course resolution, one would give up significant benefit of sub-cellular chemical detail otherwise available from Raman scattering. Spontaneous Raman imaging could be useful in a guided subsampling mode, where only regions of special interest were imaged. On the other hand, even with spatial multiplexing and 90 Hz spectral acquisition, only 0.003\% of the total tissue area could be imaged at high resolution in the 2 minute time frame required to obtain a 1 $cm^2$ whole slide brightfield image. 

At this point it is appropriate to consider that the 2 minute / slide imaging time is the standard for single-contrast image formation, whereas Raman spectral images could contain most or all of the potentially desired contrasts in a single, label-free image. Nonetheless, it seems clear that some trade-off between coverage and resolution would be necessary if spontaneous Raman scattering is used. The severity of such a trade-off could be reduced by use of coherent Raman techniques, which provide equivalent spectroscopic information with imaging throughput that is closer to that of WSI.

\section{Coherent Raman Imaging}
\begin{figure}[h]
	\centering
	\includegraphics[width=9cm]{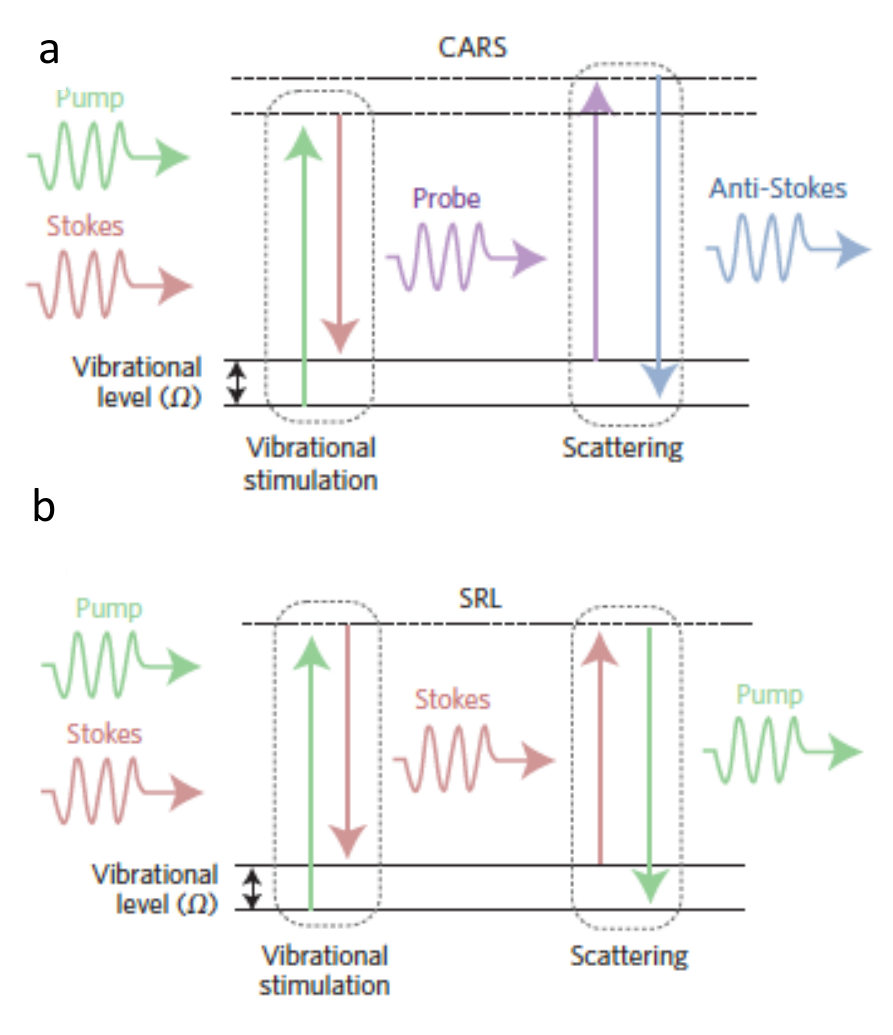}
	\caption{Jablonski diagrams for CARS and SRL.}
	\label{fgr:Jablonski}
\end{figure}

\subsection{Introduction and Mechanism}
Coherent Raman scattering was first predicted in 1962\cite{Armstrong:PR:1962} and first measured in 1964.\cite{Maker:PR:1965} In 1982, Duncan et al.\cite{Duncan1982} first demonstrated the coherent Raman scattering effect in a microscope, but the cross-beam (phase-matching) optical arrangement they used was challenging. In 1999 Zumbusch et al.\cite{Zumbusch1999} demonstrated that the CARS effect could be achieved with a simple, collinear beam geometry. This discovery gave rise to the now burgeoning field of coherent Raman imaging (CRI). Currently, the most prominent bioimaging CRI methods are stimulated Raman scattering (SRS) and coherent anti-Stokes Raman scattering (CARS). It is not our purpose to review the CRI field in its entirety; excellent reviews have been given recently.\cite{Volkmer:JPD:2005,Evans:AnalRev:2008,Pezacki:NCB:2011,Yue:BPJ:2012,CampJr:NatPhoton:2015} Instead, we focus on applications of CRI to histopathology, and associated technical aspects that facilitate or hinder such applications.

We can gain insight into coherent Raman scattering by considering that an oscillating electric field of light induces a polarization in a substance that can be described as:
\begin{equation}
P(t)=N\alpha(t)E\sin(\omega t)\label{Eq:GenPol}
\end{equation}
where $P$ is an induced polarization, $N$ is the number of molecules in the sample volume, $\alpha$ is the molecular polarizability, and $E$ is the electric field amplitude of the light. As with spontaneous Raman scattering, the coherent Raman cross-section depends on changes of $\alpha$ with position along the normal mode coordinate, $Q$. We consider $\alpha$ in terms of its Taylor expansion about $Q=0$.

\begin{equation}
\alpha(t)\approx \alpha_0 +\sum_{m}\sum_{n}\frac{1}{n!}\left.\frac{\partial^n\alpha}{\partial^n Q_m}\right|_0 Q_m^n(t)\label{Eq:alpha_series}
\end{equation}
where $n$ is the expansion order and $m$ indexes normal modes.

In coherent scattering, vibrational modes are driven by external light fields. When the difference in energy between a pump (p) and Stokes (S) field is equal to the energy difference ($\Omega_v$) between two adjacent quantum levels of the vibrational mode, i.e., when $\Omega_m=\Delta\omega=\omega_p-\omega_S$, energy can be transferred to that vibrational mode. While the primary fields, $E(t)=E_je^{-i\omega_j t}$, (j=p,S), $\omega_p>\omega_S\gg\Omega_m$, are at much too high a frequency to directly induce significant molecular response, electrons are able to follow the fields adiabatically. Through its nonlinearity, the electron response will contain some amplitude at the difference frequency, $\Delta\omega$. This component of the response will in turn exert an oscillatory force $F(t)$ on the molecule:\cite{Potma:JBook:2013}

\begin{equation}
F(t)=\left.\frac{\partial \alpha}{\partial Q_m}\right|_0\left [E_pE_S^\star e^{-i\Delta\omega t}+c.c.\right]\label{Eq:F(t)}
\end{equation}
which will drive motion along the coordinate $Q$ that can be described by: 
\begin{equation}
\frac{d^2Q_m(t)}{dt^2}+2\gamma \frac{dQ_m(t)}{dt}+\Omega_mQ_m(t)=\frac{F(t)}{\mu}
\end{equation}
where $\Omega_m$ is the resonant frequency, $\mu$ is the reduced mass of the vibrational mode (not to be confused with the absorption dipole described previously), and $\gamma$ is a damping constant. $Q(t)$ is then given as:
\begin{equation}
Q_m(t)=\frac{1}{\mu_m}\left.\frac{\partial \alpha}{\partial Q_m}\right|_0\frac{E_pE_S^\star e^{-i\Omega_m t}}{\Omega_m^2-\Delta\omega^2-2i\Delta\omega\gamma_m}\label{Eq:Q(t)}
\end{equation}
Substituting Eq.(\ref{Eq:Q(t)}) and a first-order truncation of Eq.(\ref{Eq:alpha_series}) into Eq.(\ref{Eq:GenPol}), we obtain an expression for the polarizability of a given mode in presence of a pump and Stokes field: 
\begin{equation}
\alpha_m(\Delta\omega)=\frac{1}{\mu_m}\frac{(\partial \alpha /\partial Q_m)^2_0E_pE_S^\star}{\Omega_m^2-\Delta\omega^2-2i\Delta\omega\gamma_m}\label{Eq:Amp_Pcoh}
\end{equation}
and the resonant susceptibility is given by the following sum over normal modes: 
\begin{equation}
\chi^{(3)}_{R}(\Delta\omega)=\frac{N_0}{6\epsilon_0}\sum_m\frac{1}{\mu_m}\frac{(\partial \alpha /\partial Q_m)^2_0}{\Omega_m-(\Delta\omega)-2i\Delta\omega\gamma_m},
\end{equation}
where, by convention, $N_0$ is the number density of molecular oscillators in a pure substance. We see that $\chi^{(3)}_R$ is complex and dissipative, so will have an overall spectral phase that varies with $\Omega$. The imaginary component of $\chi^{(3)}_R$ has essentially identical\cite{Tolles:AppSpec:1977} line shape to the spontaneous Raman response. Molecular systems have an additional, nonresonant, susceptibility, $\chi^{(3)}_{NR}$, that arises directly through nonlinearity in the electronic response to the driving laser fields. The overall third order susceptibility is then given by:
\begin{equation}
\chi^{(3)}(\Delta\omega)=\chi^{(3)}_{NR}(\Delta\omega)+\chi^{(3)}_{R}(\Delta\omega)
\end{equation}

In all third order coherent Raman processes, a pump and a Stokes field interact with a medium according to:
\begin{equation}
C(\Delta\omega)=\chi^{(3)}(\Delta\omega)\left[E_S \star E_p\right](\Delta\omega)\label{Eq:coh_gen}
\end{equation}
where $E_p$, $E_S$ are the frequency-domain pump, and Stokes fields, respectively and $\star$ is the cross-correlation operator.

\subsection{Coherent Anti-Stokes Raman Scattering (CARS)}
Coherent anti-Stokes Raman scattering is generated by interaction of a probe field with the vibrational coherence generated through the process described by Eq.(\ref{Eq:coh_gen}), as represented by the Jablonski diagram in Fig. \ref{fgr:Jablonski} a, and by the following expression:

\begin{equation}
I_{CARS}(\omega)\propto \left| \left [ \left \{ \chi^{(3)}\left[E_S\star E_p\right]\right \} \ast E_{pr}\right ] (\omega)\right|^2
\end{equation}
where $E_{pr}$ is the frequency-domain probe field, `$\ast$' is the convolution operator. For pump, Stokes, and probe sources with center frequencies of $\omega_{p}$, $\omega_{S}$, and $\omega_{pr}$, respectively, the output anti-Stokes frequency $\omega_{as}=\omega=\omega_{pr} + \omega_p -\omega_S = \omega_{pr} + \Delta\omega$. In the limit that the probe pulse is spectrally narrow, we can treat it as a delta function, so the CARS signal contains the same spectral information that is contained in the spontaneous Raman signal. However, there are important differences between CARS and spontaneous Raman scattering. One of these is that the CARS signal is blue-shifted with respect to the excitation light, so does not compete with autofluorescence from the sample. Also, the CARS signal is forward-scattered, and contained in a narrow cone angle, making it a simple matter to collect a large fraction of the CARS light in a transmission geometry. 

Another important difference is that CARS signal contains a nonresonant background (NRB), arising from $\chi_{NR}^{(3)}$. It is termed "nonresonant" because it arises from the electronic response, and is not sensitive to vibrational resonances. Since the electrons adiabatically follow the driving field of the laser, the NRB is always in phase with the driving field, and retains a fixed phase relationship with the resonant response. The overall CARS signal has a component that is a coherent mixing between these two terms:
\begin{equation}\label{Eq:Heterodyne3Terms}
I_{CARS}(\omega)\propto\left|\chi^{(3)}_{NR}(\omega)\right|^2 +\left|\chi^{(3)}_{R}(\omega)\right|^2+2 \chi^{(3)}_{NR}(\omega)Re\left[\chi^{(3)}_{R}(\omega)\right ]
\end{equation}

Owing to the three terms in Eq. (\ref{Eq:Heterodyne3Terms}), the qualitative nature of the raw CARS signal depends on the ratio of resonant to nonresonant response. $\chi^{(3)}_{NR}$ is of similar magnitude for most biological materials but $\chi^{(3)}_R$ may vary widely owing to the density of molecular oscillators, $\tilde{N}$ (fraction of $N_0$ within the focal volume) and the Raman scattering cross-section of individual vibrational modes. Typically, $\chi^{(3)}_{NR}\leq 0.1 \tilde{N}\chi^{(3)}_R$ for CH stretch resonances and $I_{CARS}\propto \tilde{N}^2$. However, for many fingerprint peaks,  $\chi^{(3)}_{NR}\ll \tilde{N}\chi^{(3)}_R$ and the resonant contribution to the $I_{CARS}$ is linear in $\tilde{N}$, but NRB dominates the signal. This is why non-spectroscopic CARS is performed almost exclusively with CH stretch resonances as contrast. In fact, considering only the term $|\tilde{N}\chi^{(3)}_{R}|^2$, Cui et al.\cite{Cui:OptLett:2009} demonstrated that the resonant CARS signal amplitude is greater than spontaneous Raman \textit{only} when oscillators are very concentrated, such as in bulk phase, and samples rich in lipid or structural protein.

Another important effect of the coherent interaction between resonant and non-resonant signal components arises from the Gouy phase shift, wherein a phase shift accrues with field propagation beginning at the focus of a Gaussian beam compared to propagation of a plane wave. This becomes a problem at sharp interfaces of features smaller than the Rayleigh range when resonant signal is generated only by light on one side of the focus while the non-resonant signal is generated at positions on both sides of the focus. When these mix, there is a spatial position-dependent spectral phase error\cite{Popov:OE:2011} causing spectral distortion. However, the Gouy phase is not a problem when $\chi^{(3)}_{NR}\ll \tilde{N}\chi^{(3)}_R$, or when there are no sharp spatial discontinuities in resonant signal contribution.\cite{Barlow:OE:2013}

\subsubsection{CARS With CH-Stretch Contrast}
In the CH-stretch spectral range there are four main peaks, located at 2850 $cm^{-1}$, 2880 $cm^{-1}$, 2920 $cm^{-1}$, and 2960 $cm^{-1}$. These correspond to symmetric and assymetric stretching of \ce{CH2} ($\nu_{\ce{CH2},sym}$, $\nu_{\ce{CH2},asym}$), and symmetric and assymetric stretching of \ce{CH3} ($\nu_{\ce{CH3},sym}$, $\nu_{\ce{CH3},asym}$), respectively. Because proteins have primarily \ce{CH3} moieties and lipids have many more \ce{CH2} groups, these peaks are convenient for estimating the relative abundance of protein and lipid. The information content in (symmetric, asymmetric) peak pairs is largely redundant, so only one peak from each pair is typically used. The water \ce{OH} stretch has a broad resonance between 3000 $cm^{-1}$ and 3600 $cm^{-1}$ that is also used in some studies.

CARS imaging with contrast from the CH stretch region can be performed by narrowband,\cite{Zumbusch1999} spectral focusing,\cite{Hellerer:APL:2004} and spectroscopic techniques.\cite{Chen:JPCB:2002,Marks2004,Kee2004,Kano:APL:2004} In narrowband CARS, a pair of picosecond pulses is used to excite coherence of a single selected vibrational band, and the higher frequency field is used to probe the coherence. Meyer et al.\cite{Meyer:AnalChem:2013} found that, for strong resonant signals ($\chi^{(3)}_{R}\geq 10\, N\chi^{(3)}_{NR}$), simple subtraction of NRB is sufficient to render a quantitative signal with no significant distortions to the CARS peak positions and amplitudes. Wu et al.\cite{Wu:AnalChem:2009} have shown quantitative correspondence between $\nu_{\ce{CH2},sym}$ signals and lipid content in liver tissue. Further, large resonant signals in the CH-stretch enables rapid image acquisition, with pixel rates as high as 6 MHz,\cite{Evans:PNAS:2005} but more typically range form 200 kHz\cite{Veilleux:IEEE:2008} to 1 MHz\cite{Meyer:AnalChem:2013} for tissue samples. Repeated images can be acquired at several discrete Raman shifts to map scenes with some chemical complexity. For for discriminating species that are spectrally distinct, like water and a protein or a lipid, a single additional scan is generally sufficient for each component.\cite{Lin:AJP:2003} A larger number of spectral contrast points may be necessary to discriminate species with only subtle differences.\cite{Meyer:AnalChem:2013} Given the high pixel acquisition rates, the necessity of acquiring several scans may not itself be onerous, but latency between scans for wavelength tuning can add significant time to image acquisition for most CARS systems. Rapid spectral scan approaches have been demonstrated for narrowband CARS with only 100 $\mu s$ required to scan from one frequency to another in the CH stretch region, making inter-scan latency a non-issue for such systems.\cite{Begin:BOE:2011}

Spectral focusing is another approach that facilitates spectroscopic scanning with narrowband CARS. In spectral focusing CARS,\cite{Hellerer:APL:2004} a pair of pulses that are spectrally broad and centered at distinct wavelengths are temporally broadened with nominally identical chirp. Although the absolute frequencies of the two light fields impinging on the sample changes significantly with time, variation in the difference of frequencies is minimized so a single vibrational state can be excited. Using this approach, it is feasible to collect single-frequency image pixels at 250 kHz.\cite{Pegoraro:OpticsExpress:2009} This approach to CARS can be performed with sub-picosecond pulses and spectral tuning over a limited range (typically 500 $cm^{-1}$) can be accomplished by simply changing the delay of one pulse relative to the other. When spectral focusing or another CARS approach is performed at a series of Raman shifts, the effects of the NRB can be separated through spectral phase retrieval approaches.\cite{Vartiainen:JOSAB:1992,Liu2009} 

Spectroscopic CARS differs from narrowband or spectral focusing CARS in that a broad swath of spectrum is obtained in each laser shot. Such an approach in the CH-stretch region has been demonstrated with picosecond / femtosecond CARS\cite{Muller:JCPB:2004,Chen:JPCB:2002,Kee2004,Kano:APL:2004} and  and with nonlinear interferometric vibrational imaging (NIVI).\cite{Benalcazar:IEEE:2010} Histological studies focused on CH-stretch using spectroscopic CARS have been performed primarily with  NIVI. From tissues, NIVI typically yields CH-stretch spectra with a bandwidth of $\approx250\, cm^{-1}$ at a 20 Hz spectral acquisition rate (1 kHz effective rate for a single frequency assuming 5 $cm^{-1}$ resolution). These are undistorted Raman spectra, linear in $N$.\cite{Benalcazar:IEEE:2010}

Several studies have evaluated the utility of CARS in the CH stretch region for generating contrast equivalent to H\&E stain. Evans et al.\cite{Evans:OpEx:2007} compared CARS images to H\&E stained images, finding that CARS could provide similar image information when $\nu_{\ce{CH2},sym}$ and $\nu_{\ce{CH3},sym}$ or $\nu_{\ce{CH3},asym}$ were used. Similar "pseudo H\&E" obtained through intravital CARS imaging\cite{Lee:Intravital:2015} enabled inspection of tumor environments and margins without staining, and no obvious evidence of photodamage even after 300 frames.

Quantitative tissue morphology metrics can be obtained from CARS contrast, similar to what is done with H\&E-contrast images. Meyer et al.\cite{Meyer:JBO:2012} compare CARS with Raman in brain, demonstrating ability to extract nucleus-to-cytoplasm ratio, cell density, nucleus size and shape. They found, however, that since nuclei have negative $\nu_{\ce{CH2},sym}$ contrast, nuclear density and shape metrics may not be as reliable as H\&E. On the other hand, the Wong group have used primarily $\nu_{\ce{CH2},sym}$ CARS contrast and image segmentation \cite{Gao:JBO:2011,Gao:ArchPathLabMed:2012,Yang:BME:2011} to determine nuclear size, cell volume, and cell-cell distance and other metrics to classify cancer subtypes in lung and breast tissue, producing encouraging results for differential diagnosis of cancer in these tissues. Uckermann et al.\cite{Uckermann:PloS:2014} also compared metrics from CARS and H\&E, finding that CARS provided additional relevant information via the signal amplitude.

In addition to conveying morphological information through imitating H\&E stain, CARS $\nu_{\ce{CH2}}$ and $\nu_{\ce{CH3}}$ contrast has been used to study lipid metabolism in cancer,\cite{Le2009} and to discriminate cell type based on lipid content. Excess lipids were found as a potential marker for circulating tumor cells,\cite{Mitra:BmcCancer:2012} and Evans et al.\cite{Evans:OpEx:2007} found contrast in $\nu_{\ce{CH2},sym}$ to evince replacement of normal white matter with lipid deficient astrocytic glioma tissue. Chowdary et al.\cite{Chowdary2010} found spectral differences arising from variation in lipid to protein ratio in a rat breast cancer model, and were able to use SVD to automatically segment normal and diseased tissue regions at a 99\% confidence level. Similarly, Uckermann et al.\cite{Uckermann:PloS:2014} also found a $\nu_{\ce{CH2},sym}$ contrast gradient in the infiltrative zone around glioblastoma, with rather distinct contrast levels in the normal region. They also found that levels of $\nu_{\ce{CH2},sym}$ contrast could be used to distinguish between glioblastoma, and metasticized breast tumor in the brain. In a similar finding, Meyer et al.\cite{Meyer:JBO:2011} showed that CARS $\nu_{\ce{CH2},sym}$ contrast in the brain could highlight metastatic tumor regions of lung origin as the metastases retain a chemical profile similar to the originating tissue. Their finding was supported by Raman and IR imaging.   

While narrowband CARS contrast from $\nu_{\ce{CH2}}$ and $\nu_{\ce{CH3}}$ alone is useful in many circumstances, it can also be combined synergistically with contrast from other methods. Yue et al. used CARS signal from $\nu_{\ce{CH3},sym}$ to visualize mammary acini, providing guidance for acquisition of spontaneous Raman spectra. From subsequent spectral analysis, they observe changes in spatial distribution of lipids concomitant with a loss of basoapical polarity in mammary acuinus, a transformation linked to early stages of carcinogens.\cite{Yue:BPJ:2012} The Popp group have taken similar approaches.\cite{Meyer:JBO:2011,Heuke:BJD:2013,Meyer:AnalChem:2013,Popp:AGIE:2016} In comparing CARS to spontaneous Raman imaging of colon sections, Krafft et al.\cite{Krafft2009} found that CARS images were consistent with expectations from spontaneous Raman scattering in the same spatial sample regions, and that with just a few CARS scans in the range 3000 to 1000 $cm^{-1}$, there was sufficient information to discriminate a couple of cell types and identify regions of normal and diseased tissue. The CARS images were acquired $10^6$ times faster than spontaneous Raman (10 $\mu s$ vs 30 s / pixel). Subsequently, they have used additional CARS peaks, in conjunction with other contrast mechanisms to characterize tissues. The same pulses that generate the CARS signal can also give rise to second harmonic generation (SHG) in collagen fibrils, and to two-photon excited fluorescence (TPEF) from intrinsic fluorophores such as NADPH, flavins, collagen, and elastin. Meyer et al.\cite{Meyer:JBO:2011,Meyer:AnalChem:2013} used CARS contrast at multiple Raman shifts along with intrinsic contrast from TPEF and SHG to detect white matter, the granule layer, elastin, ordered collagen, and  NAD(P)H in brain, and elastic fibers, triglycerides, collagen, myelin, cellular cytoplasm, and lipid droplets in perivascular tissue.   

\subsubsection{CARS With Fingerprint Contrast}
As we have seen above, CH-stretch contrast is useful for morphology and some species quantification. However, contrast from fingerprint spectra can provide abundance information for many more species of diagnostic value. Some examples are shown in Table \ref{tbl:VibMarkers}, including carotenoids, cholesterol esters, glycogen, collagen, keratin and elastin. Additionally, fingerprint spectra can positively identify cell phenotype.\cite{Chan:AnalChem:2008,Chan:AnalChem:2009} 

Because the fingerprint peaks are typically much weaker than the nonresonant signal, it is not possible to obtain quantitative spectral information by simply subtracting the NRB. Because typically $\chi^{(3)}_R\ll\chi^{(3)}_{NR}$ for fingerprint resonances, the third (cross) term in Eq. (\ref{Eq:Heterodyne3Terms}) dominates the resonant contribution to the CARS signal, and the overall spectral phase must be retrieved in order to extract a Raman spectrum. CARS phase retrieval is possible with an approach based on maximum entropy methods\cite{Vartiainen:JOSAB:1992} or one based on time-domain Kramers-Kronig.\cite{Liu2009} Both provide functionally equivalent results, but the computations for the Kramers-Kronig approach require considerably less time.\cite{Cicerone:JRS:2012} It has recently been appreciated that the estimate of the NRB shape estimate required by both of these methods can have a profound impact on extracted Raman peak amplitudes. Camp et al.\cite{Camp2015} showed that symmetry relations between the NRB and resonant signal can be used to unambiguously determine the NRB shape, and thus the relative peak amplitudes of the retrieved Raman spectrum. In this way, the NRB acts as an internal calibration for spectroscopic CARS, and using this internal calibration, retrieved peak ratios are absolute. 

Spectral acquisition of CARS signal is required for recovery of Raman fingerprint signals through phase retrieval, and also facilitates system noise removal via singular value decomposition.\cite{Rajwade:IEEE:2013} Spectrally focused CARS\cite{Pegoraro:OpticsExpress:2009} can be delay-scanned used to generate spectroscopic images that are amenable to phase retrieval\cite{Pegoraro:JBP:2011} Additionally, several inherently spectroscopic CARS microscopy approaches have been developed, but most are too inefficient for biological imaging. Here we will exclusively consider approaches that have been used to generate images from cells and tissues. The first of these was based on a combination of spectrally broad but uncompressed Stokes pulses, and spectrally narrow probe pulses.\cite{Kee2004,Kano:APL:2004} This approach generates broadband CARS spectra with up to 3000 $cm^{-1}$ bandwidth at $\approx$20 Hz spectral acquisition rate (12 kHz single spectral element, assuming 5 $cm^{-1}$ resolution), with strong CH stretch but quite weak fingerprint spectra.\cite{Parekh:BJ:2010} The quality of fingerprint spectra was improved significantly in one case by implementation of a narrow-bandwidth sub-nanosecond probe,\cite{Okuno:ACIE:2010} and in another by using a temporally compressed Stokes source.\cite{CampJr2014} In the latter work, a nearly transform-limited broadband pulse acts as both pump and Stokes, exciting vibrational coherence much more efficiently than separate pulses. This approach yielded CARS spectra in both CH and fingerprint regions with uniformly high SNR across the spectrum, and 280 Hz spectral acquisition rate (corresponding to 170 kHz single spectral element acquisition, assuming 5 $cm^{-1}$ resolution).\cite{CampJr2014}

Perhaps because it has only recently become possible to acquire high quality fingerprint spectra in an imaging modality, there is very little work on histological samples using fingerprint CARS. With relatively weak fingerprint signal, Pohling:\cite{Pohling:BMOE:2011} was able to use principal component analysis to discriminate between gray and white matter, and identify layers of granule and Purkinje cells in mouse brain tissue. Their structural assignments agreed with those obtained from H\&E staining. Also in murine brain tissue, but with significantly better SNR in the fingerprint due to impulsive coherence generation, Camp et al.\cite{CampJr2014} were able to identify many diagnostically important species and features directly from spectral peaks, without multivariate analysis techniques. These included red blood cells through the 1548 $cm^{-1}$ and 1565 $cm^{-1}$ hemoglobin peaks (C-C stretch), as well as collagen, elastin, and nucleotides through peaks listed in Table \ref{tbl:VibMarkers} for each of these species. They were able to identify tumor margins, as there were spectral differences between normal and  invasive glioblastoma image regions, such as at 785 $cm^{-1}$ (nucleotide), 1004 $cm^{-1}$ (phenylalanine ring breathing), and 2956 $cm^{-1}$ ($CH_3$ stretch). In both the Pohling and Camp work, a tumor-associated reduction in $\nu_{\ce{CH2}}$  was observed, as in CARS studies that focused on CH-stretch. 

In the mouse glioblastoma work, Camp et al.\cite{CampJr2014} obtained SHG and TPEF signals simultaneously with fingerprint CARS spectra. Although SHG and TPEF are typically used to identify collagen and elastin\cite{Meyer:JBO:2011,Lee2007} the multi-peak spectral identification of these two proteins did not agree particularly well with the SHG and TFEP images. They noted that SHG and TPEF provide uncertain chemical specificity, as other biologically
relevant molecular species (such as NADH) are also known to generate a TPEF response, and collagen generates a SHG response only when it is coiled in a triple helix.

\begin{figure*}
	\includegraphics[width=18cm]{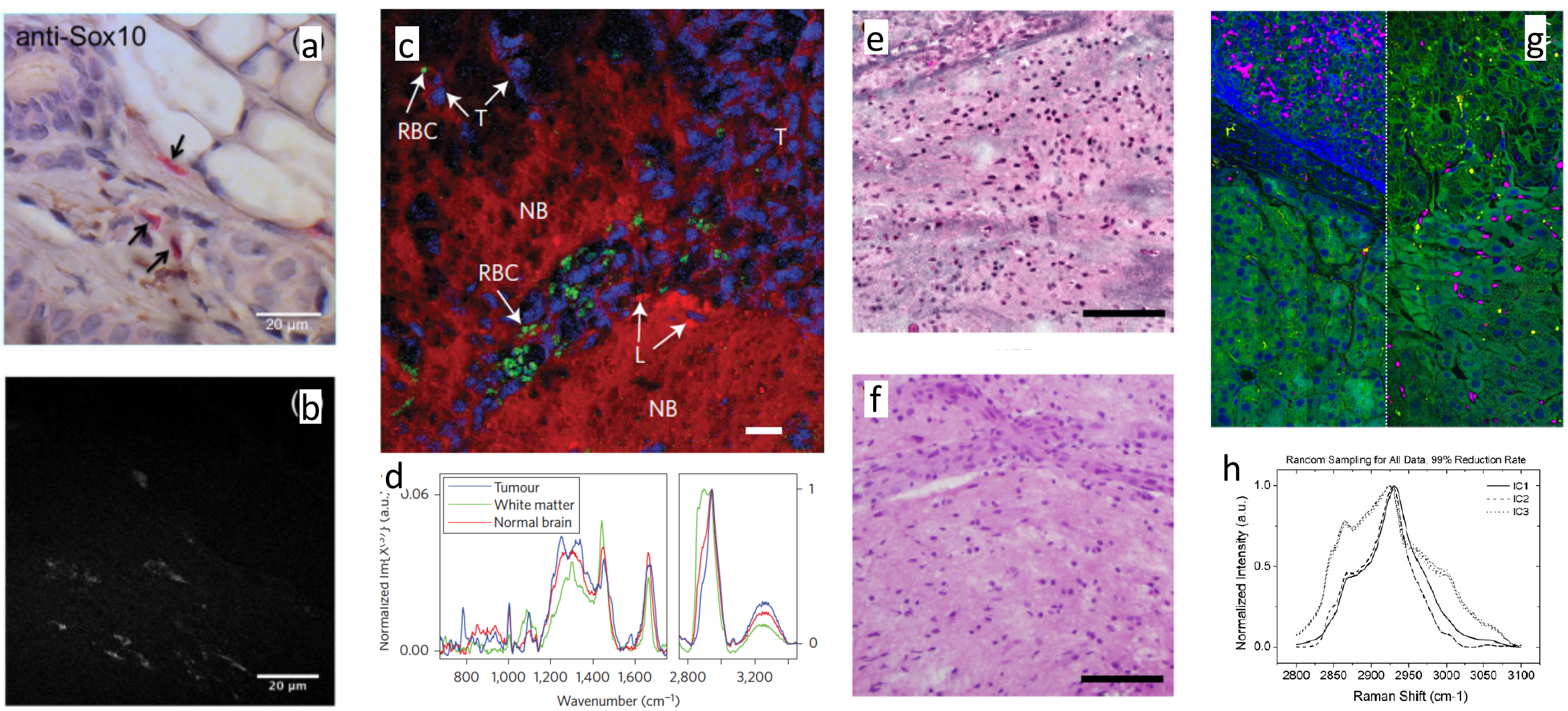}
	\caption{a \& b:\cite{Wang:SciRep:2016} (a) Immunohistochemical stain for Sox-10 in mouse ear with haematoxylin counterstain, revealing melanocytes in red. (b) CARS image of adjacent slice to (a), showing pigmented regions, consistent with positive staining in (a). c\& d:\cite{CampJr2014} (c) Pseudocolour BCARS image of tumor and normal brain tissue, with nuclei highlighted in blue, lipid content in red and red blood cells in green. (d) Raman spectra retrieved from (c), with each spectrum taken from one pixel, and acquired in 3.5 ms. e \& f\cite{Orringer:NatBME:2017} SRH image (e) is compared to a similar section of tumor imaged after formalin-fixation, paraffin-embedding and H\&E staining (f). Comparison was made using a lookup table to match SRH "staining" with H\&E. g \& h\cite{Otsuka:BiOS:2014} (g) Spectroscopic SRS images of pancreas and liver (left and right respectively). (d) Independent component spectra retrieved from (g) with 99\% data reduction}
	\label{fgr:Coherent}
\end{figure*}

\subsection{Stimulated Raman Scattering (SRS)}
In stimulated Raman scattering (SRS), energy transfers from a higher frequency (pump) field to a lower frequency (Stokes) field when the frequency differences between the fields is equal to that of a vibrational resonance in the material being probed. This effect can be monitored as stimulate Raman loss (SRL) in the pump field, or stimulated Raman gain (SRG) in the Stokes field. An expression for SRL can be given as:  
\begin{equation}
I_{SRL}(\omega)\propto\left| \left [ \left \{ \chi^{(3)} \left[E_S \star E_p\right]\right \} \ast E_{S}\right ] (\omega)+E_p(\omega)\right|^2
\end{equation}

The theoretically achievable signal-to-noise ratio for SRS is the same as that of CARS,\cite{Ozeki:OptEx:2009} and a related value, the sensitivity limit, seems to be similar for both approaches.\cite{CampJr2014}
One important difference between SRS and CARS is that the SRS signal contains only the imaginary component of the complex coherent Raman response. Thus, the nonresonant component that is problematic for narrowband CARS is not detected, and one directly obtains a signal proportional to the spontaneous Raman scattering cross section. (See Eq.(\ref{Eq:Amp_Pcoh}))

\begin{equation}
I_{SRL}(\omega)\propto \Im\left\{\chi^{(3)}(\omega)\right\}I_p I_S
\end{equation}

For field strengths permissible in tissue microscopy, the relative change in pump or Stokes intensity is on the order of $10^{-6}$ to $10^{-8}$, even for the strong peaks in the CH stretch region. These small signals cannot be spectrally separated from the intense excitation light as in CARS, so must be discriminated by other means, such as lock-in or balanced detection methods. Under favorable conditions, pixel acquisition rates up to 7 MHz have been demonstrated with contrast from Raman peaks in the CH stretch region.\cite{Saar:Science:2011} However, 30 - 40 kHz pixel rates are more common for single spectral bands in tissue samples.\cite{Freudiger2012,Orringer:NatBME:2017} Furthermore, even the strongest fingerprint peaks, such as found at 1440 \& 1660 $cm^{-1}$ (\ce{CH2} deformation \& \ce{C=C} stretch) are challenging to use for image contrast with SRS. Ozeki et al.\cite{Ozeki:NatPhot:2012} were unable to directly use the 1660 $cm^{-1}$ peak for contrast, even when pixel dwell times were 10$\times$ longer than necessary for good SNR in the CH-stretch region.

While the NRB does not interfere with the SRS signal, other background signals can. These include two photon absorption (TPA), cross phase modulation and thermal lensing. Signals from these sources can have amplitude comparable to or greater than even the largest SRS signals. For example, in skin, the TPA signal exceeds that of CH stretch SRS by almost 100-fold.\cite{Wang:SciRep:2016} Not surprisingly, these background signals can easily mask weak contrast from the fingerprint region, and careful measures may be taken to suppress their effects and recover some of the stronger fingerprint signals.\cite{Berto:PRL:2014,Fu:JACS:2016} 

Because the SRS signal must be extracted directly from the laser field, lasers with very low residual intensity noise (RIN) are preferred. These are typically bulk lasers. However, fiber lasers have a number of qualities that make them desirable in a clinical setting, including being robust to physical movement, vibration, and temperature shifts, and requiring very little maintenance. A low-noise fiber source has been developed specifically for SRS\cite{Freudiger:NatPhot:2014} although even for this laser, it is necessary to use both balanced detection and lock-in approaches to recover the SRS signal.

\subsubsection{SRS With CH-Stretch Contrast}

As with CARS, SRS imaging in the CH-stretch region has been performed with several approaches, including dual narrowband pulses,\cite{Nandakumar:NJP:2009,Freudiger2008} spectral focusing,\cite{Fu:JPCB:2013} rapid spectral scanning,\cite{Ozeki:NatPhot:2012,Alshaykh:OptLett:2017} and simultaneous multi-band imaging.\cite{Liao:Light:2015} The same chemical information can be obtained with CARS and SRS in the CH-stretch region. Accordingly, the types of histology studies with SRS are quite similar to those of CARS. 


Using only contrast from the $\nu_{\ce{CH3},sym}$ band and morphological clues, Mittal et al.\cite{Mittal:LasersInSurgeryAndMedicine:2013} were able to identify cell membranes, nuclei, collagen and keratin in tissues containing squamous cell carcinoma. In that study, the authors found that SRS images from this single band provided similar information to that found in H\&E images.

Lu et al.\cite{Lu:CancerRes:2016} compared images from H\&E staining to those generated with SRS contrast from $\nu_{\ce{CH2},sym}$ and $\nu_{\ce{CH3},sym}$. They found that cell density counts from SRS and H\&E images were indistinguishable. However, using the lipid/protein ratio from the SRS signal, and confirming this with luxol fast blue staining for the myelin sheath, they found that the two SRS bands provided a better distinction between white and gray matter than could be achieved with H\&E staining. They were also able to identify red blood cells and vessels through morphological features of images with contrast at 2800 $cm^{-1}$, in the nominally quiescent region. The contrast there may have arisen from TPA of heme. They also found that necrotic regions were characterized by lack of nuclei and lower protein signal, and banded regions of high protein signal appeared to be collagen.


Ji et al.\cite{Ji:SciTransMed:2015} used SRS contrast for $\nu_{\ce{CH2}, symm}$ and $\nu_{\ce{CH3}, symm}$ in fresh surgical brain tissues from 22 patients to create a classifier based on cellularity, axnoal density, and protein to lipid ratios, which provided tumor infiltration detection with 97.5\% sensitivity and 98.5\% specificity. They found that SRS images were superior to H\&E images for detecting infiltrating glioma, and that SRS microscopy could detect invasive tumor cells in peri-tumor brain tissue that appears grossly normal.

Orringer et al.\cite{Orringer:NatBME:2017} used $\nu_{\ce{CH2}, symm}$ and $\nu_{\ce{CH3}, symm}$ bands from SRS to recapitulate image contrast from H\&E, but on unprocessed tissue, with an eye towards improving intrasurgical tissue sampling.  Pathologist diagnosis based on proximally located SRS and H\&E tissue samples from 30 patients was compared and found to agree 92\% of the time. They also used computer algorithm-derived image attributes (based on SRS contrast) to successfully distinguish between diagnostic classes of tumors with 90\% accuracy.


Cui et al.\cite{Cui:SPIE:2017} used SRS spectra in the region (2800 to 3000) $cm^{-1}$ and multivariate curve resolution to find a correlation between cholesterol ester content in intracellular lipid droplets and Gleason scores of prostate tissue. Using a calibrated surrogate mixture for cellular lipid content, they found that the ratio between a band assigned to cholesterol at 2870 $cm^{-1}$ and the $\nu_{\ce{CH2},sym}$ band was proportional to the fraction of total lipids due to cholesterol ester.

While full spectral information in the CH-stretch region is sometimes useful, a small number of peaks often provide sufficient discrimination. Otsuka et al.\cite{Otsuka:BiOS:2014} examined tissues from a tumor-graft mouse model using spectrally scanned SRS spectrally scanned over the range (2800 to 3100) $cm^{-1}$. They found that through independent component analysis, they could reduce their spectral data set by $>$ 95\%, and still obtain images in which nucleus, zymogen granules, lipid droplets, and red blood cells could be visualized, and irregular glandular cell alignments detected. They found that images from the reduced spectral data was qualitatively equivalent to H\&E stained images. Similarly, Egawa et al.\cite{Egawa:JBO:2016} used SRS spectroscopy in the range (2800 to 3100) $cm^{-1}$ to map spatially resolved keratinocyte differentiation, and distinguish these from  Langerhans cells in the spinous layer of the epidermis. The key differences were primarily morphological, as both cell shape and nuclear prominence differ between the cell types, so the important contrast came at $\nu_{\ce{CH2},symm}$ and $\nu_{\ce{CH3},symm}$ bands. 


In other studies, vibrational signal from SRS has been used in concert with other nonlinear contrast mechanisms. Freudiger et al.\cite{Freudiger2012} combined $\nu_{\ce{CH2},sym}$ and $\nu_{\ce{CH3},sym}$ bands with hemoglobin TPA to derive contrast of roughly equal quality to H\&E-stained sections, recapitulating major histologic features in tissue from several brain regions. Galli et al.\cite{Galli:JBPhot:2016} used $\nu_{\ce{CH2}}$ in conjunction with TPEF of an extrinsic label, and SHG from structured collagen to observe significant differences in nuclear area of normal and cancerous regions in human tumors and murine tumor models. Yue et al.\cite{Yue:CellMet:2014} used SRS $\nu_{\ce{CH2},sym}$ contrast for morphological imaging in liver and prostate tissue, and for guidance in identifying lipid droplets (LDs). They then used spontaneous Raman spectroscopy at LD positions to determine that the cholesterol ester content of the LDs was correlated with the ratio of 702 and 1442 $cm^{-1}$ peak intensity ratios, and with the aggressiveness of the cancer. 

\subsubsection{SRS With Fingerprint Contrast}

As with CARS, little histology CRI has been performed with contrast in the fingerprint region. Ozeki\cite{Ozeki:NatPhot:2012} showed that, for typical modulation schemes, it is difficult to acquire image contrast using fingerprint Raman peaks. In CARS, the problem of weak fingerprint signal was overcome with a highly efficient signal generation mechanism.\cite{CampJr2014} While no analogous enhanced-efficiency signal generation approach has been found for SRS, modulation schemes have been demonstrated which discriminate against parasitic background signals sufficiently that fingerprint peaks can be observed. Berto et al.\cite{Berto:PRL:2014} proposed a scheme that exploits the even modulation symmetry of various background signals, providing sufficient discrimination to acquire fingerprint spectra from mouse skin at a 20 Hz spectral acquisition rate (720 Hz spectral pixel). Zhang et al\cite{Zhang:PCP:2012} were able to image cell nuclei using SRS with contrast from DNA at 785 $cm^{-1}$, at pixel acquisition rates on the order of 10 kHz. Fu et al.\cite{Fu:JACS:2016} demonstrated an SRS modulation approach that appears to 3 - 4 $\times$ more effective for rejecting background signals that vary slowly with excitation wavelength than amplitude modulation. They used delay-modulated spectral focusing achieve $\approx10\,mmole/L$ sensitivity from 720 $cm^{-1}$ peak of acetylcholine in neuronal tissue with at a 250 kHz pixel acquisition rate.

Colon tissues from 9 patients were imaged by spectrally scanned SRS in conjunction with SHG, THG, TPEF.\cite{Wang:ProcSPIE:2016} They found that an increase in protein ($\nu_{\ce{CH3},sym}$ and Amide I), and water (3250 $cm^{-1}$), and a decrease in lipid ($\nu_{\ce{CH2},sym}$, 1445 $cm^{-1}$, 1745 $cm^{-1}$) all correlated with presence of adenocarcinoma.  They reported similarity in TPEF of NADH and 1665 $cm^{-1}$ contrast images, suggesting that increased protein could be due to increased metabolic activity. They also suggested that increased water in cancerous regions could be due to increased aquaporin protein production.

\section{A Future Role For Coherent Raman Imaging in Histopathology}

We expect that CRI methods for histopathology will be adopted to the extent they meet specific needs, and fit well into the established histology work flow, possibly streamlining it. As early applications are explored, and as physicians become increasingly comfortable with the technology, it is possible that more complex and potentially disruptive adaptations of this technology may be adopted. In this section, we briefly discuss classes of problems that CRI can address, both in the near and longer terms.

\subsection{Immediacy of Information}
It is not uncommon to have 24 hour latency between tissue excision and histology analysis. While this is often acceptable, there are other situations, such as sampling a suspected tumor, and intra-operative margin evaluation, in which immediate characterization is optimal or even required. In these cases, label-free and \textit{in situ} histology modalities are desirable.

Tissue involved in and surrounding a tumor is typically sampled and analyzed by histology or for biomarkers at various time points to guide therapy by establishing prognosis and following disease progression. Accordingly, inappropriate tumor sampling is thought to be a leading problem in both prognosis and evaluating treatment response.\cite{Tam:JVIR:2016} 

Tissue sampling by needle biopsy or exploratory surgery is often guided by palpitation, preparative imaging, or visual inspection, and is notoriously inefficient.\cite{Tam:JVIR:2016} Neoplasms may be difficult to properly sample at early stages, and the ability to categorize tissue as normal or neoplastic during the biopsy process would go a long way towards ensuring high-quality samples for subsequent analysis. Spontaneous Raman was demonstrated decades ago as an potential guidance tool for needle biopsy,\cite{Frank:AnalChem:1995} and progress has been made toward clinical adaptation, but challenges remain with regards to low signal levels, and typically, Raman needle probes must be coupled with other detection modalities.\cite{Stevens:ChemSocRev:2016} Coherent Raman methods could easily provide sufficient signal levels for this application. Narrowband CRI approaches have been implemented in fiber,\cite{Balu:OptEx:2010,Saar:OptLett:2011,Hammoudi:OptEx:2011} so could be adapted to needle biopsy, but are probably not suited for this application since disease detection would be based strictly on spectra, and thus benefit greatly from sensitivity to multiple peaks in the fingerprint spectral region. Current broadband spectroscopic CRI approaches based on impulsive excitation or spectral focusing will be difficult to implement in a fiber delivery, but other approaches may be possible in the future.

While narrowband CRI may not be appropriate for purely spectroscopic disease detection, in needle biopsy, it is effective in a microscopy modality for \textit{in vivo} detection and analysis of skin carcinomas, which are present in the optically accessible epidermis layer of the skin.\cite{Heuke:Healthcare:2013,Heuke:BJD:2013,Mittal:LasersInSurgeryAndMedicine:2013,Wang:SciRep:2016} New approaches to signal detection may push the useful range of \textit{in vivo} CRI to slightly deeper tissues,\cite{Liao:SciAdv:2015} and to detection of fingerprint signals.\cite{Fu:JACS:2016}

Another widely recognized clinical problem is the need for intraoperative analysis of tissue at a resection margin. Strong correlations exist between presence of cancer cells in surgical margins and recurrence rates of cancer,\cite{Spivack:AOS:1994,Lacroix:JNeuro:2001} and incomplete margin resection is held as a major cause for recurrence of cancers that call for tissue conservation, such as in brain and breast.

A hand-held probe is required for \textit{in situ} margin analysis, and appropriate equipment has already been described for narrowband CRI\cite{Balu:OptEx:2010,Saar:OptLett:2011,Hammoudi:OptEx:2011}  although, to our knowledge, clinical studies using these probes have not yet been published. Use of a hand-held Raman probe for identification of tumor tissue in surgical margins was clinically demonstrated for brain surgery by Jermyn et al.\cite{Jermyn:SciTransMed:2015} and proof of principle for breast was demonstrated Kong et al.\cite{Kong:PMB:2014} As with Raman needle probes, tumor detection in these reports was based on spectroscopy alone, thus suffering somewhat from low signal and precluding imaging. This difficulty was avoided in the breast application by using autofluorescence levels to identify regions of interest, and those regions were imaged sparsely with Raman. In the brain application, the surgeon visually selected regions of interest.

Although CRI technology has not yet progressed to \textit{in situ} margin analysis, it has been demonstrated as an effective tool for characterizing freshly excised, unprocessed tissue.\cite{Orringer:NatBME:2017} Typically, margin tissue must be frozen and stained for imaging. This process can add 20 minutes to a surgery each time a margin is to be investigated. The label-free nature of CRI, and the ability to image in a backscattering mode makes it possible to obtain histology-quality images from surface regions of thick, unprocessed tissue, significantly reducing the delay between tissue removal and analysis.\cite{Orringer:NatBME:2017}

Another area in which speed of information availability would be useful is in initial evaluation of histology slides. The majority of tissue in most slides is normal, so a pathologist may spend a significant portion of their time searching for potential diseased regions. One particularly strong benefit of chemical imaging is that it can provide clear indications of tissue type and regions of normal and potentially abnormal tissue. This regional tissue typing has already been demonstrated with IR on slides from which high-resolution H\&E images have been obtained, and the two images overlaid.\cite{Kwak:BmcCancer:2011} CRI methods can provide both the regional tissue typing\cite{Evans:OpEx:2007,Uckermann:PloS:2014,Chowdary2010} and the high-resolution H\&E mimic without the need for staining or imaging in multiple contrast modes.\cite{Evans:OpEx:2007,Lee:Intravital:2015,Orringer:NatBME:2017}

\subsection{Better Quantification of Morphological Parameters}
A survey of pathologists by the National Collation on Health Care and Best Doctors\cite{BestDoctors:2013} gave lack of sub-specialty expertise the most often suspected reason for cancer misdiagnoses from histology. In that study, physicians most frequently indicated that improved pathology resources (i.e., innovations in histopathology) is the thing most needed to improve diagnostic accuracy. One particular approach to improving pathology resources, namely CAD, aims to bridge the sub-specialty expertise gap. 

An important example of how CAD can bridge this gap is through providing basic quantification of image features. As mentioned previously, non-specialist pathologists were found to be much less precise in cell counts and classification than sub-specialists,\cite{Fuchs:CMIG:2011} no doubt contributing significantly to overall diagnostic uncertainty. Many of the same factors contributing to non-specialist variability, including the fact that cells are often overlaid in H\&E images, with no clear delineation between nuclei.\cite{Fuchs:CMIG:2011} are also challenges to the families of morphological image analysis on which CAD is based. Further, factors such as variability in staining protocols, and in image field illumination, and issues with nonspecific staining contribute to overall image variability, adversely impacting reliability in thresholding and feature identification for CAD.  These, and related issues are expected to be significant barriers to widespread adoption of CAD. In recognition of this, the US Food and Drug Administration has issued guidelines for performance of WHI instrumentation.\cite{WSI:FDA:2016}

CRI methods can ameliorate or resolve many of the reproducibility and reliability issues in image formation. For example, obtaining reliable cell counts will be easier because CRI has intrinsically high axial resolution, making it feasible to resolve stacked nuclei. Additionally, where imaged nuclei are spatially close, there will be a nuclear and cellular membrane signal between them, strengthening the ability of CRI-based analysis to arrive at accurate cell counts. Yang et al.\cite{Yang:BME:2011} have already demonstrated 90\% accuracy and reproducibility in automated cell counting using coherent Raman imaging with only $\nu_{\ce{CH2},sym}$, and thus poor nuclear contrast.\cite{Meyer:JBO:2012} This is an important near-term opportunity for coherent Raman methods. 

In addition to intrinsic z-sectioning, CRI can provide a large number of contrasts through various vibrational bands. Since increased image dimensionality generally leads to higher precision segmentation and feature identification,\cite{Fernandez:NatBio:2005,Singh:AnalQuantCyto:2004,Weaver:ModernPath:2003} one could expect improved performance in these aspects with these multi-contrast images. The Wong group have derived morphological parameters familiar to pathologists, such as cell shape and density, from CARS images and based on these were able to identify with 80 to 95 \% accuracy breast tumor types.\cite{Yang:BME:2011}  

Although CRI requires little or no sample preparation, it seems likely that most early implementations will include fixation and paraffin embedding of samples, simply for consistency with protocols for staining. In fact, both of these steps have potential to introduce variability in CRI signals. Paraffin has a strong CH stretch signal, and fixing agents can liberate diagnostically important molecules, particularly lipids.\cite{Shim:PCPB:1996} For example, compared to fresh frozen tissue, formalin-fixed brain showed modification in $\nu_{\ce{CH2},assym}$ and $\nu_{\ce{CH3},assym}$ amplitudes from CARS, but not in $\nu_{\ce{CH2},sym}$. In the fingerprint, only choline groups of phospholipids at 717 $cm^{-1}$ dropped slightly after formalin fixing. However, methanol or acetone fixation severely reduced all these peaks and others.\cite{Galli:JBO:2013}

\subsection{Increased Chemical Information}

Morphological analysis of an image from a single stain such as H\&E is sufficient for diagnosis in most cases. However, difficult cases often require multiple additional stains to provide needed auxiliary information, but many of these desirable stains are rare or unavailable,\cite{Titford:JHistotech:2009} and it seems that this lack of access is a clinically significant issue.\cite{BestDoctors:2013} CRI with good fingerprint contrast has potential to make an important contribution by providing a considerable range of contrasts. 

Fingerprint CRI provides chemical specificity identical to that of spontaneous Raman spectroscopy,\cite{CampJr2014} and therefore seems likely to be a potential substitute for most or all stains, and perhaps many IHC agents. Raman spectra can be used to recapitulate many image contrast agents used commonly in biology.\cite{Klein:BJ:2012} Also, as shown by the partial listing in Table \ref{tbl:VibMarkers}, fingerprint Raman has already been used to detect many diagnostic markers. With fingerprint CRI, these will be available as label-free image contrasts. An important added benefit of spectroscopic CARS is that it provides an internal reference based on NRB,\cite{Camp2015} so that spectroscopic CARS images can serve as quantitative, rather than qualitative maps of markers. This level of quantification cannot generally be achieved with stains, fluorescence, or even spontaneous Raman. The quantitative imaging and absence of non-specific staining will greatly simplify image analysis and increase confidence in image metrics and associated diagnostic decisions. A few specific examples of potential important contributions are given below.

The number density of mitotic cells plays an important role in diagnosis and grading of many cancer types,\cite{StanfordHistoCrit} and is usually determined by IHC or cell morphology analysis. The best morphologically-derived automated detection of mitoses achieve F-scores (the geometric mean of precision and accuracy) between 75\% and 80\%.\cite{Cirean:MICCAI:2013} It seems likely that fingerprint CRI imaging could yield significantly better results. By mapping the intranuclear distribution of DNA, RNA, and chromatin, Raman spectroscopic imaging can not only positively identify whether a cell is mitotic, but which stage of mitosis it is in.\cite{Matthaus:ApplSpec:2006, Boydston:VibSpec:2005,Boydston:BioSpec:1999} The chemical distribution information is essentially orthogonal to morphological analysis, so can only serve to increase precision and accuracy in determining mitosis counts.  

Accurate characterization of pleomorphism (atypical cell shape) is also important to proper diagnosis of many cancers.\cite{StanfordHistoCrit} Pleomorphic cells of one type may be mistakenly identified as normal cells of a different type, and this was part of a 42\% cell count and classification disagreement among non-specialists found by Fuchs et al.\cite{Fuchs:CMIG:2011} It is well established that Raman spectroscopy can be used to positively identify cell type.\cite{Chan:AnalChem:2009} Applying spectral cell identification as an orthogonal metric will facilitate much better recognition of pleomorphism. 

Invasion of immune cells into a tumor site, and their state of activation is an key factor in diagnosis of some cancers.\cite{StanfordHistoCrit} As mentioned above, morphology-based cell identification can be difficult, and determination of immune cell activation is currently done by assessing the presence of cell surface markers, typically by IHC. For this important problem, fingerprint spectroscopic imaging will not only allow immune cell identification, but also whether they are activated.\cite{Chan:AnalChem:2008,Diem:JPB:2013} 

Identifying the tissue from which of CUP (cancer unknown primary) metastases originate is useful for identifying appropriate therapeutic strategies, but presently requires involved analysis.\cite{Groschel:CSH:2016,Mountzios:CROH:2010,Navin:CSH:2010} Fingerprint spectroscopic CRI may be able to positively identify metastatic tumor origins simply from chemical content of the tumor cells. Even narrowband CRI studies were able to discriminate primary (metastasized) cancer cells from cells belonging to the target tissue simply on basis of contrast at $\nu_{\ce{CH2}, symm}$ and $\nu_{\ce{CH3}, symm}$.\cite{Meyer:JBO:2011,Uckermann:PloS:2014}

In recent years it has become clear that genomic subclasses of cancers can have distinct clinical outcomes,\cite{Curtis:Nature:2012} and thus require varied interventions. While spectroscopic Raman imaging has not been evaluated as a tool for identifying these subclasses, it seems reasonable that it may have power to do so in some cases. CRI will not directly detection genome changes, but it is likely to detect downstream consequences of those changes. For example, cell phenotype\cite{Chan:AnalChem:2009} and function\cite{Chan:AnalChem:2009,Diem:JPB:2013} can be discriminated using Raman spectra. These spectral changes come not from DNA sequence or methylation modifications directly, but from the intra- and extra-cellular changes in bimolecular profiles they induce. Furthermore, microenvironment properties determined through histology have recently been found to correlate with the genomic cancer subtypes, and these microenvironment changes should be detectable with CRI in most cases.\cite{Natrajan:PLOS:2016}. 

\subsubsection{Speed \& Information Trade-off}

Single frequency CRI approaches can provide imaging nearly as fast to WSI instrumentation. CARS approaches typically provide tissue imaging at pixel rates of 200 kHz to 1 MHz,\cite{Meyer:AnalChem:2013,Veilleux:IEEE:2008} as little as a factor of three slower than commercial WSI. SRS instruments seem to run a bit more slowly, typically at $\approx$50 kHz.\cite{Orringer:NatBME:2017,Freudiger2012} However, a recent SRS demonstration achieves fingerprint contrast at 250 kHz, between 10 and 60 times slower than WSI. Although these speeds may still be too low for widespread use, they present only a minimal imaging rate-related barrier to applying narrowband CRI to histopathology problems in a research mode. Accordingly, reports of narrowband histology studies are beginning to appear. On the other hand, many of CRI's potential contributions to histopathology will require chemical information beyond that attainable from a single Raman peak. Based on current technology, there will be a significant speed penalty for increased chemical information.

Broadband CRI tissue imaging using CH stretch has been demonstrated at 20 Hz for CARS\cite{Marks2004} and for 4 kHz for SRS\cite{Otsuka:BiOS:2014,Egawa:JBO:2016} over 300 $cm^{-1}$. Fingerprint CARS spectro-microscopy has been demonstrated at 300 Hz over 3000 $cm^{-1}$ in tissue.\cite{CampJr2014} Overall, these range in speed from $10^5$ to $10^3$ times slower than WSI. This difference may be difficult to surmount. Assuming the 0.5 $cm^2$/s imaging speed is indeed an appropriate mark, it seems that present spectral CRI methods are not good candidates, although some newer spectroscopic CRI approaches may hold promise. a spectral focusing approach has been demonstrated at 77 kHz for a 200 $cm^{-1}$ bandwidth,\cite{Liao:SciAdv:2015} although not yet in tissue, and a family of approaches featuring $\approx$ 25 kHz spectroscopic CARS acquisition has been demonstrated,\cite{Ideguchi:Nature:2013,Hashimoto:SciRep:2016} but with laser pulses incompatible with tissue imaging. Broadband SRS\cite{Ploetz:APB:2007,Rock:OE:2013} has been demonstrated with potentially fast imaging, but seem to suffer from insufficient modulation sensitivity.

Assuming that signal generation and collection efficiency of current CRI instruments is nearly optimal, and there is no way to intrinsically speed up image acquisition, there are at least two non-exclusive options for applying spectroscopic CRI. One is to use a faster imaging modality (such as narrowband CRI) to generate a guide image from which regions of interest can be determined, and use spectroscopic CRI over ROIs covering a small fraction of the specimen. The other approach would be to use spatial multiplexing. 

Provided sufficient excitation laser power, spatial multiplexing approaches can yield speed increases proportional to the number of detection elements. For spectroscopically detected signal, such as in spontaneous Raman scattering, 1-D (line) spatial multiplexing can realistically lead to 20$\times$ - 100$\times$ imaging speed increase.\cite{Okada:PNAS:2012} This approach could also be used for spectroscopic CARS. 2-D multiplexing has been demonstrated with narrowband CARS\cite{Shi:PRL:2010,Heinrich:OpEx:2008,Toytman:OptLett:2007,Lei:JBMO:2011,Berto:PRL:2012} though not in ways suited for clinical applications. Spectroscopic approaches using single element detectors, such as time-domain\cite{Ideguchi:Nature:2013} or spectrally scanned\cite{Pegoraro:OpticsExpress:2009} CARS could also be 2-D multiplexed, so $10^4 \times$ speed increase is conceivable, provided laser sources of sufficient pulse energy can be found to spread light over a larger area and still obtain signal. SRS will be more challenging to implement in spatially multiplexed modes since it requires very high sensitivity to small amplitude modulations. Lock-in and balanced detection are not commonly available in an array format, although examples do exist. Slipchenko et al.\cite{Slipchenko:JB:2012} introduced an array of 32 LC circuits for simultaneously detecting SRS signal in as many channels, and CMOS cameras can be used in differential or lock-in mode, but these have yet to demonstrate needed sensitivity.\cite{Ploetz:APB:2007,Rock:OE:2013}

Finally, we note that, using spectroscopic CRI, information storage could be an issue. Storing single contrast images from WSI instruments is already a challenge. If all histology images were acquired in full spectral mode, this would increase the data storage load by $10^3$ times. This does not seem tenable. It is conceivable that, ultimately, only a small fraction of overall tissue surface will be imaged spectroscopically. Also, it is possible that not all spectral dimensions will be necessary in every case,\cite{Tiwari:AnalChem:2016} and that acceptable protocols for reduction of spectral information can be found.

\section{Conclusions}
We expect that coherent Raman imaging is most likely to find acceptance in the clinic as a drop-in replacement for current technologies as it is able to help resolve recognized problems related to detection, diagnosis, and treatment of disease. Such problems are likely to include fast and reliable intra-surgical margin evaluation, either through direct analysis of the surgical site or through rapid analysis of freshly excised tissue. Similarly, CRI's ability to provide label-free molecularly specific signal quickly may be useful for \textit{in-situ} detection of small tumors or neoplasms in exploratory surgery or needle biopsy.

Intra-surgical microscopic analysis of excised tissue has already been demonstrated. \textit{In-situ} use will likely require a hand-held probe with fiber-delivered excitation light and signal collection. Whether \textit{in-situ} tissue analysis will be best done by narrow-band coherent Raman imaging or by coherent Raman spectroscopic analysis is unclear at this time.

Another significant opportunity for CRI is facilitating improvements in histology-based diagnoses. The art of histopathology emerged near the end of the $19^{th}$ century in nearly its present form. Significant new knowledge of disease mechanisms has improved interpretation of histological features, but, apart from the introduction of IHC in the 1940's, histopathology tools in use today are largely the same as they were in the early $20^{th}$ century. Efforts to modernize histopathology practice are making headway, however, fully transforming it from a subjective art to a quantitative and objective science remains a challenge. Most of these challenges involve variability and contrast reliability in image formation. It appears that many of these issues, could be largely resolved with CRI. Furthermore, spectroscopic CRI may provide a significantly enhanced contrast space that will open new options for quickly stratifying cancer and other diseases, leading to a more "personalized" diagnosis.

While opportunities for CRI to improve clinical practice are clear, important barriers remain for widespread adoption in histology practice. An increase in imaging speed for narrow-band CARS and a slightly larger increase for SRS would be helpful. For spectroscopic CRI, substantial imaging speed increases will be necessary if they are to be used to a significant extent in an imaging mode. Fortunately, the CRI field is still vibrant with new innovations for increased imaging speed and chemical selectivity and sensitivity. Much work remains to demonstrate the utility and reliability of these methods to the medical community. Finally, potential regulatory issues must be considered and resolved to the extent that they are found.





\bibliographystyle{rsc} 
\bibliography{ARev}

\providecommand*{\mcitethebibliography}{\thebibliography}
\csname @ifundefined\endcsname{endmcitethebibliography}
{\let\endmcitethebibliography\endthebibliography}{}
\begin{mcitethebibliography}{257}
\providecommand*{\natexlab}[1]{#1}
\providecommand*{\mciteSetBstSublistMode}[1]{}
\providecommand*{\mciteSetBstMaxWidthForm}[2]{}
\providecommand*{\mciteBstWouldAddEndPuncttrue}
  {\def\EndOfBibitem{\unskip.}}
\providecommand*{\mciteBstWouldAddEndPunctfalse}
  {\let\EndOfBibitem\relax}
\providecommand*{\mciteSetBstMidEndSepPunct}[3]{}
\providecommand*{\mciteSetBstSublistLabelBeginEnd}[3]{}
\providecommand*{\EndOfBibitem}{}
\mciteSetBstSublistMode{f}
\mciteSetBstMaxWidthForm{subitem}
{(\emph{\alph{mcitesubitemcount}})}
\mciteSetBstSublistLabelBeginEnd{\mcitemaxwidthsubitemform\space}
{\relax}{\relax}

\bibitem[Milestones2009()]{Milestones2009}
\emph{Nature Cell Biology}, 2009, \textbf{11}, S6--S22\relax
\mciteBstWouldAddEndPuncttrue
\mciteSetBstMidEndSepPunct{\mcitedefaultmidpunct}
{\mcitedefaultendpunct}{\mcitedefaultseppunct}\relax
\EndOfBibitem
\bibitem[Titford(2009)]{Titford:JHistotech:2009}
M.~Titford, \emph{Journal of Histotechnology}, 2009, \textbf{32}, 9--19\relax
\mciteBstWouldAddEndPuncttrue
\mciteSetBstMidEndSepPunct{\mcitedefaultmidpunct}
{\mcitedefaultendpunct}{\mcitedefaultseppunct}\relax
\EndOfBibitem
\bibitem[Hunninghake \emph{et~al.}(2001)Hunninghake, Zimmerman, Schwartz, King,
  Lynch, Hegele, Waldron, Colby, Muller, Lynch, Galvin, Gross, Hogg, Toews,
  Helmers, Cooper, Baugman, Strange, and Millard]{Hunninghake2001}
G.~Hunninghake, M.~B. Zimmerman, D.~Schwartz, T.~King, J.~Lynch, R.~Hegele,
  J.~Waldron, T.~Colby, N.~Muller, D.~Lynch, J.~Galvin, B.~Gross, J.~Hogg,
  G.~Toews, R.~Helmers, J.~A. Cooper, R.~Baugman, C.~Strange and M.~Millard,
  \emph{Am J Respir Crit Care Med}, 2001, \textbf{164}, 193--196\relax
\mciteBstWouldAddEndPuncttrue
\mciteSetBstMidEndSepPunct{\mcitedefaultmidpunct}
{\mcitedefaultendpunct}{\mcitedefaultseppunct}\relax
\EndOfBibitem
\bibitem[van Rhijn \emph{et~al.}(2010)van Rhijn, van Leenders, Ooms, Kirkels,
  Zlotta, Boev\'{e}, and J\"{o}bsis]{VanRhijn2010}
B.~W.~G. van Rhijn, G.~J. L.~H. van Leenders, B.~C.~M. Ooms, W.~J. Kirkels,
  A.~R. Zlotta, E.~R. Boev\'{e} and A.~C. J\"{o}bsis, \emph{European Urology},
  2010, \textbf{57}, 1052--1057\relax
\mciteBstWouldAddEndPuncttrue
\mciteSetBstMidEndSepPunct{\mcitedefaultmidpunct}
{\mcitedefaultendpunct}{\mcitedefaultseppunct}\relax
\EndOfBibitem
\bibitem[Allsbrook~Jr \emph{et~al.}(2001)Allsbrook~Jr, Mangold, Johnson, Lane,
  Lane, and Epstein]{AllsbrookJr2001}
W.~C. Allsbrook~Jr, K.~A. Mangold, M.~H. Johnson, R.~B. Lane, C.~G. Lane and
  J.~I. Epstein, \emph{Human Pathology}, 2001, \textbf{32}, 81--88\relax
\mciteBstWouldAddEndPuncttrue
\mciteSetBstMidEndSepPunct{\mcitedefaultmidpunct}
{\mcitedefaultendpunct}{\mcitedefaultseppunct}\relax
\EndOfBibitem
\bibitem[Costantini \emph{et~al.}(2003)Costantini, Sciallero, Giannini,
  Gatteschi, Rinaldi, Lanzanova, Bonelli, Casetti, Bertinelli, Giuliani,
  Castiglione, Mantellini, Naldoni, and Bruzzi]{Costantini2003}
M.~Costantini, S.~Sciallero, A.~Giannini, B.~Gatteschi, P.~Rinaldi,
  G.~Lanzanova, L.~Bonelli, T.~Casetti, E.~Bertinelli, O.~Giuliani,
  G.~Castiglione, P.~Mantellini, C.~Naldoni and P.~Bruzzi, \emph{Journal of
  Clinical Epidemiology}, 2003, \textbf{56}, 209--214\relax
\mciteBstWouldAddEndPuncttrue
\mciteSetBstMidEndSepPunct{\mcitedefaultmidpunct}
{\mcitedefaultendpunct}{\mcitedefaultseppunct}\relax
\EndOfBibitem
\bibitem[Longacre \emph{et~al.}(2005)Longacre, Ennis, Quenneville, Bane,
  Bleiweiss, Carter, Catelano, Hendrickson, Hibshoosh, Layfield, Memeo, Wu, and
  O'Malley]{Longacre2005}
T.~A. Longacre, M.~Ennis, L.~A. Quenneville, A.~L. Bane, I.~J. Bleiweiss, B.~A.
  Carter, E.~Catelano, M.~R. Hendrickson, H.~Hibshoosh, L.~J. Layfield,
  L.~Memeo, H.~Wu and F.~P. O'Malley, \emph{Mod Pathol}, 2005, \textbf{19},
  195--207\relax
\mciteBstWouldAddEndPuncttrue
\mciteSetBstMidEndSepPunct{\mcitedefaultmidpunct}
{\mcitedefaultendpunct}{\mcitedefaultseppunct}\relax
\EndOfBibitem
\bibitem[StanfordHistoCrit()]{StanfordHistoCrit}
\emph{Stanford Medicine: Surgical Pathology Criteria},
  \url{www.surgpathcriteria.stanford.edu}\relax
\mciteBstWouldAddEndPuncttrue
\mciteSetBstMidEndSepPunct{\mcitedefaultmidpunct}
{\mcitedefaultendpunct}{\mcitedefaultseppunct}\relax
\EndOfBibitem
\bibitem[Coons \emph{et~al.}(1941)Coons, Creech, and Jones]{Coons1941}
A.~H. Coons, H.~J. Creech and R.~N. Jones, \emph{Experimental Biology and
  Medicine}, 1941, \textbf{47}, 200--202\relax
\mciteBstWouldAddEndPuncttrue
\mciteSetBstMidEndSepPunct{\mcitedefaultmidpunct}
{\mcitedefaultendpunct}{\mcitedefaultseppunct}\relax
\EndOfBibitem
\bibitem[Fletcher \emph{et~al.}(2002)Fletcher, Berman, Corless, Gorstein,
  Lasota, Longley, Miettinen, O'Leary, Remotti, Rubin, Shmookler, Sobin, and
  Weiss]{Fletcher2002}
C.~D. Fletcher, J.~J. Berman, C.~Corless, F.~Gorstein, J.~Lasota, B.~J.
  Longley, M.~Miettinen, T.~J. O'Leary, H.~Remotti, B.~P. Rubin, B.~Shmookler,
  L.~H. Sobin and S.~W. Weiss, \emph{Human Pathology}, 2002, \textbf{33},
  459--465\relax
\mciteBstWouldAddEndPuncttrue
\mciteSetBstMidEndSepPunct{\mcitedefaultmidpunct}
{\mcitedefaultendpunct}{\mcitedefaultseppunct}\relax
\EndOfBibitem
\bibitem[Werner \emph{et~al.}(2000)Werner, Chott, Fabiano, and
  Battifora]{Werner2000}
M.~Werner, A.~Chott, A.~Fabiano and H.~Battifora, \emph{The American Journal of
  Surgical Pathology}, 2000, \textbf{24}, 1016\relax
\mciteBstWouldAddEndPuncttrue
\mciteSetBstMidEndSepPunct{\mcitedefaultmidpunct}
{\mcitedefaultendpunct}{\mcitedefaultseppunct}\relax
\EndOfBibitem
\bibitem[Sabah \emph{et~al.}(2003)Sabah, Leader, and Kay]{Sabah2003}
M.~Sabah, M.~Leader and E.~Kay, \emph{Applied Immunohistochemistry \& Molecular
  Morphology}, 2003, \textbf{11}, 56--61\relax
\mciteBstWouldAddEndPuncttrue
\mciteSetBstMidEndSepPunct{\mcitedefaultmidpunct}
{\mcitedefaultendpunct}{\mcitedefaultseppunct}\relax
\EndOfBibitem
\bibitem[Kirkegaard \emph{et~al.}(2006)Kirkegaard, Edwards, Tovey, McGlynn,
  Krishna, Mukherjee, Tam, Munro, Dunne, and Bartlett]{Kirkegaard2006}
T.~Kirkegaard, J.~Edwards, S.~Tovey, L.~M. McGlynn, S.~N. Krishna,
  R.~Mukherjee, L.~Tam, A.~F. Munro, B.~Dunne and J.~M.~S. Bartlett,
  \emph{Histopathology}, 2006, \textbf{48}, 787--794\relax
\mciteBstWouldAddEndPuncttrue
\mciteSetBstMidEndSepPunct{\mcitedefaultmidpunct}
{\mcitedefaultendpunct}{\mcitedefaultseppunct}\relax
\EndOfBibitem
\bibitem[Weinstein \emph{et~al.}(2009)Weinstein, Graham, Richter, Barker,
  Krupinski, Lopez, Erps, Bhattacharyya, Yagi, and
  Gilbertson]{Weinstein:HumPath:2009}
R.~S. Weinstein, A.~R. Graham, L.~C. Richter, G.~P. Barker, E.~A. Krupinski,
  A.~M. Lopez, K.~A. Erps, A.~K. Bhattacharyya, Y.~Yagi and J.~R. Gilbertson,
  \emph{Hum Pathol}, 2009, \textbf{40}, 1057--1069\relax
\mciteBstWouldAddEndPuncttrue
\mciteSetBstMidEndSepPunct{\mcitedefaultmidpunct}
{\mcitedefaultendpunct}{\mcitedefaultseppunct}\relax
\EndOfBibitem
\bibitem[Woernley(1952)]{Woernley:Cancer:1952}
D.~L. Woernley, \emph{Cancer Research}, 1952, \textbf{12}, 516--523\relax
\mciteBstWouldAddEndPuncttrue
\mciteSetBstMidEndSepPunct{\mcitedefaultmidpunct}
{\mcitedefaultendpunct}{\mcitedefaultseppunct}\relax
\EndOfBibitem
\bibitem[Kendall \emph{et~al.}(2003)Kendall, Stone, Shepherd, Geboes, Warren,
  Bennett, and Barr]{Kendall:TheJournalOfPathology:2003}
C.~Kendall, N.~Stone, N.~Shepherd, K.~Geboes, B.~Warren, R.~Bennett and
  H.~Barr, \emph{The Journal of Pathology}, 2003, \textbf{200}, 602--609\relax
\mciteBstWouldAddEndPuncttrue
\mciteSetBstMidEndSepPunct{\mcitedefaultmidpunct}
{\mcitedefaultendpunct}{\mcitedefaultseppunct}\relax
\EndOfBibitem
\bibitem[Bhargava and Levin(2004)]{Bhargava:ApplSpectrosc:2004}
R.~Bhargava and I.~W. Levin, \emph{Appl Spectrosc}, 2004, \textbf{58},
  995--1000\relax
\mciteBstWouldAddEndPuncttrue
\mciteSetBstMidEndSepPunct{\mcitedefaultmidpunct}
{\mcitedefaultendpunct}{\mcitedefaultseppunct}\relax
\EndOfBibitem
\bibitem[Haka \emph{et~al.}(2005)Haka, Shafer-Peltier, Fitzmaurice, Crowe,
  Dasari, and Feld]{Haka2005}
A.~S. Haka, K.~E. Shafer-Peltier, M.~Fitzmaurice, J.~Crowe, R.~R. Dasari and
  M.~S. Feld, \emph{Proceedings of the National Academy of Sciences of the
  United States of America}, 2005, \textbf{102}, 12371\relax
\mciteBstWouldAddEndPuncttrue
\mciteSetBstMidEndSepPunct{\mcitedefaultmidpunct}
{\mcitedefaultendpunct}{\mcitedefaultseppunct}\relax
\EndOfBibitem
\bibitem[Zumbusch \emph{et~al.}(1999)Zumbusch, Holtom, and Xie]{Zumbusch1999}
A.~Zumbusch, G.~R. Holtom and X.~S. Xie, \emph{Physical Review Letters}, 1999,
  \textbf{82}, 4142--4145\relax
\mciteBstWouldAddEndPuncttrue
\mciteSetBstMidEndSepPunct{\mcitedefaultmidpunct}
{\mcitedefaultendpunct}{\mcitedefaultseppunct}\relax
\EndOfBibitem
\bibitem[Nasse \emph{et~al.}(2011)Nasse, Walsh, Mattson, Reininger,
  Kajdacsy-Balla, Macias, Bhargava, and Hirschmugl]{Nasse:NatMeth:2011}
M.~J. Nasse, M.~J. Walsh, E.~C. Mattson, R.~Reininger, A.~Kajdacsy-Balla,
  V.~Macias, R.~Bhargava and C.~J. Hirschmugl, \emph{Nat Meth}, 2011,
  \textbf{8}, 413--416\relax
\mciteBstWouldAddEndPuncttrue
\mciteSetBstMidEndSepPunct{\mcitedefaultmidpunct}
{\mcitedefaultendpunct}{\mcitedefaultseppunct}\relax
\EndOfBibitem
\bibitem[Yeh \emph{et~al.}(2015)Yeh, Kenkel, Liu, and
  Bhargava]{Yeh:AnalChem:2015}
K.~Yeh, S.~Kenkel, J.-N. .~N. Liu and R.~Bhargava, \emph{Anal Chem}, 2015,
  \textbf{87}, 485--493\relax
\mciteBstWouldAddEndPuncttrue
\mciteSetBstMidEndSepPunct{\mcitedefaultmidpunct}
{\mcitedefaultendpunct}{\mcitedefaultseppunct}\relax
\EndOfBibitem
\bibitem[ProtAtlas()]{ProtAtlas}
\emph{Protein Atlas}, \url{www.proteinatlas.org}\relax
\mciteBstWouldAddEndPuncttrue
\mciteSetBstMidEndSepPunct{\mcitedefaultmidpunct}
{\mcitedefaultendpunct}{\mcitedefaultseppunct}\relax
\EndOfBibitem
\bibitem[Fridman \emph{et~al.}(2012)Fridman, Pag\`{e}s, Saut\`{e}s-Fridman, and
  Galon]{Fridman2012}
W.~H. Fridman, F.~Pag\`{e}s, C.~Saut\`{e}s-Fridman and J.~Galon, \emph{Nature
  Reviews Cancer}, 2012, \textbf{12}, 298--306\relax
\mciteBstWouldAddEndPuncttrue
\mciteSetBstMidEndSepPunct{\mcitedefaultmidpunct}
{\mcitedefaultendpunct}{\mcitedefaultseppunct}\relax
\EndOfBibitem
\bibitem[Schubert \emph{et~al.}(2006)Schubert, Bonnekoh, Pommer, Philipsen,
  B\"{o}ckelmann, Malykh, Gollnick, Friedenberger, Bode, and
  Dress]{Schubert2006}
W.~Schubert, B.~Bonnekoh, A.~J. Pommer, L.~Philipsen, R.~B\"{o}ckelmann,
  Y.~Malykh, H.~Gollnick, M.~Friedenberger, M.~Bode and A.~W.~M. Dress,
  \emph{Nat Biotechnol}, 2006, \textbf{24}, 1270--1278\relax
\mciteBstWouldAddEndPuncttrue
\mciteSetBstMidEndSepPunct{\mcitedefaultmidpunct}
{\mcitedefaultendpunct}{\mcitedefaultseppunct}\relax
\EndOfBibitem
\bibitem[Humphrey(2007)]{Humphrey:JClinPath:2007}
P.~A. Humphrey, \emph{Journal of Clinical Pathology}, 2007, \textbf{60},
  35--42\relax
\mciteBstWouldAddEndPuncttrue
\mciteSetBstMidEndSepPunct{\mcitedefaultmidpunct}
{\mcitedefaultendpunct}{\mcitedefaultseppunct}\relax
\EndOfBibitem
\bibitem[Kronz \emph{et~al.}(1999)Kronz, Westra, and Epstein]{Kronz1999}
J.~D. Kronz, W.~H. Westra and J.~I. Epstein, \emph{Cancer}, 1999, \textbf{86},
  2426--2435\relax
\mciteBstWouldAddEndPuncttrue
\mciteSetBstMidEndSepPunct{\mcitedefaultmidpunct}
{\mcitedefaultendpunct}{\mcitedefaultseppunct}\relax
\EndOfBibitem
\bibitem[Raab \emph{et~al.}(2005)Raab, Nakhleh, and Ruby]{Raab2005}
S.~S. Raab, R.~E. Nakhleh and S.~G. Ruby, \emph{Archives of Pathology \&
  Laboratory Medicine}, 2005, \textbf{129}, 459--466\relax
\mciteBstWouldAddEndPuncttrue
\mciteSetBstMidEndSepPunct{\mcitedefaultmidpunct}
{\mcitedefaultendpunct}{\mcitedefaultseppunct}\relax
\EndOfBibitem
\bibitem[Ho \emph{et~al.}(2006)Ho, Parwani, Jukic, Yagi, Anthony, and
  Gilbertson]{Ho2006}
J.~Ho, A.~V. Parwani, D.~M. Jukic, Y.~Yagi, L.~Anthony and J.~R. Gilbertson,
  \emph{Human Pathology}, 2006, \textbf{37}, 322--331\relax
\mciteBstWouldAddEndPuncttrue
\mciteSetBstMidEndSepPunct{\mcitedefaultmidpunct}
{\mcitedefaultendpunct}{\mcitedefaultseppunct}\relax
\EndOfBibitem
\bibitem[Renshaw \emph{et~al.}(2001)Renshaw, Lezon, and Wilbur]{Renshaw2001}
A.~A. Renshaw, K.~M. Lezon and D.~C. Wilbur, \emph{Cancer Cytopathology}, 2001,
  \textbf{93}, 106--110\relax
\mciteBstWouldAddEndPuncttrue
\mciteSetBstMidEndSepPunct{\mcitedefaultmidpunct}
{\mcitedefaultendpunct}{\mcitedefaultseppunct}\relax
\EndOfBibitem
\bibitem[Lurkin \emph{et~al.}(2010)Lurkin, Ducimeti\`{e}re, Vince,
  Decouvelaere, Cellier, Gilly, Salameire, Biron, de~Laroche, Blay, and
  Ray-Coquard]{Lurkin:BmcCancer:2010}
A.~Lurkin, F.~Ducimeti\`{e}re, D.~R. Vince, A.-V. . V. .~V. Decouvelaere,
  D.~Cellier, F.~N. Gilly, D.~Salameire, P.~Biron, G.~de~Laroche, J.~Y. Blay
  and I.~Ray-Coquard, \emph{BMC Cancer}, 2010, \textbf{10}, 150--\relax
\mciteBstWouldAddEndPuncttrue
\mciteSetBstMidEndSepPunct{\mcitedefaultmidpunct}
{\mcitedefaultendpunct}{\mcitedefaultseppunct}\relax
\EndOfBibitem
\bibitem[Bruner \emph{et~al.}(1997)Bruner, Inouye, Fuller, and
  Langford]{Bruner1997}
J.~M. Bruner, L.~Inouye, G.~N. Fuller and L.~A. Langford, \emph{Cancer}, 1997,
  \textbf{79}, 796--803\relax
\mciteBstWouldAddEndPuncttrue
\mciteSetBstMidEndSepPunct{\mcitedefaultmidpunct}
{\mcitedefaultendpunct}{\mcitedefaultseppunct}\relax
\EndOfBibitem
\bibitem[Nguyen \emph{et~al.}(2004)Nguyen, Schultz, Renshaw, Vollmer, Welch,
  Cote, DAmico, and D'Amico]{Nguyen2004}
P.~L. Nguyen, D.~Schultz, A.~A. Renshaw, R.~T. Vollmer, W.~R. Welch, K.~Cote,
  A.~V. DAmico and A.~V. D'Amico, \emph{Urologic Oncology: Seminars and
  Original Investigations}, 2004, \textbf{22}, 295--299\relax
\mciteBstWouldAddEndPuncttrue
\mciteSetBstMidEndSepPunct{\mcitedefaultmidpunct}
{\mcitedefaultendpunct}{\mcitedefaultseppunct}\relax
\EndOfBibitem
\bibitem[Staradub(2002)]{Staradub2002}
V.~L. Staradub, \emph{Annals of Surgical Oncology}, 2002, \textbf{9},
  982--987\relax
\mciteBstWouldAddEndPuncttrue
\mciteSetBstMidEndSepPunct{\mcitedefaultmidpunct}
{\mcitedefaultendpunct}{\mcitedefaultseppunct}\relax
\EndOfBibitem
\bibitem[Coblentz \emph{et~al.}(2001)Coblentz, Mills, and
  Theodorescu]{Coblentz2001}
T.~R. Coblentz, S.~E. Mills and D.~Theodorescu, \emph{Cancer}, 2001,
  \textbf{91}, 1284--1290\relax
\mciteBstWouldAddEndPuncttrue
\mciteSetBstMidEndSepPunct{\mcitedefaultmidpunct}
{\mcitedefaultendpunct}{\mcitedefaultseppunct}\relax
\EndOfBibitem
\bibitem[Selman \emph{et~al.}(1999)Selman, Niemann, Fowler, and
  Copeland]{Selman1999}
A.~E. Selman, T.~H. Niemann, J.~M. Fowler and L.~J. Copeland, \emph{Obstetrics
  and Gynecology}, 1999, \textbf{94}, 302--306\relax
\mciteBstWouldAddEndPuncttrue
\mciteSetBstMidEndSepPunct{\mcitedefaultmidpunct}
{\mcitedefaultendpunct}{\mcitedefaultseppunct}\relax
\EndOfBibitem
\bibitem[BestDoctors:2013()]{BestDoctors:2013}
\emph{MisdiagnosisSurvey\_FINALiv.pdf},
  \url{http://www.bestdoctors.com/about-best-doctors/news-and-events/nchc-misdiagnosis-survey}\relax
\mciteBstWouldAddEndPuncttrue
\mciteSetBstMidEndSepPunct{\mcitedefaultmidpunct}
{\mcitedefaultendpunct}{\mcitedefaultseppunct}\relax
\EndOfBibitem
\bibitem[Hollensead \emph{et~al.}(2004)Hollensead, Lockwood, and
  Elin]{Hollensead:JSurgOncol:2004}
S.~C. Hollensead, W.~B. Lockwood and R.~J. Elin, \emph{J Surg Oncol}, 2004,
  \textbf{88}, 161--81\relax
\mciteBstWouldAddEndPuncttrue
\mciteSetBstMidEndSepPunct{\mcitedefaultmidpunct}
{\mcitedefaultendpunct}{\mcitedefaultseppunct}\relax
\EndOfBibitem
\bibitem[Mello \emph{et~al.}(2010)Mello, Chandra, Gawande, and
  Studdert]{Mello:HealthAff:2010}
M.~M. Mello, A.~Chandra, A.~A. Gawande and D.~M. Studdert, \emph{Health Aff
  (Millwood)}, 2010, \textbf{29}, 1569--77\relax
\mciteBstWouldAddEndPuncttrue
\mciteSetBstMidEndSepPunct{\mcitedefaultmidpunct}
{\mcitedefaultendpunct}{\mcitedefaultseppunct}\relax
\EndOfBibitem
\bibitem[CancerFacts:2016()]{CancerFacts:2016}
\emph{ACS\_2016\_FactsFigures.pdf}, \url{www.cancer.org}\relax
\mciteBstWouldAddEndPuncttrue
\mciteSetBstMidEndSepPunct{\mcitedefaultmidpunct}
{\mcitedefaultendpunct}{\mcitedefaultseppunct}\relax
\EndOfBibitem
\bibitem[CancerCost2012()]{CancerCost2012}
\emph{Cost of Cancer Care}, 2012\relax
\mciteBstWouldAddEndPuncttrue
\mciteSetBstMidEndSepPunct{\mcitedefaultmidpunct}
{\mcitedefaultendpunct}{\mcitedefaultseppunct}\relax
\EndOfBibitem
\bibitem[Huang \emph{et~al.}(2003)Huang, McWilliams, Lui, McLean, Lam, and
  Zeng]{Huang2003}
Z.~Huang, A.~McWilliams, H.~Lui, D.~I. McLean, S.~Lam and H.~Zeng,
  \emph{International Journal of Cancer}, 2003, \textbf{107}, 1047--1052\relax
\mciteBstWouldAddEndPuncttrue
\mciteSetBstMidEndSepPunct{\mcitedefaultmidpunct}
{\mcitedefaultendpunct}{\mcitedefaultseppunct}\relax
\EndOfBibitem
\bibitem[Stone \emph{et~al.}(2004)Stone, Kendall, Smith, Crow, and
  Barr]{Stone2004}
N.~Stone, C.~Kendall, J.~Smith, P.~Crow and H.~Barr, \emph{Faraday Discuss},
  2004, \textbf{126}, 141--157\relax
\mciteBstWouldAddEndPuncttrue
\mciteSetBstMidEndSepPunct{\mcitedefaultmidpunct}
{\mcitedefaultendpunct}{\mcitedefaultseppunct}\relax
\EndOfBibitem
\bibitem[Bakker~Schut \emph{et~al.}(2006)Bakker~Schut, Maquelin, van~der Kwast,
  Bangma, Kok, and Puppels]{BakkerSchut2006}
T.~C. Bakker~Schut, K.~Maquelin, T.~van~der Kwast, C.~H. Bangma, D.-J. .~J. Kok
  and G.~J. Puppels, \emph{Anal Chem}, 2006, \textbf{78}, 7761--7769\relax
\mciteBstWouldAddEndPuncttrue
\mciteSetBstMidEndSepPunct{\mcitedefaultmidpunct}
{\mcitedefaultendpunct}{\mcitedefaultseppunct}\relax
\EndOfBibitem
\bibitem[Gniadecka \emph{et~al.}(2004)Gniadecka, Philipsen, Sigurdsson, Wessel,
  Nielsen, Christensen, Hercogova, Rossen, Thomsen, Gniadecki, Hansen, and
  Wulf]{Gniadecka2004}
M.~Gniadecka, P.~A. Philipsen, S.~Sigurdsson, S.~Wessel, O.~F. Nielsen, D.~H.
  Christensen, J.~Hercogova, K.~Rossen, H.~K. Thomsen, R.~Gniadecki, L.~K.
  Hansen and H.~C. Wulf, \emph{J Invest Dermatol}, 2004, \textbf{122},
  443--449\relax
\mciteBstWouldAddEndPuncttrue
\mciteSetBstMidEndSepPunct{\mcitedefaultmidpunct}
{\mcitedefaultendpunct}{\mcitedefaultseppunct}\relax
\EndOfBibitem
\bibitem[Uhl\'{e}n \emph{et~al.}(2005)Uhl\'{e}n, Bj\"{o}rling, Agaton,
  Szigyarto, Amini, Andersen, Andersson, Angelidou, Asplund, Asplund, Berglund,
  Bergstr\"{o}m, Brumer, Cerjan, Ekstr\"{o}m, Elobeid, Eriksson, Fagerberg,
  Falk, Fall, Forsberg, Bj\"{o}rklund, Gumbel, Halimi, Hallin, Hamsten,
  Hansson, Hedhammar, Hercules, Kampf, Larsson, Lindskog, Lodewyckx, Lund,
  Lundeberg, Magnusson, Malm, Nilsson, Odling, Oksvold, Olsson, Oster,
  Ottosson, Paavilainen, Persson, Rimini, Rockberg, Runeson, Sivertsson,
  Sk\"{o}llermo, Steen, Stenvall, Sterky, Str\"{o}mberg, Sundberg, Tegel,
  Tourle, Wahlund, Wald\'{e}n, Wan, Wern\'{e}rus, Westberg, Wester, Wrethagen,
  Xu, Hober, and Pont\'{e}n]{Uhlen2005}
M.~Uhl\'{e}n, E.~Bj\"{o}rling, C.~Agaton, C.~A.-K. .~K. Szigyarto, B.~Amini,
  E.~Andersen, A.-C. .~C. Andersson, P.~Angelidou, A.~Asplund, C.~Asplund,
  L.~Berglund, K.~Bergstr\"{o}m, H.~Brumer, D.~Cerjan, M.~Ekstr\"{o}m,
  A.~Elobeid, C.~Eriksson, L.~Fagerberg, R.~Falk, J.~Fall, M.~Forsberg, M.~G.
  Bj\"{o}rklund, K.~Gumbel, A.~Halimi, I.~Hallin, C.~Hamsten, M.~Hansson,
  M.~Hedhammar, G.~Hercules, C.~Kampf, K.~Larsson, M.~Lindskog, W.~Lodewyckx,
  J.~Lund, J.~Lundeberg, K.~Magnusson, E.~Malm, P.~Nilsson, J.~Odling,
  P.~Oksvold, I.~Olsson, E.~Oster, J.~Ottosson, L.~Paavilainen, A.~Persson,
  R.~Rimini, J.~Rockberg, M.~Runeson, A.~Sivertsson, A.~Sk\"{o}llermo,
  J.~Steen, M.~Stenvall, F.~Sterky, S.~Str\"{o}mberg, M.~A. Sundberg, H.~Tegel,
  S.~Tourle, E.~Wahlund, A.~Wald\'{e}n, J.~Wan, H.~Wern\'{e}rus, J.~Westberg,
  K.~Wester, U.~Wrethagen, L.~L. Xu, S.~Hober and F.~Pont\'{e}n, \emph{Mol Cell
  Proteomics}, 2005, \textbf{4}, 1920--1932\relax
\mciteBstWouldAddEndPuncttrue
\mciteSetBstMidEndSepPunct{\mcitedefaultmidpunct}
{\mcitedefaultendpunct}{\mcitedefaultseppunct}\relax
\EndOfBibitem
\bibitem[Welsh \emph{et~al.}(2003)Welsh, Sapinoso, Kern, Brown, Liu, Bauskin,
  Ward, Hawkins, Quinn, Russell, Sutherland, Breit, Moskaluk, Frierson, and
  Hampton]{Welsh:PNAS:2003}
J.~B. Welsh, L.~M. Sapinoso, S.~G. Kern, D.~A. Brown, T.~Liu, A.~R. Bauskin,
  R.~L. Ward, N.~J. Hawkins, D.~I. Quinn, P.~J. Russell, R.~L. Sutherland,
  S.~N. Breit, C.~A. Moskaluk, H.~F. Frierson and G.~M. Hampton,
  \emph{Proceedings of the National Academy of Sciences}, 2003, \textbf{100},
  3410--3415\relax
\mciteBstWouldAddEndPuncttrue
\mciteSetBstMidEndSepPunct{\mcitedefaultmidpunct}
{\mcitedefaultendpunct}{\mcitedefaultseppunct}\relax
\EndOfBibitem
\bibitem[Honda \emph{et~al.}(1998)Honda, Yamada, Endo, Ino, Gotoh, Tsuda,
  Yamada, Chiba, and Hirohashi]{Honda:JCB:1998}
K.~Honda, T.~Yamada, R.~Endo, Y.~Ino, M.~Gotoh, H.~Tsuda, Y.~Yamada, H.~Chiba
  and S.~Hirohashi, \emph{The Journal of Cell Biology}, 1998, \textbf{140},
  1383--1393\relax
\mciteBstWouldAddEndPuncttrue
\mciteSetBstMidEndSepPunct{\mcitedefaultmidpunct}
{\mcitedefaultendpunct}{\mcitedefaultseppunct}\relax
\EndOfBibitem
\bibitem[Mauri \emph{et~al.}(2005)Mauri, Scarpa, Nascimbeni, Benazzi,
  Parmagnani, Mafficini, Della~Peruta, Bassi, Miyazaki, and
  Sorio]{Mauri:PASEB:2005}
P.~Mauri, A.~Scarpa, A.~C. Nascimbeni, L.~Benazzi, E.~Parmagnani, A.~Mafficini,
  M.~Della~Peruta, C.~Bassi, K.~Miyazaki and C.~Sorio, \emph{The FASEB
  Journal}, 2005\relax
\mciteBstWouldAddEndPuncttrue
\mciteSetBstMidEndSepPunct{\mcitedefaultmidpunct}
{\mcitedefaultendpunct}{\mcitedefaultseppunct}\relax
\EndOfBibitem
\bibitem[Alfonso \emph{et~al.}(2005)Alfonso, N\'{u}\~{n}ez, Madoz-Gurpide,
  Lombardia, S\'{a}nchez, and Casal]{Alfonso:Proteo:2005}
P.~Alfonso, A.~N\'{u}\~{n}ez, J.~Madoz-Gurpide, L.~Lombardia, L.~S\'{a}nchez
  and J.~I. Casal, \emph{Proteomics}, 2005, \textbf{5}, 2602--11\relax
\mciteBstWouldAddEndPuncttrue
\mciteSetBstMidEndSepPunct{\mcitedefaultmidpunct}
{\mcitedefaultendpunct}{\mcitedefaultseppunct}\relax
\EndOfBibitem
\bibitem[Levenson and Mansfield(2006)]{Levenson:Cytometry:2006}
R.~M. Levenson and J.~R. Mansfield, \emph{Cytometry Part A}, 2006,
  \textbf{69A}, 748--758\relax
\mciteBstWouldAddEndPuncttrue
\mciteSetBstMidEndSepPunct{\mcitedefaultmidpunct}
{\mcitedefaultendpunct}{\mcitedefaultseppunct}\relax
\EndOfBibitem
\bibitem[Zrazhevskiy \emph{et~al.}(2013)Zrazhevskiy, True, and
  Gao]{Zrazhevskiy:NatProt:2013}
P.~Zrazhevskiy, L.~D. True and X.~Gao, \emph{Nature protocols}, 2013,
  \textbf{8}, 1852--1869\relax
\mciteBstWouldAddEndPuncttrue
\mciteSetBstMidEndSepPunct{\mcitedefaultmidpunct}
{\mcitedefaultendpunct}{\mcitedefaultseppunct}\relax
\EndOfBibitem
\bibitem[Gerdes \emph{et~al.}(2013)Gerdes, Sevinsky, Sood, Adak, Bello,
  Bordwell, Can, Corwin, Dinn, Filkins, Hollman, Kamath, Kaanumalle, Kenny,
  Larsen, Lazare, Li, Lowes, McCulloch, McDonough, Montalto, Pang, Rittscher,
  Santamaria-Pang, Sarachan, Seel, Seppo, Shaikh, Sui, Zhang, and
  Ginty]{Gerdes2013}
M.~J. Gerdes, C.~J. Sevinsky, A.~Sood, S.~Adak, M.~O. Bello, A.~Bordwell,
  A.~Can, A.~Corwin, S.~Dinn, R.~J. Filkins, D.~Hollman, V.~Kamath,
  S.~Kaanumalle, K.~Kenny, M.~Larsen, M.~Lazare, Q.~Li, C.~Lowes, C.~C.
  McCulloch, E.~McDonough, M.~C. Montalto, Z.~Pang, J.~Rittscher,
  A.~Santamaria-Pang, B.~D. Sarachan, M.~L. Seel, A.~Seppo, K.~Shaikh, Y.~Sui,
  J.~Zhang and F.~Ginty, \emph{Proceedings of the National Academy of
  Sciences}, 2013, \textbf{110}, 11982--11987\relax
\mciteBstWouldAddEndPuncttrue
\mciteSetBstMidEndSepPunct{\mcitedefaultmidpunct}
{\mcitedefaultendpunct}{\mcitedefaultseppunct}\relax
\EndOfBibitem
\bibitem[Riordan \emph{et~al.}(2015)Riordan, Varma, West, and
  Brown]{Riordan2015}
D.~P. Riordan, S.~Varma, R.~B. West and P.~O. Brown, \emph{PLoS ONE}, 2015,
  \textbf{10}, e0128975--\relax
\mciteBstWouldAddEndPuncttrue
\mciteSetBstMidEndSepPunct{\mcitedefaultmidpunct}
{\mcitedefaultendpunct}{\mcitedefaultseppunct}\relax
\EndOfBibitem
\bibitem[Schubert \emph{et~al.}(1994)Schubert, Gross, Siderits, Deckenbaugh,
  He, and Becich]{Schubert1994}
E.~Schubert, W.~Gross, R.~H. Siderits, L.~Deckenbaugh, F.~He and M.~J. Becich,
  \emph{Seminars in diagnostic pathology}, 1994, \textbf{11}, 263--273\relax
\mciteBstWouldAddEndPuncttrue
\mciteSetBstMidEndSepPunct{\mcitedefaultmidpunct}
{\mcitedefaultendpunct}{\mcitedefaultseppunct}\relax
\EndOfBibitem
\bibitem[Pantanowitz \emph{et~al.}(2015)Pantanowitz, Farahani, and
  Parwani]{Pantanowitz2015}
L.~Pantanowitz, N.~Farahani and A.~Parwani, \emph{Pathology and Laboratory
  Medicine International}, 2015,  23\relax
\mciteBstWouldAddEndPuncttrue
\mciteSetBstMidEndSepPunct{\mcitedefaultmidpunct}
{\mcitedefaultendpunct}{\mcitedefaultseppunct}\relax
\EndOfBibitem
\bibitem[Pantanowitz \emph{et~al.}(2011)Pantanowitz, Valenstein, Evans, Kaplan,
  Pfeifer, Wilbur, Collins, and Colgan]{Pantanowitz:JPatholInform:2011}
L.~Pantanowitz, P.~N. Valenstein, A.~J. Evans, K.~J. Kaplan, J.~D. Pfeifer,
  D.~C. Wilbur, L.~C. Collins and T.~J. Colgan, \emph{J Pathol Inform}, 2011,
  \textbf{2}, 36\relax
\mciteBstWouldAddEndPuncttrue
\mciteSetBstMidEndSepPunct{\mcitedefaultmidpunct}
{\mcitedefaultendpunct}{\mcitedefaultseppunct}\relax
\EndOfBibitem
\bibitem[WSI:FDA:2016()]{WSI:FDA:2016}
\emph{Guidance for Industry and Food and Drug Administration Staff},
  \url{fda.gov/downloads/MedicalDevices/DeviceRegulationandGuidance/GuidanceDocuments/UCM435355.pdf}\relax
\mciteBstWouldAddEndPuncttrue
\mciteSetBstMidEndSepPunct{\mcitedefaultmidpunct}
{\mcitedefaultendpunct}{\mcitedefaultseppunct}\relax
\EndOfBibitem
\bibitem[M\'{e}ndez \emph{et~al.}(1998)M\'{e}ndez, Tahoces, Lado, Souto, and
  Vidal]{Mendez1998}
A.~J. M\'{e}ndez, P.~G. Tahoces, M.~J. Lado, M.~Souto and J.~J. Vidal,
  \emph{Med Phys}, 1998, \textbf{25}, 957--964\relax
\mciteBstWouldAddEndPuncttrue
\mciteSetBstMidEndSepPunct{\mcitedefaultmidpunct}
{\mcitedefaultendpunct}{\mcitedefaultseppunct}\relax
\EndOfBibitem
\bibitem[Orcutt(2015)]{Orcutt2015}
M.~Orcutt, \emph{MIT Technology Review}, 2015\relax
\mciteBstWouldAddEndPuncttrue
\mciteSetBstMidEndSepPunct{\mcitedefaultmidpunct}
{\mcitedefaultendpunct}{\mcitedefaultseppunct}\relax
\EndOfBibitem
\bibitem[He \emph{et~al.}(2010)He, Long, Antani, and Thoma]{He:SGAMA:2010}
L.~He, L.~R. Long, S.~Antani and G.~Thoma, \emph{Sequence and genome analysis:
  methods and applications}, 2010, \textbf{510}, 271--287\relax
\mciteBstWouldAddEndPuncttrue
\mciteSetBstMidEndSepPunct{\mcitedefaultmidpunct}
{\mcitedefaultendpunct}{\mcitedefaultseppunct}\relax
\EndOfBibitem
\bibitem[Mousavi \emph{et~al.}(2015)Mousavi, Monga, Rao, and
  Rao]{Mousavi:JPathInform:2015}
H.~Mousavi, V.~Monga, G.~Rao and A.~Rao, \emph{J Pathol Inform}, 2015,
  \textbf{6}, 15--15\relax
\mciteBstWouldAddEndPuncttrue
\mciteSetBstMidEndSepPunct{\mcitedefaultmidpunct}
{\mcitedefaultendpunct}{\mcitedefaultseppunct}\relax
\EndOfBibitem
\bibitem[Barker \emph{et~al.}(2016)Barker, Hoogi, Depeursinge, and
  Rubin]{Barker:MedicalImageAnalysis:2016}
J.~Barker, A.~Hoogi, A.~Depeursinge and D.~L. Rubin, \emph{Medical Image
  Analysis}, 2016, \textbf{30}, 60--71\relax
\mciteBstWouldAddEndPuncttrue
\mciteSetBstMidEndSepPunct{\mcitedefaultmidpunct}
{\mcitedefaultendpunct}{\mcitedefaultseppunct}\relax
\EndOfBibitem
\bibitem[Bengtsson \emph{et~al.}(2017)Bengtsson, Danielsen, Treanor, Gurcan,
  MacAulay, and Moln\'{a}r]{Bengtsson:CytA:2017}
E.~Bengtsson, H.~Danielsen, D.~Treanor, M.~N. Gurcan, C.~MacAulay and
  B.~Moln\'{a}r, \emph{Cytometry A}, 2017, \textbf{91}, 551--554\relax
\mciteBstWouldAddEndPuncttrue
\mciteSetBstMidEndSepPunct{\mcitedefaultmidpunct}
{\mcitedefaultendpunct}{\mcitedefaultseppunct}\relax
\EndOfBibitem
\bibitem[Gurcan \emph{et~al.}(2009)Gurcan, Boucheron, Can, Madabhushi, Rajpoot,
  and Yener]{Gurcan:IEEE:2009}
M.~N. Gurcan, L.~E. Boucheron, A.~Can, A.~Madabhushi, N.~M. Rajpoot and
  B.~Yener, \emph{IEEE Reviews in Biomedical Engineering}, 2009, \textbf{2},
  147--171\relax
\mciteBstWouldAddEndPuncttrue
\mciteSetBstMidEndSepPunct{\mcitedefaultmidpunct}
{\mcitedefaultendpunct}{\mcitedefaultseppunct}\relax
\EndOfBibitem
\bibitem[Irshad \emph{et~al.}(2014)Irshad, Veillard, Roux, and
  Racoceanu]{Irshad:IEEEReview:2014}
H.~Irshad, A.~Veillard, L.~Roux and D.~Racoceanu, \emph{IEEE Reviews in
  Biomedical Engineering}, 2014, \textbf{7}, 97--114\relax
\mciteBstWouldAddEndPuncttrue
\mciteSetBstMidEndSepPunct{\mcitedefaultmidpunct}
{\mcitedefaultendpunct}{\mcitedefaultseppunct}\relax
\EndOfBibitem
\bibitem[Marty(2007)]{Marty:BioTechnique:2007}
G.~D. Marty, \emph{BioTechniques}, 2007, \textbf{42}, 716\relax
\mciteBstWouldAddEndPuncttrue
\mciteSetBstMidEndSepPunct{\mcitedefaultmidpunct}
{\mcitedefaultendpunct}{\mcitedefaultseppunct}\relax
\EndOfBibitem
\bibitem[Ruifrok and Johnston(2001)]{Ruifrok:AQCH:2001}
A.~C. Ruifrok and D.~A. Johnston, \emph{Analytical and quantitative cytology
  and histology}, 2001, \textbf{23}, 291--299\relax
\mciteBstWouldAddEndPuncttrue
\mciteSetBstMidEndSepPunct{\mcitedefaultmidpunct}
{\mcitedefaultendpunct}{\mcitedefaultseppunct}\relax
\EndOfBibitem
\bibitem[Macenko \emph{et~al.}(2009)Macenko, Niethammer, Marron, Borland,
  Woosley, Guan, Schmitt, and Thomas]{Macenko:IEEE:2009}
M.~Macenko, M.~Niethammer, J.~S. Marron, D.~Borland, J.~T. Woosley, X.~Guan,
  C.~Schmitt and N.~E. Thomas, Biomedical Imaging: From Nano to Macro, 2009.
  ISBI'09. IEEE International Symposium on, 2009, pp. 1107--1110\relax
\mciteBstWouldAddEndPuncttrue
\mciteSetBstMidEndSepPunct{\mcitedefaultmidpunct}
{\mcitedefaultendpunct}{\mcitedefaultseppunct}\relax
\EndOfBibitem
\bibitem[Khan \emph{et~al.}(2014)Khan, Rajpoot, Treanor, and
  Magee]{Khan:IEEE:2014}
A.~M. Khan, N.~Rajpoot, D.~Treanor and D.~Magee, \emph{IEEE Transactions on
  Biomedical Engineering}, 2014, \textbf{61}, 1729--1738\relax
\mciteBstWouldAddEndPuncttrue
\mciteSetBstMidEndSepPunct{\mcitedefaultmidpunct}
{\mcitedefaultendpunct}{\mcitedefaultseppunct}\relax
\EndOfBibitem
\bibitem[Otsu(1975)]{Otsu:Auto:1975}
N.~Otsu, \emph{Automatica}, 1975, \textbf{11}, 23--27\relax
\mciteBstWouldAddEndPuncttrue
\mciteSetBstMidEndSepPunct{\mcitedefaultmidpunct}
{\mcitedefaultendpunct}{\mcitedefaultseppunct}\relax
\EndOfBibitem
\bibitem[Chang \emph{et~al.}(2000)Chang, Yu, and Vetterli]{Chang:IEEE:2000}
S.~G. Chang, B.~Yu and M.~Vetterli, \emph{IEEE Transactions on image
  Processing}, 2000, \textbf{9}, 1522--1531\relax
\mciteBstWouldAddEndPuncttrue
\mciteSetBstMidEndSepPunct{\mcitedefaultmidpunct}
{\mcitedefaultendpunct}{\mcitedefaultseppunct}\relax
\EndOfBibitem
\bibitem[Roerdink and Meijster(2000)]{Roerdink:FundaInform:2000}
J.~B. Roerdink and A.~Meijster, \emph{Fundamenta Informaticae}, 2000,
  \textbf{41}, 187--228\relax
\mciteBstWouldAddEndPuncttrue
\mciteSetBstMidEndSepPunct{\mcitedefaultmidpunct}
{\mcitedefaultendpunct}{\mcitedefaultseppunct}\relax
\EndOfBibitem
\bibitem[Bezdek \emph{et~al.}(1984)Bezdek, Ehrlich, and
  Full]{Bezdek:CompGeo:1984}
J.~C. Bezdek, R.~Ehrlich and W.~Full, \emph{Computers \& Geosciences}, 1984,
  \textbf{10}, 191--203\relax
\mciteBstWouldAddEndPuncttrue
\mciteSetBstMidEndSepPunct{\mcitedefaultmidpunct}
{\mcitedefaultendpunct}{\mcitedefaultseppunct}\relax
\EndOfBibitem
\bibitem[Moon(1996)]{Moon:IEEE:1996}
T.~K. Moon, \emph{IEEE Signal processing magazine}, 1996, \textbf{13},
  47--60\relax
\mciteBstWouldAddEndPuncttrue
\mciteSetBstMidEndSepPunct{\mcitedefaultmidpunct}
{\mcitedefaultendpunct}{\mcitedefaultseppunct}\relax
\EndOfBibitem
\bibitem[Kass \emph{et~al.}(1988)Kass, Witkin, and Terzopoulos]{Kass:IJCV:1988}
M.~Kass, A.~Witkin and D.~Terzopoulos, \emph{International journal of computer
  vision}, 1988, \textbf{1}, 321--331\relax
\mciteBstWouldAddEndPuncttrue
\mciteSetBstMidEndSepPunct{\mcitedefaultmidpunct}
{\mcitedefaultendpunct}{\mcitedefaultseppunct}\relax
\EndOfBibitem
\bibitem[Suri \emph{et~al.}(2002)Suri, Liu, Singh, Laxminarayan, Zeng, and
  Reden]{Suri:IEEE:2002}
J.~S. Suri, K.~Liu, S.~Singh, S.~N. Laxminarayan, X.~Zeng and L.~Reden,
  \emph{IEEE Transactions on information technology in biomedicine}, 2002,
  \textbf{6}, 8--28\relax
\mciteBstWouldAddEndPuncttrue
\mciteSetBstMidEndSepPunct{\mcitedefaultmidpunct}
{\mcitedefaultendpunct}{\mcitedefaultseppunct}\relax
\EndOfBibitem
\bibitem[Naik \emph{et~al.}(2008)Naik, Doyle, Agner, Madabhushi, Feldman, and
  Tomaszewski]{Naik:IEEE:2008}
S.~Naik, S.~Doyle, S.~Agner, A.~Madabhushi, M.~Feldman and J.~Tomaszewski, 2008
  5th IEEE International Symposium on Biomedical Imaging: From Nano to Macro,
  2008, pp. 284--287\relax
\mciteBstWouldAddEndPuncttrue
\mciteSetBstMidEndSepPunct{\mcitedefaultmidpunct}
{\mcitedefaultendpunct}{\mcitedefaultseppunct}\relax
\EndOfBibitem
\bibitem[Alilou \emph{et~al.}(2013)Alilou, Kovalev, and
  Taimouri]{Alilou:CMIAG:2013}
M.~Alilou, V.~Kovalev and V.~Taimouri, \emph{Computerized Medical Imaging and
  Graphics}, 2013, \textbf{37}, 488--499\relax
\mciteBstWouldAddEndPuncttrue
\mciteSetBstMidEndSepPunct{\mcitedefaultmidpunct}
{\mcitedefaultendpunct}{\mcitedefaultseppunct}\relax
\EndOfBibitem
\bibitem[Jolliffe(2002)]{Jolliffe:PCA:2002}
I.~Jolliffe, \emph{Principal component analysis}, Springer, New York, 2nd edn,
  2002\relax
\mciteBstWouldAddEndPuncttrue
\mciteSetBstMidEndSepPunct{\mcitedefaultmidpunct}
{\mcitedefaultendpunct}{\mcitedefaultseppunct}\relax
\EndOfBibitem
\bibitem[Joliffe and Morgan(1992)]{Joliffe:StatMeth:1992}
I.~T. Joliffe and B.~J.~T. Morgan, \emph{Statistical methods in medical
  research}, 1992, \textbf{1}, 69--95\relax
\mciteBstWouldAddEndPuncttrue
\mciteSetBstMidEndSepPunct{\mcitedefaultmidpunct}
{\mcitedefaultendpunct}{\mcitedefaultseppunct}\relax
\EndOfBibitem
\bibitem[Hyv\"{a}rinen \emph{et~al.}(2001)Hyv\"{a}rinen, Karhunen, and
  Oja]{Hyvarinen:ICA:2001}
A.~Hyv\"{a}rinen, J.~Karhunen and E.~Oja, \emph{Independent component
  analysis}, John Wiley \& Sons, New York, 2001, vol.~46\relax
\mciteBstWouldAddEndPuncttrue
\mciteSetBstMidEndSepPunct{\mcitedefaultmidpunct}
{\mcitedefaultendpunct}{\mcitedefaultseppunct}\relax
\EndOfBibitem
\bibitem[Mart\'{i}nez and Kak(2001)]{Martinez:IEEE:2001}
A.~M. Mart\'{i}nez and A.~C. Kak, \emph{IEEE transactions on pattern analysis
  and machine intelligence}, 2001, \textbf{23}, 228--233\relax
\mciteBstWouldAddEndPuncttrue
\mciteSetBstMidEndSepPunct{\mcitedefaultmidpunct}
{\mcitedefaultendpunct}{\mcitedefaultseppunct}\relax
\EndOfBibitem
\bibitem[Doyle \emph{et~al.}(2007)Doyle, Hwang, Shah, Madabhushi, Feldman, and
  Tomaszeweski]{Doyle:ISBI:2007}
S.~Doyle, M.~Hwang, K.~Shah, A.~Madabhushi, M.~Feldman and J.~Tomaszeweski,
  Biomedical imaging: from nano to macro, 2007. ISBI 2007. 4th IEEE
  international symposium on, 2007, pp. 1284--1287\relax
\mciteBstWouldAddEndPuncttrue
\mciteSetBstMidEndSepPunct{\mcitedefaultmidpunct}
{\mcitedefaultendpunct}{\mcitedefaultseppunct}\relax
\EndOfBibitem
\bibitem[Breiman(2001)]{Breiman:MachineLearning:2001}
L.~Breiman, \emph{Machine Learning}, 2001, \textbf{45}, 5--32\relax
\mciteBstWouldAddEndPuncttrue
\mciteSetBstMidEndSepPunct{\mcitedefaultmidpunct}
{\mcitedefaultendpunct}{\mcitedefaultseppunct}\relax
\EndOfBibitem
\bibitem[Valkonen \emph{et~al.}(2017)Valkonen, Kartasalo, Liimatainen, Nykter,
  Latonen, and Ruusuvuori]{Valkonen:CytA:2017}
M.~Valkonen, K.~Kartasalo, K.~Liimatainen, M.~Nykter, L.~Latonen and
  P.~Ruusuvuori, \emph{Cytometry A}, 2017, \textbf{91}, 555--565\relax
\mciteBstWouldAddEndPuncttrue
\mciteSetBstMidEndSepPunct{\mcitedefaultmidpunct}
{\mcitedefaultendpunct}{\mcitedefaultseppunct}\relax
\EndOfBibitem
\bibitem[Cirean \emph{et~al.}(2013)Cirean, Giusti, Gambardella, and
  Schmidhuber]{Cirean:MICCAI:2013}
D.~C. Cirean, A.~Giusti, L.~M. Gambardella and J.~Schmidhuber, \emph{Medical
  Image Computing and Computer-Assisted Intervention MICCAI 2013: 16th
  International Conference, Nagoya, Japan, September 22-26, 2013, Proceedings,
  Part II}, Springer Berlin Heidelberg, Berlin, Heidelberg, 2013, pp.
  411--418\relax
\mciteBstWouldAddEndPuncttrue
\mciteSetBstMidEndSepPunct{\mcitedefaultmidpunct}
{\mcitedefaultendpunct}{\mcitedefaultseppunct}\relax
\EndOfBibitem
\bibitem[Fuchs and Buhmann(2011)]{Fuchs:CMIG:2011}
T.~J. Fuchs and J.~M. Buhmann, \emph{Computerized Medical Imaging and
  Graphics}, 2011, \textbf{35}, 515--530\relax
\mciteBstWouldAddEndPuncttrue
\mciteSetBstMidEndSepPunct{\mcitedefaultmidpunct}
{\mcitedefaultendpunct}{\mcitedefaultseppunct}\relax
\EndOfBibitem
\bibitem[Fernandez \emph{et~al.}(2005)Fernandez, Bhargava, Hewitt, and
  Levin]{Fernandez:NatBio:2005}
D.~C. Fernandez, R.~Bhargava, S.~M. Hewitt and I.~W. Levin, \emph{Nat Biotech},
  2005, \textbf{23}, 469--474\relax
\mciteBstWouldAddEndPuncttrue
\mciteSetBstMidEndSepPunct{\mcitedefaultmidpunct}
{\mcitedefaultendpunct}{\mcitedefaultseppunct}\relax
\EndOfBibitem
\bibitem[Singh \emph{et~al.}(2004)Singh, Qaqish, Johnson, Ford~3rd, Foley,
  Maygarden, and Mohler]{Singh:AnalQuantCyto:2004}
S.~S. Singh, B.~Qaqish, J.~L. Johnson, O.~H. Ford~3rd, J.~F. Foley, S.~J.
  Maygarden and J.~L. Mohler, \emph{Analytical and quantitative cytology and
  histology/the International Academy of Cytology [and] American Society of
  Cytology}, 2004, \textbf{26}, 194--200\relax
\mciteBstWouldAddEndPuncttrue
\mciteSetBstMidEndSepPunct{\mcitedefaultmidpunct}
{\mcitedefaultendpunct}{\mcitedefaultseppunct}\relax
\EndOfBibitem
\bibitem[Weaver \emph{et~al.}(2003)Weaver, Krag, Manna, Ashikaga, Harlow, and
  Bauer]{Weaver:ModernPath:2003}
D.~L. Weaver, D.~N. Krag, E.~A. Manna, T.~Ashikaga, S.~P. Harlow and K.~D.
  Bauer, \emph{Modern pathology}, 2003, \textbf{16}, 1159--1163\relax
\mciteBstWouldAddEndPuncttrue
\mciteSetBstMidEndSepPunct{\mcitedefaultmidpunct}
{\mcitedefaultendpunct}{\mcitedefaultseppunct}\relax
\EndOfBibitem
\bibitem[Diem \emph{et~al.}(2013)Diem, Mazur, Lenau, Schubert, Bird,
  Miljkovi\'{c}, Krafft, and Popp]{Diem:JPB:2013}
M.~Diem, A.~Mazur, K.~Lenau, J.~Schubert, B.~Bird, M.~Miljkovi\'{c}, C.~Krafft
  and J.~Popp, \emph{Journal of Biophotonics}, 2013, \textbf{6}, 855--886\relax
\mciteBstWouldAddEndPuncttrue
\mciteSetBstMidEndSepPunct{\mcitedefaultmidpunct}
{\mcitedefaultendpunct}{\mcitedefaultseppunct}\relax
\EndOfBibitem
\bibitem[Baker \emph{et~al.}(2014)Baker, Trevisan, Bassan, Bhargava, Butler,
  Dorling, Fielden, Fogarty, Fullwood, Heys, Hughes, Lasch, Martin-Hirsch,
  Obinaju, Sockalingum, Sul\'{e}-Suso, Strong, Walsh, Wood, Gardner, and
  Martin]{Baker:NatProt:2014}
M.~J. Baker, J.~Trevisan, P.~Bassan, R.~Bhargava, H.~J. Butler, K.~M. Dorling,
  P.~R. Fielden, S.~W. Fogarty, N.~J. Fullwood, K.~A. Heys, C.~Hughes,
  P.~Lasch, P.~L. Martin-Hirsch, B.~Obinaju, G.~D. Sockalingum,
  J.~Sul\'{e}-Suso, R.~J. Strong, M.~J. Walsh, B.~R. Wood, P.~Gardner and F.~L.
  Martin, \emph{Nat Protoc}, 2014, \textbf{9}, 1771--91\relax
\mciteBstWouldAddEndPuncttrue
\mciteSetBstMidEndSepPunct{\mcitedefaultmidpunct}
{\mcitedefaultendpunct}{\mcitedefaultseppunct}\relax
\EndOfBibitem
\bibitem[Matthews \emph{et~al.}(2011)Matthews, Brolo, Lum, Duan, and
  Jirasek]{Matthews:PMB:2011}
Q.~Matthews, A.~Brolo, J.~Lum, X.~Duan and A.~Jirasek, \emph{Phys Med Biol},
  2011, \textbf{56}, 19--38\relax
\mciteBstWouldAddEndPuncttrue
\mciteSetBstMidEndSepPunct{\mcitedefaultmidpunct}
{\mcitedefaultendpunct}{\mcitedefaultseppunct}\relax
\EndOfBibitem
\bibitem[Krafft \emph{et~al.}(2006)Krafft, Sobottka, Schackert, and
  Salzer]{Krafft:JRS:2006}
C.~Krafft, S.~B. Sobottka, G.~Schackert and R.~Salzer, \emph{Journal of Raman
  Spectroscopy}, 2006, \textbf{37}, 367--375\relax
\mciteBstWouldAddEndPuncttrue
\mciteSetBstMidEndSepPunct{\mcitedefaultmidpunct}
{\mcitedefaultendpunct}{\mcitedefaultseppunct}\relax
\EndOfBibitem
\bibitem[Austin \emph{et~al.}(2016)Austin, Osseiran, and
  Evans]{Austin:Analyst:2016}
L.~A. Austin, S.~Osseiran and C.~L. Evans, \emph{Analyst}, 2016, \textbf{141},
  476--503\relax
\mciteBstWouldAddEndPuncttrue
\mciteSetBstMidEndSepPunct{\mcitedefaultmidpunct}
{\mcitedefaultendpunct}{\mcitedefaultseppunct}\relax
\EndOfBibitem
\bibitem[Eberhardt \emph{et~al.}(2015)Eberhardt, Stiebing, Matth\"{a}us,
  Schmitt, and Popp]{Eberhardt:ERMD:2015}
K.~Eberhardt, C.~Stiebing, C.~Matth\"{a}us, M.~Schmitt and J.~Popp,
  \emph{Expert Review of Molecular Diagnostics}, 2015, \textbf{15},
  773--787\relax
\mciteBstWouldAddEndPuncttrue
\mciteSetBstMidEndSepPunct{\mcitedefaultmidpunct}
{\mcitedefaultendpunct}{\mcitedefaultseppunct}\relax
\EndOfBibitem
\bibitem[Stevens \emph{et~al.}(2016)Stevens, Iping~Petterson, Day, and
  Stone]{Stevens:ChemSocRev:2016}
O.~Stevens, I.~E. Iping~Petterson, J.~C.~C. Day and N.~Stone, \emph{Chem Soc
  Rev}, 2016, \textbf{45}, 1919--34\relax
\mciteBstWouldAddEndPuncttrue
\mciteSetBstMidEndSepPunct{\mcitedefaultmidpunct}
{\mcitedefaultendpunct}{\mcitedefaultseppunct}\relax
\EndOfBibitem
\bibitem[Montgomery \emph{et~al.}(2001)Montgomery, Bronner, Goldblum, Greenson,
  Haber, Hart, Lamps, Lauwers, Lazenby, Lewin, Robert, Toledano, Shyr, and
  Washington]{Montgomery2001}
E.~Montgomery, M.~P. Bronner, J.~R. Goldblum, J.~K. Greenson, M.~M. Haber,
  J.~Hart, L.~W. Lamps, G.~Y. Lauwers, A.~J. Lazenby, D.~N. Lewin, M.~E.
  Robert, A.~Y. Toledano, Y.~Shyr and K.~Washington, \emph{Human Pathology},
  2001, \textbf{32}, 368--378\relax
\mciteBstWouldAddEndPuncttrue
\mciteSetBstMidEndSepPunct{\mcitedefaultmidpunct}
{\mcitedefaultendpunct}{\mcitedefaultseppunct}\relax
\EndOfBibitem
\bibitem[Jermyn \emph{et~al.}(2014)Jermyn, Mok, Desroches, Mercier,
  Saint-Arnaud, Bernstein, Guiot, Petrecca, and Leblond]{Jermyn:OSA:2014}
M.~Jermyn, K.~Mok, J.~Desroches, J.~Mercier, K.~Saint-Arnaud, L.~Bernstein,
  M.-C. .~C. Guiot, K.~Petrecca and F.~Leblond, Biomedical Optics 2014, Miami,
  Florida, 2014, pp. BS5A.4--\relax
\mciteBstWouldAddEndPuncttrue
\mciteSetBstMidEndSepPunct{\mcitedefaultmidpunct}
{\mcitedefaultendpunct}{\mcitedefaultseppunct}\relax
\EndOfBibitem
\bibitem[Hollon \emph{et~al.}(2016)Hollon, Lewis, Freudiger, Sunney~Xie, and
  Orringer]{Hollon2016}
T.~Hollon, S.~Lewis, C.~W. Freudiger, X.~Sunney~Xie and D.~A. Orringer,
  \emph{Neurosurgical Focus}, 2016, \textbf{40}, E9--\relax
\mciteBstWouldAddEndPuncttrue
\mciteSetBstMidEndSepPunct{\mcitedefaultmidpunct}
{\mcitedefaultendpunct}{\mcitedefaultseppunct}\relax
\EndOfBibitem
\bibitem[Manoharan \emph{et~al.}(1996)Manoharan, Wang, and Feld]{Manoharan1996}
R.~Manoharan, Y.~Wang and M.~S. Feld, \emph{Spectrochimica Acta Part A:
  Molecular and Biomolecular Spectroscopy}, 1996, \textbf{52}, 215--249\relax
\mciteBstWouldAddEndPuncttrue
\mciteSetBstMidEndSepPunct{\mcitedefaultmidpunct}
{\mcitedefaultendpunct}{\mcitedefaultseppunct}\relax
\EndOfBibitem
\bibitem[Petibois and Deleris(2006)]{Petibois:TrendBiotech:2006}
C.~Petibois and G.~Deleris, \emph{Trends Biotechnol}, 2006, \textbf{24},
  455--62\relax
\mciteBstWouldAddEndPuncttrue
\mciteSetBstMidEndSepPunct{\mcitedefaultmidpunct}
{\mcitedefaultendpunct}{\mcitedefaultseppunct}\relax
\EndOfBibitem
\bibitem[Meyer \emph{et~al.}(2011)Meyer, Bergner, Bielecki, Krafft, Akimov,
  Romeike, Reichart, Kalff, Dietzek, and Popp]{Meyer:JBO:2011}
T.~Meyer, N.~Bergner, C.~Bielecki, C.~Krafft, D.~Akimov, B.~F.~M. Romeike,
  R.~Reichart, R.~Kalff, B.~Dietzek and J.~Popp, \emph{Journal of Biomedical
  Optics}, 2011, \textbf{16}, 21110--21113\relax
\mciteBstWouldAddEndPuncttrue
\mciteSetBstMidEndSepPunct{\mcitedefaultmidpunct}
{\mcitedefaultendpunct}{\mcitedefaultseppunct}\relax
\EndOfBibitem
\bibitem[Pence and Mahadevan-Jansen(2016)]{Pence:ChemSocRev:2016}
I.~Pence and A.~Mahadevan-Jansen, \emph{Chem Soc Rev}, 2016, \textbf{45},
  1958--79\relax
\mciteBstWouldAddEndPuncttrue
\mciteSetBstMidEndSepPunct{\mcitedefaultmidpunct}
{\mcitedefaultendpunct}{\mcitedefaultseppunct}\relax
\EndOfBibitem
\bibitem[Shipp \emph{et~al.}(2017)Shipp, Sinjab, and Notingher]{Shipp:AOP:2017}
D.~Shipp, F.~Sinjab and I.~Notingher, \emph{Advances in Optics and Photonics},
  2017, \textbf{9}, year\relax
\mciteBstWouldAddEndPuncttrue
\mciteSetBstMidEndSepPunct{\mcitedefaultmidpunct}
{\mcitedefaultendpunct}{\mcitedefaultseppunct}\relax
\EndOfBibitem
\bibitem[Fukuda(1966)]{Fukuda:Histo:1966}
K.~Fukuda, \emph{Histochemie}, 1966, \textbf{6}, 127--130\relax
\mciteBstWouldAddEndPuncttrue
\mciteSetBstMidEndSepPunct{\mcitedefaultmidpunct}
{\mcitedefaultendpunct}{\mcitedefaultseppunct}\relax
\EndOfBibitem
\bibitem[De~Mul \emph{et~al.}(1984)De~Mul, Buiteveld, Lankester, Mud, and
  Greve]{DelMul:HumPath:1984}
F.~F.~M. De~Mul, H.~Buiteveld, J.~Lankester, J.~Mud and J.~Greve, \emph{Human
  Pathology}, 1984, \textbf{15}, 1062--1068\relax
\mciteBstWouldAddEndPuncttrue
\mciteSetBstMidEndSepPunct{\mcitedefaultmidpunct}
{\mcitedefaultendpunct}{\mcitedefaultseppunct}\relax
\EndOfBibitem
\bibitem[Stone \emph{et~al.}(2000)Stone, Stavroulaki, Kendall, Birchall, and
  Barr]{Stone:Lrng:2000}
N.~Stone, P.~Stavroulaki, C.~Kendall, M.~Birchall and H.~Barr, \emph{The
  Laryngoscope}, 2000, \textbf{110}, 1756--1763\relax
\mciteBstWouldAddEndPuncttrue
\mciteSetBstMidEndSepPunct{\mcitedefaultmidpunct}
{\mcitedefaultendpunct}{\mcitedefaultseppunct}\relax
\EndOfBibitem
\bibitem[Beljebbar \emph{et~al.}(2009)Beljebbar, Bouch\'{e}, Di\'{e}bold,
  Guillou, Palot, Eudes, and Manfait]{Beljebbar:CRO:2009}
A.~Beljebbar, O.~Bouch\'{e}, M.~D. Di\'{e}bold, P.~J. Guillou, J.~P. Palot,
  D.~Eudes and M.~Manfait, \emph{Critical Reviews in Oncology/Hematology},
  2009, \textbf{72}, 255--264\relax
\mciteBstWouldAddEndPuncttrue
\mciteSetBstMidEndSepPunct{\mcitedefaultmidpunct}
{\mcitedefaultendpunct}{\mcitedefaultseppunct}\relax
\EndOfBibitem
\bibitem[Gniadecka \emph{et~al.}(1997)Gniadecka, Wulf, Nielsen, Christensen,
  and Hercogova]{Gniadecka:PCPB:1997}
M.~Gniadecka, H.~C. Wulf, O.~F. Nielsen, D.~H. Christensen and J.~Hercogova,
  \emph{Photochemistry and Photobiology}, 1997, \textbf{66}, 418--423\relax
\mciteBstWouldAddEndPuncttrue
\mciteSetBstMidEndSepPunct{\mcitedefaultmidpunct}
{\mcitedefaultendpunct}{\mcitedefaultseppunct}\relax
\EndOfBibitem
\bibitem[Koljenovi\'{c} \emph{et~al.}(2002)Koljenovi\'{c}, Choo-Smith,
  Bakker~Schut, Kros, van~den Berge, and Puppels]{Koljenovic:LI:2002}
S.~Koljenovi\'{c}, L.-P. Choo-Smith, T.~C. Bakker~Schut, J.~M. Kros, H.~J.
  van~den Berge and G.~J. Puppels, \emph{Laboratory Investigation}, 2002,
  \textbf{82}, 1265--1277\relax
\mciteBstWouldAddEndPuncttrue
\mciteSetBstMidEndSepPunct{\mcitedefaultmidpunct}
{\mcitedefaultendpunct}{\mcitedefaultseppunct}\relax
\EndOfBibitem
\bibitem[Rehman \emph{et~al.}(2007)Rehman, Movasaghi, Tucker, Joel, Darr,
  Ruban, and Rehman]{Rehman:JRS:2007}
S.~Rehman, Z.~Movasaghi, A.~T. Tucker, S.~P. Joel, J.~A. Darr, A.~V. Ruban and
  I.~U. Rehman, \emph{Journal of Raman Spectroscopy}, 2007, \textbf{38},
  1345--1351\relax
\mciteBstWouldAddEndPuncttrue
\mciteSetBstMidEndSepPunct{\mcitedefaultmidpunct}
{\mcitedefaultendpunct}{\mcitedefaultseppunct}\relax
\EndOfBibitem
\bibitem[Chen \emph{et~al.}(2016)Chen, Shimada, Yabumoto, Okajima, Ando, Chang,
  Lee, Wong, Chiou, and Hamaguchi]{Chen:SciRep:2016}
P.-H.~H. Chen, R.~Shimada, S.~Yabumoto, H.~Okajima, M.~Ando, C.-T.~T. Chang,
  L.-T.~T. Lee, Y.-K.~K. Wong, A.~Chiou and H.-o.~O. Hamaguchi, \emph{Sci Rep},
  2016, \textbf{6}, 20097\relax
\mciteBstWouldAddEndPuncttrue
\mciteSetBstMidEndSepPunct{\mcitedefaultmidpunct}
{\mcitedefaultendpunct}{\mcitedefaultseppunct}\relax
\EndOfBibitem
\bibitem[Kaminaka \emph{et~al.}(2001)Kaminaka, Yamazaki, Ito, Kohda, and
  Hamaguchi]{Kaminaka:JRS:2001}
S.~Kaminaka, H.~Yamazaki, T.~Ito, E.~Kohda and H.-o. .~O. Hamaguchi,
  \emph{Journal of Raman Spectroscopy}, 2001, \textbf{32}, 139--141\relax
\mciteBstWouldAddEndPuncttrue
\mciteSetBstMidEndSepPunct{\mcitedefaultmidpunct}
{\mcitedefaultendpunct}{\mcitedefaultseppunct}\relax
\EndOfBibitem
\bibitem[Nijssen \emph{et~al.}(2007)Nijssen, Maquelin, Santos, Caspers,
  Bakker~Schut, den Hollander, Neumann, and Puppels]{Nijssen:JBO:2007}
A.~Nijssen, K.~Maquelin, L.~F. Santos, P.~J. Caspers, T.~C. Bakker~Schut, J.~C.
  den Hollander, M.~H.~A. Neumann and G.~J. Puppels, \emph{J Biomed Opt}, 2007,
  \textbf{12}, 034004\relax
\mciteBstWouldAddEndPuncttrue
\mciteSetBstMidEndSepPunct{\mcitedefaultmidpunct}
{\mcitedefaultendpunct}{\mcitedefaultseppunct}\relax
\EndOfBibitem
\bibitem[Mizuno \emph{et~al.}(1994)Mizuno, Kitajima, Kawauchi, Muraishi, and
  Ozaki]{Mizuno:JRS:1994}
A.~Mizuno, H.~Kitajima, K.~Kawauchi, S.~Muraishi and Y.~Ozaki, \emph{Journal of
  Raman Spectroscopy}, 1994, \textbf{25}, 25--29\relax
\mciteBstWouldAddEndPuncttrue
\mciteSetBstMidEndSepPunct{\mcitedefaultmidpunct}
{\mcitedefaultendpunct}{\mcitedefaultseppunct}\relax
\EndOfBibitem
\bibitem[Kwak \emph{et~al.}(2015)Kwak, Kajdacsy-Balla, Macias, Walsh, Sinha,
  and Bhargava]{Kwak2015}
J.~T. Kwak, A.~Kajdacsy-Balla, V.~Macias, M.~Walsh, S.~Sinha and R.~Bhargava,
  \emph{Sci Rep}, 2015, \textbf{5}, 8758\relax
\mciteBstWouldAddEndPuncttrue
\mciteSetBstMidEndSepPunct{\mcitedefaultmidpunct}
{\mcitedefaultendpunct}{\mcitedefaultseppunct}\relax
\EndOfBibitem
\bibitem[Manoharan \emph{et~al.}(1998)Manoharan, Shafer, Perelman, Wu, Chen,
  Deinum, Fitzmaurice, Myles, Crowe, Dasarl, and Feld]{Manoharan:PCPB:1998}
R.~Manoharan, K.~Shafer, L.~Perelman, J.~Wu, K.~Chen, G.~Deinum,
  M.~Fitzmaurice, J.~Myles, J.~Crowe, R.~R. Dasarl and M.~S. Feld,
  \emph{Photochemistry and Photobiology}, 1998, \textbf{67}, 15--22\relax
\mciteBstWouldAddEndPuncttrue
\mciteSetBstMidEndSepPunct{\mcitedefaultmidpunct}
{\mcitedefaultendpunct}{\mcitedefaultseppunct}\relax
\EndOfBibitem
\bibitem[Liu \emph{et~al.}(1992)Liu, Das, Glassman, Tang, Yoo, Zhu, Akins,
  Lubicz, Cleary, Prudente, Celmer, Caron, and Alfano]{Liu:JPPB:1992}
C.~H. Liu, B.~B. Das, W.~L.~S. Glassman, G.~C. Tang, K.~M. Yoo, H.~R. Zhu,
  D.~L. Akins, S.~S. Lubicz, J.~Cleary, R.~Prudente, E.~Celmer, A.~Caron and
  R.~R. Alfano, \emph{Journal of Photochemistry and Photobiology B: Biology},
  1992, \textbf{16}, 187--209\relax
\mciteBstWouldAddEndPuncttrue
\mciteSetBstMidEndSepPunct{\mcitedefaultmidpunct}
{\mcitedefaultendpunct}{\mcitedefaultseppunct}\relax
\EndOfBibitem
\bibitem[Utzinger \emph{et~al.}(2001)Utzinger, Heintzelman, Mahadevan-Jansen,
  Malpica, Follen, and Richards-Kortum]{Utzinger:ApplSpec:2001}
U.~Utzinger, D.~L. Heintzelman, A.~Mahadevan-Jansen, A.~Malpica, M.~Follen and
  R.~Richards-Kortum, \emph{Appl Spectrosc}, 2001, \textbf{55}, 955--959\relax
\mciteBstWouldAddEndPuncttrue
\mciteSetBstMidEndSepPunct{\mcitedefaultmidpunct}
{\mcitedefaultendpunct}{\mcitedefaultseppunct}\relax
\EndOfBibitem
\bibitem[You \emph{et~al.}(2016)You, Tu, Zhao, Liu, Chaney, Marjanovic, and
  Boppart]{You:SciRep:2016}
S.~You, H.~Tu, Y.~Zhao, Y.~Liu, E.~J. Chaney, M.~Marjanovic and S.~A. Boppart,
  \emph{Sci Rep}, 2016, \textbf{6}, 32922\relax
\mciteBstWouldAddEndPuncttrue
\mciteSetBstMidEndSepPunct{\mcitedefaultmidpunct}
{\mcitedefaultendpunct}{\mcitedefaultseppunct}\relax
\EndOfBibitem
\bibitem[Kast \emph{et~al.}(2008)Kast, Serhatkulu, Cao, Pandya, Dai, Thakur,
  Naik, Naik, Klein, Auner, and Rabah]{Kast:Biopolymers:2008}
R.~E. Kast, G.~K. Serhatkulu, A.~Cao, A.~K. Pandya, H.~Dai, J.~S. Thakur, V.~M.
  Naik, R.~Naik, M.~D. Klein, G.~W. Auner and R.~Rabah, \emph{Biopolymers},
  2008, \textbf{89}, 235--241\relax
\mciteBstWouldAddEndPuncttrue
\mciteSetBstMidEndSepPunct{\mcitedefaultmidpunct}
{\mcitedefaultendpunct}{\mcitedefaultseppunct}\relax
\EndOfBibitem
\bibitem[Miljkovi\'{c} \emph{et~al.}(2010)Miljkovi\'{c}, Chernenko, Romeo,
  Bird, Matth\"{a}us, and Diem]{Miljkovic:Anal:2010}
M.~Miljkovi\'{c}, T.~Chernenko, M.~J. Romeo, B.~Bird, C.~Matth\"{a}us and
  M.~Diem, \emph{The Analyst}, 2010, \textbf{135}, 2002--13\relax
\mciteBstWouldAddEndPuncttrue
\mciteSetBstMidEndSepPunct{\mcitedefaultmidpunct}
{\mcitedefaultendpunct}{\mcitedefaultseppunct}\relax
\EndOfBibitem
\bibitem[Vajna \emph{et~al.}(2011)Vajna, Patyi, Nagy, B{\'{o}}dis, Farkas, and
  Marosi]{Vajna2011}
B.~Vajna, G.~Patyi, Z.~Nagy, A.~B{\'{o}}dis, A.~Farkas and G.~Marosi,
  \emph{Journal of Raman Spectroscopy}, 2011, \textbf{42}, 1977--1986\relax
\mciteBstWouldAddEndPuncttrue
\mciteSetBstMidEndSepPunct{\mcitedefaultmidpunct}
{\mcitedefaultendpunct}{\mcitedefaultseppunct}\relax
\EndOfBibitem
\bibitem[Butler \emph{et~al.}(2016)Butler, Ashton, Bird, Cinque, Curtis,
  Esmonde-white, Fullwood, Gardner, Martin, Walsh, Mcainsh, Stone, Martin,
  Butler, and Martin-hirsch]{Butler2016}
H.~J. Butler, L.~Ashton, B.~Bird, G.~Cinque, K.~Curtis, K.~Esmonde-white, N.~J.
  Fullwood, B.~Gardner, P.~L. Martin, M.~J. Walsh, M.~R. Mcainsh, N.~Stone,
  F.~L. Martin, H.~J. Butler and P.~L. Martin-hirsch, \emph{Nature Protocols},
  2016, \textbf{11}, 1--47\relax
\mciteBstWouldAddEndPuncttrue
\mciteSetBstMidEndSepPunct{\mcitedefaultmidpunct}
{\mcitedefaultendpunct}{\mcitedefaultseppunct}\relax
\EndOfBibitem
\bibitem[Masia \emph{et~al.}(2013)Masia, Glen, Stephens, Borri, and
  Langbein]{Masia2013}
F.~Masia, A.~Glen, P.~Stephens, P.~Borri and W.~Langbein, \emph{Analytical
  Chemistry}, 2013, \textbf{85}, 10820--10828\relax
\mciteBstWouldAddEndPuncttrue
\mciteSetBstMidEndSepPunct{\mcitedefaultmidpunct}
{\mcitedefaultendpunct}{\mcitedefaultseppunct}\relax
\EndOfBibitem
\bibitem[Andrew and Hancewicz(1998)]{Andrew1998}
J.~J. Andrew and T.~M. Hancewicz, \emph{Applied Spectroscopy}, 1998,
  \textbf{52}, 797--807\relax
\mciteBstWouldAddEndPuncttrue
\mciteSetBstMidEndSepPunct{\mcitedefaultmidpunct}
{\mcitedefaultendpunct}{\mcitedefaultseppunct}\relax
\EndOfBibitem
\bibitem[Patel \emph{et~al.}(2011)Patel, Trevisan, Evans, Llabjani,
  Martin-Hirsch, Stringfellow, and Martin]{Patel2011}
I.~I. Patel, J.~Trevisan, G.~Evans, V.~Llabjani, P.~L. Martin-Hirsch, H.~F.
  Stringfellow and F.~L. Martin, \emph{The Analyst}, 2011, \textbf{136},
  4950--9\relax
\mciteBstWouldAddEndPuncttrue
\mciteSetBstMidEndSepPunct{\mcitedefaultmidpunct}
{\mcitedefaultendpunct}{\mcitedefaultseppunct}\relax
\EndOfBibitem
\bibitem[Zhang \emph{et~al.}(2013)Zhang, Wang, Slipchenko, Ben-Amotz, Weiner,
  and Cheng]{Zhang2013}
D.~Zhang, P.~Wang, M.~N. Slipchenko, D.~Ben-Amotz, A.~M. Weiner and J.~X.
  Cheng, \emph{Analytical Chemistry}, 2013, \textbf{85}, 98--106\relax
\mciteBstWouldAddEndPuncttrue
\mciteSetBstMidEndSepPunct{\mcitedefaultmidpunct}
{\mcitedefaultendpunct}{\mcitedefaultseppunct}\relax
\EndOfBibitem
\bibitem[Alfonso-Garc{\'{i}}a \emph{et~al.}(2017)Alfonso-Garc{\'{i}}a, Paugh,
  Farid, Garg, Jester, and Potma]{Alfonso-Garcia2017}
A.~Alfonso-Garc{\'{i}}a, J.~Paugh, M.~Farid, S.~Garg, J.~Jester and E.~Potma,
  \emph{Journal of Raman Spectroscopy}, 2017\relax
\mciteBstWouldAddEndPuncttrue
\mciteSetBstMidEndSepPunct{\mcitedefaultmidpunct}
{\mcitedefaultendpunct}{\mcitedefaultseppunct}\relax
\EndOfBibitem
\bibitem[Berger \emph{et~al.}(1999)Berger, Koo, Itzkan, Horowitz, and
  Feld]{Berger1999}
A.~J. Berger, T.~W. Koo, I.~Itzkan, G.~Horowitz and M.~S. Feld, \emph{Applied
  optics}, 1999, \textbf{38}, 2916--26\relax
\mciteBstWouldAddEndPuncttrue
\mciteSetBstMidEndSepPunct{\mcitedefaultmidpunct}
{\mcitedefaultendpunct}{\mcitedefaultseppunct}\relax
\EndOfBibitem
\bibitem[Bergholt \emph{et~al.}(2014)Bergholt, Zheng, Ho, Teh, Yeoh, Yan~So,
  Shabbir, and Huang]{Bergholt:Gastroenterology:2014}
M.~S. Bergholt, W.~Zheng, K.~Y. Ho, M.~Teh, K.~G. Yeoh, J.~B. Yan~So,
  A.~Shabbir and Z.~Huang, \emph{Gastroenterology}, 2014, \textbf{146},
  27--32\relax
\mciteBstWouldAddEndPuncttrue
\mciteSetBstMidEndSepPunct{\mcitedefaultmidpunct}
{\mcitedefaultendpunct}{\mcitedefaultseppunct}\relax
\EndOfBibitem
\bibitem[Widjaja \emph{et~al.}(2008)Widjaja, Zheng, and Huang]{Widjaja2008}
E.~Widjaja, W.~Zheng and Z.~Huang, \emph{International Journal of Oncology},
  2008, \textbf{32}, 653--662\relax
\mciteBstWouldAddEndPuncttrue
\mciteSetBstMidEndSepPunct{\mcitedefaultmidpunct}
{\mcitedefaultendpunct}{\mcitedefaultseppunct}\relax
\EndOfBibitem
\bibitem[R\"{o}sch \emph{et~al.}(2005)R\"{o}sch, Harz, Schmitt, Ronneberger,
  Burkhardt, Lankers, Hofer, Thiele, Ro, Peschke, and Motzkus]{Rosch2005}
P.~R\"{o}sch, M.~Harz, M.~Schmitt, O.~Ronneberger, H.~Burkhardt, M.~Lankers,
  S.~Hofer, H.~Thiele, P.~Ro, K.-d. .~D. Peschke and H.-w. .~W. Motzkus,
  \emph{Applied and Environmental Microbiology}, 2005, \textbf{71},
  1626--1637\relax
\mciteBstWouldAddEndPuncttrue
\mciteSetBstMidEndSepPunct{\mcitedefaultmidpunct}
{\mcitedefaultendpunct}{\mcitedefaultseppunct}\relax
\EndOfBibitem
\bibitem[Teh \emph{et~al.}(2009)Teh, Zheng, Lau, and Huang]{Teh2009}
S.~K. Teh, W.~Zheng, D.~P. Lau and Z.~Huang, \emph{The Analyst}, 2009,
  \textbf{134}, 1232\relax
\mciteBstWouldAddEndPuncttrue
\mciteSetBstMidEndSepPunct{\mcitedefaultmidpunct}
{\mcitedefaultendpunct}{\mcitedefaultseppunct}\relax
\EndOfBibitem
\bibitem[Kallenbach-Thieltges \emph{et~al.}(2013)Kallenbach-Thieltges,
  Gro{\ss}er{\"{u}}schkamp, Mosig, Diem, Tannapfel, and
  Gerwert]{Kallenbach-Thieltges2013}
A.~Kallenbach-Thieltges, F.~Gro{\ss}er{\"{u}}schkamp, A.~Mosig, M.~Diem,
  A.~Tannapfel and K.~Gerwert, \emph{Journal of Biophotonics}, 2013,
  \textbf{6}, 88--100\relax
\mciteBstWouldAddEndPuncttrue
\mciteSetBstMidEndSepPunct{\mcitedefaultmidpunct}
{\mcitedefaultendpunct}{\mcitedefaultseppunct}\relax
\EndOfBibitem
\bibitem[Old \emph{et~al.}(2016)Old, Lloyd, Nallala, Isabelle, Almond,
  Shepherd, Kendall, Shore, Barr, and Stone]{Old:Analyst:2016}
O.~J. Old, G.~R. Lloyd, J.~Nallala, M.~Isabelle, L.~M. Almond, N.~A. Shepherd,
  C.~A. Kendall, A.~C. Shore, H.~Barr and N.~Stone, \emph{Analyst}, 2016\relax
\mciteBstWouldAddEndPuncttrue
\mciteSetBstMidEndSepPunct{\mcitedefaultmidpunct}
{\mcitedefaultendpunct}{\mcitedefaultseppunct}\relax
\EndOfBibitem
\bibitem[Berger \emph{et~al.}(1998)Berger, Koo, Itzkan, and Feld]{Berger1998}
A.~J. Berger, T.-w. Koo, I.~Itzkan and M.~S. Feld, \emph{Analytical Chemistry},
  1998, \textbf{70}, 623--627\relax
\mciteBstWouldAddEndPuncttrue
\mciteSetBstMidEndSepPunct{\mcitedefaultmidpunct}
{\mcitedefaultendpunct}{\mcitedefaultseppunct}\relax
\EndOfBibitem
\bibitem[Lloyd \emph{et~al.}(2014)Lloyd, Almond, Stone, Shepherd, Sanders,
  Hutchings, Barr, and Kendall]{Lloyd2014}
G.~R. Lloyd, L.~M. Almond, N.~Stone, N.~Shepherd, S.~Sanders, J.~Hutchings,
  H.~Barr and C.~Kendall, \emph{The Analyst}, 2014, \textbf{139},
  381--388\relax
\mciteBstWouldAddEndPuncttrue
\mciteSetBstMidEndSepPunct{\mcitedefaultmidpunct}
{\mcitedefaultendpunct}{\mcitedefaultseppunct}\relax
\EndOfBibitem
\bibitem[LeCun \emph{et~al.}(2015)LeCun, Yoshua, and Geoffrey]{LeCun2015}
Y.~LeCun, B.~Yoshua and H.~Geoffrey, \emph{Nature}, 2015, \textbf{521},
  436--444\relax
\mciteBstWouldAddEndPuncttrue
\mciteSetBstMidEndSepPunct{\mcitedefaultmidpunct}
{\mcitedefaultendpunct}{\mcitedefaultseppunct}\relax
\EndOfBibitem
\bibitem[Gniadecka \emph{et~al.}(1997)Gniadecka, Wulf, Mortensen, Nielsen, and
  Christensen]{Gniadecka1997}
M.~Gniadecka, H.~C. Wulf, N.~N. Mortensen, O.~F. Nielsen and D.~H. Christensen,
  \emph{J Raman spectroscopy}, 1997, \textbf{28}, 125--129\relax
\mciteBstWouldAddEndPuncttrue
\mciteSetBstMidEndSepPunct{\mcitedefaultmidpunct}
{\mcitedefaultendpunct}{\mcitedefaultseppunct}\relax
\EndOfBibitem
\bibitem[Goodacre \emph{et~al.}(1998)Goodacre, Timmins, Burton, Kaderbhai,
  Woodward, Kell, and Rooney]{Goodacre1998}
R.~Goodacre, E.~M. Timmins, R.~Burton, N.~Kaderbhai, A.~M. Woodward, D.~B. Kell
  and P.~J. Rooney, \emph{Microbiology}, 1998, \textbf{144}, 1157--1170\relax
\mciteBstWouldAddEndPuncttrue
\mciteSetBstMidEndSepPunct{\mcitedefaultmidpunct}
{\mcitedefaultendpunct}{\mcitedefaultseppunct}\relax
\EndOfBibitem
\bibitem[Manescu \emph{et~al.}(2017)Manescu, {Jong Lee}, Camp, Cicerone, Brady,
  and Bajcsy]{Manescu2017}
P.~Manescu, Y.~{Jong Lee}, C.~Camp, M.~Cicerone, M.~Brady and P.~Bajcsy,
  \emph{Medical Image Analysis}, 2017, \textbf{37}, 37--45\relax
\mciteBstWouldAddEndPuncttrue
\mciteSetBstMidEndSepPunct{\mcitedefaultmidpunct}
{\mcitedefaultendpunct}{\mcitedefaultseppunct}\relax
\EndOfBibitem
\bibitem[Pilling \emph{et~al.}(2016)Pilling, Henderson, Shanks, Brown, Clarke,
  and Gardner]{Pilling:Anal:2017}
M.~J. Pilling, A.~Henderson, J.~H. Shanks, M.~D. Brown, N.~W. Clarke and
  P.~Gardner, \emph{Analyst}, 2016\relax
\mciteBstWouldAddEndPuncttrue
\mciteSetBstMidEndSepPunct{\mcitedefaultmidpunct}
{\mcitedefaultendpunct}{\mcitedefaultseppunct}\relax
\EndOfBibitem
\bibitem[Kwak \emph{et~al.}(2011)Kwak, Hewitt, Sinha, and
  Bhargava]{Kwak:BmcCancer:2011}
J.~T. Kwak, S.~M. Hewitt, S.~Sinha and R.~Bhargava, \emph{BMC Cancer}, 2011,
  \textbf{11}, 62\relax
\mciteBstWouldAddEndPuncttrue
\mciteSetBstMidEndSepPunct{\mcitedefaultmidpunct}
{\mcitedefaultendpunct}{\mcitedefaultseppunct}\relax
\EndOfBibitem
\bibitem[Chan \emph{et~al.}(2009)Chan, Lieu, Huser, and Li]{Chan:AnalChem:2009}
J.~W. Chan, D.~K. Lieu, T.~Huser and R.~A. Li, \emph{Analytical Chemistry},
  2009, \textbf{81}, 1324--1331\relax
\mciteBstWouldAddEndPuncttrue
\mciteSetBstMidEndSepPunct{\mcitedefaultmidpunct}
{\mcitedefaultendpunct}{\mcitedefaultseppunct}\relax
\EndOfBibitem
\bibitem[Matth\"{a}us \emph{et~al.}(2006)Matth\"{a}us, Boydston-White,
  Miljkovi\'{c}, Romeo, and Diem]{Matthaus:ApplSpec:2006}
C.~Matth\"{a}us, S.~Boydston-White, M.~Miljkovi\'{c}, M.~Romeo and M.~Diem,
  \emph{Applied spectroscopy}, 2006, \textbf{60}, 1--8\relax
\mciteBstWouldAddEndPuncttrue
\mciteSetBstMidEndSepPunct{\mcitedefaultmidpunct}
{\mcitedefaultendpunct}{\mcitedefaultseppunct}\relax
\EndOfBibitem
\bibitem[Okada \emph{et~al.}(2012)Okada, Smith, Palonpon, Endo, Kawata,
  Sodeoka, and Fujita]{Okada:PNAS:2012}
M.~Okada, N.~I. Smith, A.~F. Palonpon, H.~Endo, S.~Kawata, M.~Sodeoka and
  K.~Fujita, \emph{Proceedings of the National Academy of Sciences}, 2012,
  \textbf{109}, 28--32\relax
\mciteBstWouldAddEndPuncttrue
\mciteSetBstMidEndSepPunct{\mcitedefaultmidpunct}
{\mcitedefaultendpunct}{\mcitedefaultseppunct}\relax
\EndOfBibitem
\bibitem[Schulze \emph{et~al.}(2013)Schulze, Konorov, Piret, Blades, and
  Turner]{Schulze:Anal:2013}
H.~G. Schulze, S.~O. Konorov, J.~M. Piret, M.~W. Blades and R.~F. Turner,
  \emph{Analyst}, 2013, \textbf{138}, 3416--3423\relax
\mciteBstWouldAddEndPuncttrue
\mciteSetBstMidEndSepPunct{\mcitedefaultmidpunct}
{\mcitedefaultendpunct}{\mcitedefaultseppunct}\relax
\EndOfBibitem
\bibitem[Boydston-White \emph{et~al.}(2005)Boydston-White, Chernenko, Regina,
  Miljkovi\'{c}, Matth\"{a}us, and Diem]{Boydston:VibSpec:2005}
S.~Boydston-White, T.~Chernenko, A.~Regina, M.~s. Miljkovi\'{c},
  C.~Matth\"{a}us and M.~Diem, \emph{Vibrational Spectroscopy}, 2005,
  \textbf{38}, 169--177\relax
\mciteBstWouldAddEndPuncttrue
\mciteSetBstMidEndSepPunct{\mcitedefaultmidpunct}
{\mcitedefaultendpunct}{\mcitedefaultseppunct}\relax
\EndOfBibitem
\bibitem[Boydston-White \emph{et~al.}(1999)Boydston-White, Gopen, Houser,
  Bargonetti, and Diem]{Boydston:BioSpec:1999}
S.~Boydston-White, T.~Gopen, S.~Houser, J.~Bargonetti and M.~Diem,
  \emph{Biospectroscopy}, 1999, \textbf{5}, 219--227\relax
\mciteBstWouldAddEndPuncttrue
\mciteSetBstMidEndSepPunct{\mcitedefaultmidpunct}
{\mcitedefaultendpunct}{\mcitedefaultseppunct}\relax
\EndOfBibitem
\bibitem[Chan \emph{et~al.}(2008)Chan, Taylor, Lane, Zwerdling, Tuscano, and
  Huser]{Chan:AnalChem:2008}
J.~W. Chan, D.~S. Taylor, S.~M. Lane, T.~Zwerdling, J.~Tuscano and T.~Huser,
  \emph{Anal Chem}, 2008, \textbf{80}, 2180--7\relax
\mciteBstWouldAddEndPuncttrue
\mciteSetBstMidEndSepPunct{\mcitedefaultmidpunct}
{\mcitedefaultendpunct}{\mcitedefaultseppunct}\relax
\EndOfBibitem
\bibitem[Tiwari \emph{et~al.}(2016)Tiwari, Raman, Reddy, Ghetler, Tella, Han,
  Moon, Hoke, and Bhargava]{Tiwari:AnalChem:2016}
S.~Tiwari, J.~Raman, V.~Reddy, A.~Ghetler, R.~P. Tella, Y.~Han, C.~R. Moon,
  C.~D. Hoke and R.~Bhargava, \emph{Anal Chem}, 2016, \textbf{88},
  10183--10190\relax
\mciteBstWouldAddEndPuncttrue
\mciteSetBstMidEndSepPunct{\mcitedefaultmidpunct}
{\mcitedefaultendpunct}{\mcitedefaultseppunct}\relax
\EndOfBibitem
\bibitem[Popp \emph{et~al.}(2016)Popp, Krafft, Schmitt, Schie, Cialla-May,
  Matthaeus, and Bocklitz]{Popp:AGIE:2016}
J.~Popp, C.~Krafft, M.~Schmitt, I.~Schie, D.~Cialla-May, C.~Matthaeus and
  T.~Bocklitz, \emph{Angew Chem Int Ed Engl}, 2016\relax
\mciteBstWouldAddEndPuncttrue
\mciteSetBstMidEndSepPunct{\mcitedefaultmidpunct}
{\mcitedefaultendpunct}{\mcitedefaultseppunct}\relax
\EndOfBibitem
\bibitem[Armstrong \emph{et~al.}(1962)Armstrong, Bloembergen, Ducuing, and
  Pershan]{Armstrong:PR:1962}
J.~A. Armstrong, N.~Bloembergen, J.~Ducuing and P.~S. Pershan, \emph{Phys.
  Rev.}, 1962, \textbf{127}, 1918--1939\relax
\mciteBstWouldAddEndPuncttrue
\mciteSetBstMidEndSepPunct{\mcitedefaultmidpunct}
{\mcitedefaultendpunct}{\mcitedefaultseppunct}\relax
\EndOfBibitem
\bibitem[Maker and Terhune(1965)]{Maker:PR:1965}
P.~D. Maker and R.~W. Terhune, \emph{Physical Review}, 1965, \textbf{137},
  A801--A818\relax
\mciteBstWouldAddEndPuncttrue
\mciteSetBstMidEndSepPunct{\mcitedefaultmidpunct}
{\mcitedefaultendpunct}{\mcitedefaultseppunct}\relax
\EndOfBibitem
\bibitem[Duncan \emph{et~al.}(1982)Duncan, Reintjes, and Manuccia]{Duncan1982}
M.~D. Duncan, J.~Reintjes and T.~J. Manuccia, \emph{Optics Letters}, 1982,
  \textbf{7}, 350--352\relax
\mciteBstWouldAddEndPuncttrue
\mciteSetBstMidEndSepPunct{\mcitedefaultmidpunct}
{\mcitedefaultendpunct}{\mcitedefaultseppunct}\relax
\EndOfBibitem
\bibitem[Volkmer(2005)]{Volkmer:JPD:2005}
A.~Volkmer, \emph{Journal of Physics D: Applied Physics}, 2005, \textbf{38},
  R59--\relax
\mciteBstWouldAddEndPuncttrue
\mciteSetBstMidEndSepPunct{\mcitedefaultmidpunct}
{\mcitedefaultendpunct}{\mcitedefaultseppunct}\relax
\EndOfBibitem
\bibitem[Evans and Xie(2008)]{Evans:AnalRev:2008}
C.~L. Evans and X.~S. Xie, \emph{Annu Rev Anal Chem (Palo Alto Calif)}, 2008,
  \textbf{1}, 883--909\relax
\mciteBstWouldAddEndPuncttrue
\mciteSetBstMidEndSepPunct{\mcitedefaultmidpunct}
{\mcitedefaultendpunct}{\mcitedefaultseppunct}\relax
\EndOfBibitem
\bibitem[Pezacki \emph{et~al.}(2011)Pezacki, Blake, Danielson, Kennedy, Lyn,
  and Singaravelu]{Pezacki:NCB:2011}
J.~P. Pezacki, J.~A. Blake, D.~C. Danielson, D.~C. Kennedy, R.~K. Lyn and
  R.~Singaravelu, \emph{NATURE CHEMICAL BIOLOGY}, 2011, \textbf{7},
  137--145\relax
\mciteBstWouldAddEndPuncttrue
\mciteSetBstMidEndSepPunct{\mcitedefaultmidpunct}
{\mcitedefaultendpunct}{\mcitedefaultseppunct}\relax
\EndOfBibitem
\bibitem[Yue \emph{et~al.}(2012)Yue, C\'{a}rdenas-Mora, Chaboub, Leli\`{e}vre,
  and Cheng]{Yue:BPJ:2012}
S.~Yue, J.~C\'{a}rdenas-Mora, L.~Chaboub, S.~Leli\`{e}vre and J.-X. .~X. Cheng,
  \emph{Biophysical Journal}, 2012, \textbf{102}, 1215--1223\relax
\mciteBstWouldAddEndPuncttrue
\mciteSetBstMidEndSepPunct{\mcitedefaultmidpunct}
{\mcitedefaultendpunct}{\mcitedefaultseppunct}\relax
\EndOfBibitem
\bibitem[Camp~Jr and Cicerone(2015)]{CampJr:NatPhoton:2015}
C.~H. Camp~Jr and M.~T. Cicerone, \emph{Nat Photon}, 2015, \textbf{9},
  295--305\relax
\mciteBstWouldAddEndPuncttrue
\mciteSetBstMidEndSepPunct{\mcitedefaultmidpunct}
{\mcitedefaultendpunct}{\mcitedefaultseppunct}\relax
\EndOfBibitem
\bibitem[Potma and Mukamel(2013)]{Potma:JBook:2013}
E.~O. Potma and S.~Mukamel, in \emph{Theory of Coherent Raman Scattering}, CRC
  Press, Boca Raton, FL, 2013\relax
\mciteBstWouldAddEndPuncttrue
\mciteSetBstMidEndSepPunct{\mcitedefaultmidpunct}
{\mcitedefaultendpunct}{\mcitedefaultseppunct}\relax
\EndOfBibitem
\bibitem[Tolles \emph{et~al.}(1977)Tolles, Nibler, McDonald, and
  Harvey]{Tolles:AppSpec:1977}
W.~M. Tolles, J.~W. Nibler, J.~R. McDonald and A.~B. Harvey, \emph{Applied
  Spectroscopy}, 1977, \textbf{31}, 253--271\relax
\mciteBstWouldAddEndPuncttrue
\mciteSetBstMidEndSepPunct{\mcitedefaultmidpunct}
{\mcitedefaultendpunct}{\mcitedefaultseppunct}\relax
\EndOfBibitem
\bibitem[Cui \emph{et~al.}(2009)Cui, Bachler, and Ogilvie]{Cui:OptLett:2009}
M.~Cui, B.~R. Bachler and J.~P. Ogilvie, \emph{Optics Letters}, 2009,
  \textbf{34}, 773--775\relax
\mciteBstWouldAddEndPuncttrue
\mciteSetBstMidEndSepPunct{\mcitedefaultmidpunct}
{\mcitedefaultendpunct}{\mcitedefaultseppunct}\relax
\EndOfBibitem
\bibitem[Popov \emph{et~al.}(2011)Popov, Pegoraro, Stolow, and
  Ramunno]{Popov:OE:2011}
K.~I. Popov, A.~F. Pegoraro, A.~Stolow and L.~Ramunno, \emph{Optics Express},
  2011, \textbf{19}, 5902--5911\relax
\mciteBstWouldAddEndPuncttrue
\mciteSetBstMidEndSepPunct{\mcitedefaultmidpunct}
{\mcitedefaultendpunct}{\mcitedefaultseppunct}\relax
\EndOfBibitem
\bibitem[Barlow \emph{et~al.}(2013)Barlow, Popov, Andreana, Moffatt, Ridsdale,
  Slepkov, Harden, Ramunno, and Stolow]{Barlow:OE:2013}
A.~M. Barlow, K.~Popov, M.~Andreana, D.~J. Moffatt, A.~Ridsdale, A.~D. Slepkov,
  J.~L. Harden, L.~Ramunno and A.~Stolow, \emph{Optics Express}, 2013,
  \textbf{21}, 15298--15307\relax
\mciteBstWouldAddEndPuncttrue
\mciteSetBstMidEndSepPunct{\mcitedefaultmidpunct}
{\mcitedefaultendpunct}{\mcitedefaultseppunct}\relax
\EndOfBibitem
\bibitem[Hellerer \emph{et~al.}(2004)Hellerer, Enejder, and
  Zumbusch]{Hellerer:APL:2004}
T.~Hellerer, A.~M. Enejder and A.~Zumbusch, \emph{Applied Physics Letters},
  2004, \textbf{85}, 25--27\relax
\mciteBstWouldAddEndPuncttrue
\mciteSetBstMidEndSepPunct{\mcitedefaultmidpunct}
{\mcitedefaultendpunct}{\mcitedefaultseppunct}\relax
\EndOfBibitem
\bibitem[Chen \emph{et~al.}(2002)Chen, Volkmer, Book, and Xie]{Chen:JPCB:2002}
J.~X. Chen, A.~Volkmer, L.~D. Book and X.~S. Xie, \emph{Journal of Physical
  Chemistry B}, 2002, \textbf{106}, 8493--8498\relax
\mciteBstWouldAddEndPuncttrue
\mciteSetBstMidEndSepPunct{\mcitedefaultmidpunct}
{\mcitedefaultendpunct}{\mcitedefaultseppunct}\relax
\EndOfBibitem
\bibitem[Marks and Boppart(2004)]{Marks2004}
D.~L. Marks and S.~A. Boppart, \emph{Physical Review Letters}, 2004,
  \textbf{92}, 123905--\relax
\mciteBstWouldAddEndPuncttrue
\mciteSetBstMidEndSepPunct{\mcitedefaultmidpunct}
{\mcitedefaultendpunct}{\mcitedefaultseppunct}\relax
\EndOfBibitem
\bibitem[Kee and Cicerone(2004)]{Kee2004}
T.~W. Kee and M.~T. Cicerone, \emph{Optics Letters}, 2004, \textbf{29},
  2701--2703\relax
\mciteBstWouldAddEndPuncttrue
\mciteSetBstMidEndSepPunct{\mcitedefaultmidpunct}
{\mcitedefaultendpunct}{\mcitedefaultseppunct}\relax
\EndOfBibitem
\bibitem[Kano and Hamaguchi(2004)]{Kano:APL:2004}
H.~Kano and H.~Hamaguchi, \emph{Applied Physics Letters}, 2004, \textbf{85},
  4298--4300\relax
\mciteBstWouldAddEndPuncttrue
\mciteSetBstMidEndSepPunct{\mcitedefaultmidpunct}
{\mcitedefaultendpunct}{\mcitedefaultseppunct}\relax
\EndOfBibitem
\bibitem[Meyer \emph{et~al.}(2013)Meyer, Chemnitz, Baumgartl, Gottschall,
  Pascher, Matth\"{a}us, Romeike, Brehm, Limpert, T\"{u}nnermann, Schmitt,
  Dietzek, and Popp]{Meyer:AnalChem:2013}
T.~Meyer, M.~Chemnitz, M.~Baumgartl, T.~Gottschall, T.~Pascher,
  C.~Matth\"{a}us, B.~F.~M. Romeike, B.~R. Brehm, J.~Limpert,
  A.~T\"{u}nnermann, M.~Schmitt, B.~Dietzek and J.~Popp, \emph{Anal Chem},
  2013, \textbf{85}, 6703--6715\relax
\mciteBstWouldAddEndPuncttrue
\mciteSetBstMidEndSepPunct{\mcitedefaultmidpunct}
{\mcitedefaultendpunct}{\mcitedefaultseppunct}\relax
\EndOfBibitem
\bibitem[Wu \emph{et~al.}(2009)Wu, Chen, Chang, Jhan, Lin, and
  Liau]{Wu:AnalChem:2009}
Y.-M.~M. Wu, H.-C.~C. Chen, W.-T.~T. Chang, J.-W.~W. Jhan, H.-L.~L. Lin and
  I.~Liau, \emph{Anal Chem}, 2009, \textbf{81}, 1496--504\relax
\mciteBstWouldAddEndPuncttrue
\mciteSetBstMidEndSepPunct{\mcitedefaultmidpunct}
{\mcitedefaultendpunct}{\mcitedefaultseppunct}\relax
\EndOfBibitem
\bibitem[Evans \emph{et~al.}(2005)Evans, Potma, Puoris'~haag, C\^{o}t\'{e},
  Lin, and Xie]{Evans:PNAS:2005}
C.~L. Evans, E.~O. Potma, M.~Puoris'~haag, D.~C\^{o}t\'{e}, C.~P. Lin and X.~S.
  Xie, \emph{Proceedings of the National Academy of Sciences of the United
  States of America}, 2005, \textbf{102}, 16807--16812\relax
\mciteBstWouldAddEndPuncttrue
\mciteSetBstMidEndSepPunct{\mcitedefaultmidpunct}
{\mcitedefaultendpunct}{\mcitedefaultseppunct}\relax
\EndOfBibitem
\bibitem[Veilleux \emph{et~al.}(2008)Veilleux, Spencer, Biss, Cote, and
  Lin]{Veilleux:IEEE:2008}
I.~Veilleux, J.~A. Spencer, D.~P. Biss, D.~Cote and C.~P. Lin, \emph{IEEE
  Journal of Selected Topics in Quantum Electronics}, 2008, \textbf{14},
  10--18\relax
\mciteBstWouldAddEndPuncttrue
\mciteSetBstMidEndSepPunct{\mcitedefaultmidpunct}
{\mcitedefaultendpunct}{\mcitedefaultseppunct}\relax
\EndOfBibitem
\bibitem[Lin \emph{et~al.}(2003)Lin, Jones, Li, Zhu, Whitney, Muller, and
  Pollard]{Lin:AJP:2003}
E.~Y. Lin, J.~G. Jones, P.~Li, L.~Zhu, K.~D. Whitney, W.~J. Muller and J.~W.
  Pollard, \emph{The American journal of pathology}, 2003, \textbf{163},
  2113--2126\relax
\mciteBstWouldAddEndPuncttrue
\mciteSetBstMidEndSepPunct{\mcitedefaultmidpunct}
{\mcitedefaultendpunct}{\mcitedefaultseppunct}\relax
\EndOfBibitem
\bibitem[B\'{e}gin \emph{et~al.}(2011)B\'{e}gin, Burgoyne, Mercier, Villeneuve,
  Vall\'{e}e, and C\^{o}t\'{e}]{Begin:BOE:2011}
S.~B\'{e}gin, B.~Burgoyne, V.~Mercier, A.~Villeneuve, R.~Vall\'{e}e and
  D.~C\^{o}t\'{e}, \emph{Biomed Opt Express}, 2011, \textbf{2}, 1296--306\relax
\mciteBstWouldAddEndPuncttrue
\mciteSetBstMidEndSepPunct{\mcitedefaultmidpunct}
{\mcitedefaultendpunct}{\mcitedefaultseppunct}\relax
\EndOfBibitem
\bibitem[Pegoraro \emph{et~al.}(2009)Pegoraro, Ridsdale, Moffatt, Pezacki,
  Thomas, Fu, Dong, Fermann, and Stolow]{Pegoraro:OpticsExpress:2009}
A.~F. Pegoraro, A.~Ridsdale, D.~J. Moffatt, J.~P. Pezacki, B.~K. Thomas, L.~B.
  Fu, L.~Dong, M.~E. Fermann and A.~Stolow, \emph{Optics Express}, 2009,
  \textbf{17}, 20700--20706\relax
\mciteBstWouldAddEndPuncttrue
\mciteSetBstMidEndSepPunct{\mcitedefaultmidpunct}
{\mcitedefaultendpunct}{\mcitedefaultseppunct}\relax
\EndOfBibitem
\bibitem[Vartiainen(1992)]{Vartiainen:JOSAB:1992}
E.~M. Vartiainen, \emph{JOSA B}, 1992, \textbf{9}, 1209--1214\relax
\mciteBstWouldAddEndPuncttrue
\mciteSetBstMidEndSepPunct{\mcitedefaultmidpunct}
{\mcitedefaultendpunct}{\mcitedefaultseppunct}\relax
\EndOfBibitem
\bibitem[Liu \emph{et~al.}(2009)Liu, Lee, and Cicerone]{Liu2009}
Y.~X. Liu, Y.~J. Lee and M.~T. Cicerone, \emph{Optics Letters}, 2009,
  \textbf{34}, 1363--1365\relax
\mciteBstWouldAddEndPuncttrue
\mciteSetBstMidEndSepPunct{\mcitedefaultmidpunct}
{\mcitedefaultendpunct}{\mcitedefaultseppunct}\relax
\EndOfBibitem
\bibitem[Muller and Schins(2002)]{Muller:JCPB:2004}
M.~Muller and J.~M. Schins, \emph{Journal of Physical Chemistry B}, 2002,
  \textbf{106}, 3715--3723\relax
\mciteBstWouldAddEndPuncttrue
\mciteSetBstMidEndSepPunct{\mcitedefaultmidpunct}
{\mcitedefaultendpunct}{\mcitedefaultseppunct}\relax
\EndOfBibitem
\bibitem[Benalcazar \emph{et~al.}(2010)Benalcazar, Chowdary, Jiang, Marks,
  Chaney, Gruebele, and Boppart]{Benalcazar:IEEE:2010}
W.~A. Benalcazar, P.~D. Chowdary, Z.~Jiang, D.~L. Marks, E.~J. Chaney,
  M.~Gruebele and S.~A. Boppart, \emph{IEEE J SEL TOP QUANT}, 2010,
  \textbf{16}, 824--832\relax
\mciteBstWouldAddEndPuncttrue
\mciteSetBstMidEndSepPunct{\mcitedefaultmidpunct}
{\mcitedefaultendpunct}{\mcitedefaultseppunct}\relax
\EndOfBibitem
\bibitem[Evans \emph{et~al.}(2007)Evans, Xu, Kesari, Xie, Wong, and
  Young]{Evans:OpEx:2007}
C.~L. Evans, X.~Xu, S.~Kesari, X.~S. Xie, S.~T. Wong and G.~S. Young,
  \emph{Optics express}, 2007, \textbf{15}, 12076--12087\relax
\mciteBstWouldAddEndPuncttrue
\mciteSetBstMidEndSepPunct{\mcitedefaultmidpunct}
{\mcitedefaultendpunct}{\mcitedefaultseppunct}\relax
\EndOfBibitem
\bibitem[Lee \emph{et~al.}(2015)Lee, Downes, Chau, Serrels, Hastie, Elfick,
  Brunton, Frame, and Serrels]{Lee:Intravital:2015}
M.~Lee, A.~Downes, Y.-Y. .~Y. Chau, B.~Serrels, N.~Hastie, A.~Elfick,
  V.~Brunton, M.~Frame and A.~Serrels, \emph{IntraVital}, 2015\relax
\mciteBstWouldAddEndPuncttrue
\mciteSetBstMidEndSepPunct{\mcitedefaultmidpunct}
{\mcitedefaultendpunct}{\mcitedefaultseppunct}\relax
\EndOfBibitem
\bibitem[Meyer \emph{et~al.}(2012)Meyer, Bergner, Medyukhina, Dietzek, Krafft,
  Romeike, Reichart, Kalff, and Popp]{Meyer:JBO:2012}
T.~Meyer, N.~Bergner, A.~Medyukhina, B.~Dietzek, C.~Krafft, B.~F.~M. Romeike,
  R.~Reichart, R.~Kalff and J.~Popp, \emph{J Biophotonics}, 2012, \textbf{5},
  729--33\relax
\mciteBstWouldAddEndPuncttrue
\mciteSetBstMidEndSepPunct{\mcitedefaultmidpunct}
{\mcitedefaultendpunct}{\mcitedefaultseppunct}\relax
\EndOfBibitem
\bibitem[Gao \emph{et~al.}(2011)Gao, Li, Thrall, Yang, Xing, Hammoudi, Zhao,
  Massoud, Cagle, and Fan]{Gao:JBO:2011}
L.~Gao, F.~Li, M.~J. Thrall, Y.~Yang, J.~Xing, A.~A. Hammoudi, H.~Zhao,
  Y.~Massoud, P.~T. Cagle and Y.~Fan, \emph{Journal of biomedical optics},
  2011, \textbf{16}, 096004--096004\relax
\mciteBstWouldAddEndPuncttrue
\mciteSetBstMidEndSepPunct{\mcitedefaultmidpunct}
{\mcitedefaultendpunct}{\mcitedefaultseppunct}\relax
\EndOfBibitem
\bibitem[Gao \emph{et~al.}(2012)Gao, Wang, Li, Hammoudi, Thrall, Cagle, and
  Wong]{Gao:ArchPathLabMed:2012}
L.~Gao, Z.~Wang, F.~Li, A.~A. Hammoudi, M.~J. Thrall, P.~T. Cagle and S.~T.
  Wong, \emph{Archives of pathology \& laboratory medicine}, 2012,
  \textbf{136}, 1502--1510\relax
\mciteBstWouldAddEndPuncttrue
\mciteSetBstMidEndSepPunct{\mcitedefaultmidpunct}
{\mcitedefaultendpunct}{\mcitedefaultseppunct}\relax
\EndOfBibitem
\bibitem[Yang \emph{et~al.}(2011)Yang, Li, Gao, Wang, Thrall, Shen, Wong, and
  Wong]{Yang:BME:2011}
Y.~Yang, F.~Li, L.~Gao, Z.~Wang, M.~J. Thrall, S.~S. Shen, K.~K. Wong and S.~T.
  Wong, \emph{Biomedical optics express}, 2011, \textbf{2}, 2160--2174\relax
\mciteBstWouldAddEndPuncttrue
\mciteSetBstMidEndSepPunct{\mcitedefaultmidpunct}
{\mcitedefaultendpunct}{\mcitedefaultseppunct}\relax
\EndOfBibitem
\bibitem[Uckermann \emph{et~al.}(2014)Uckermann, Galli, Tamosaityte, Leipnitz,
  Geiger, Schackert, Koch, Steiner, and Kirsch]{Uckermann:PloS:2014}
O.~Uckermann, R.~Galli, S.~Tamosaityte, E.~Leipnitz, K.~D. Geiger,
  G.~Schackert, E.~Koch, G.~Steiner and M.~Kirsch, \emph{PLoS One}, 2014,
  \textbf{9}, e107115\relax
\mciteBstWouldAddEndPuncttrue
\mciteSetBstMidEndSepPunct{\mcitedefaultmidpunct}
{\mcitedefaultendpunct}{\mcitedefaultseppunct}\relax
\EndOfBibitem
\bibitem[Le \emph{et~al.}(2009)Le, Huff, and Cheng]{Le2009}
T.~T. Le, T.~B. Huff and J.~X. Cheng, \emph{BMC CANCER}, 2009, \textbf{9},
  year\relax
\mciteBstWouldAddEndPuncttrue
\mciteSetBstMidEndSepPunct{\mcitedefaultmidpunct}
{\mcitedefaultendpunct}{\mcitedefaultseppunct}\relax
\EndOfBibitem
\bibitem[Mitra \emph{et~al.}(2012)Mitra, Chao, Urasaki, Goodman, and
  Le]{Mitra:BmcCancer:2012}
R.~Mitra, O.~Chao, Y.~Urasaki, O.~B. Goodman and T.~T. Le, \emph{BMC Cancer},
  2012, \textbf{12}, 540\relax
\mciteBstWouldAddEndPuncttrue
\mciteSetBstMidEndSepPunct{\mcitedefaultmidpunct}
{\mcitedefaultendpunct}{\mcitedefaultseppunct}\relax
\EndOfBibitem
\bibitem[Chowdary \emph{et~al.}(2010)Chowdary, Jiang, Chaney, Benalcazar,
  Marks, Gruebele, and Boppart]{Chowdary2010}
P.~D. Chowdary, Z.~Jiang, E.~J. Chaney, W.~A. Benalcazar, D.~L. Marks,
  M.~Gruebele and S.~A. Boppart, \emph{Cancer Res}, 2010, \textbf{70},
  9562--9569\relax
\mciteBstWouldAddEndPuncttrue
\mciteSetBstMidEndSepPunct{\mcitedefaultmidpunct}
{\mcitedefaultendpunct}{\mcitedefaultseppunct}\relax
\EndOfBibitem
\bibitem[Heuke \emph{et~al.}(2013)Heuke, Vogler, Meyer, Akimov, Kluschke,
  R\"{o}wert-Huber, Lademann, Dietzek, and Popp]{Heuke:BJD:2013}
S.~Heuke, N.~Vogler, T.~Meyer, D.~Akimov, F.~Kluschke, H.-J.~J.
  R\"{o}wert-Huber, J.~Lademann, B.~Dietzek and J.~Popp, \emph{Br J Dermatol},
  2013, \textbf{169}, 794--803\relax
\mciteBstWouldAddEndPuncttrue
\mciteSetBstMidEndSepPunct{\mcitedefaultmidpunct}
{\mcitedefaultendpunct}{\mcitedefaultseppunct}\relax
\EndOfBibitem
\bibitem[Krafft \emph{et~al.}(2009)Krafft, Ramoji, Bielecki, Vogler, Meyer,
  Akimov, R\"{o}sch, Schmitt, Dietzek, Petersen, Stallmach, and
  Popp]{Krafft2009}
C.~Krafft, A.~A. Ramoji, C.~Bielecki, N.~Vogler, T.~Meyer, D.~Akimov,
  P.~R\"{o}sch, M.~Schmitt, B.~Dietzek, I.~Petersen, A.~Stallmach and J.~Popp,
  \emph{J. Biophoton.}, 2009, \textbf{2}, 303--312\relax
\mciteBstWouldAddEndPuncttrue
\mciteSetBstMidEndSepPunct{\mcitedefaultmidpunct}
{\mcitedefaultendpunct}{\mcitedefaultseppunct}\relax
\EndOfBibitem
\bibitem[Cicerone \emph{et~al.}(2012)Cicerone, Aamer, Lee, and
  Vartiainen]{Cicerone:JRS:2012}
M.~T. Cicerone, K.~A. Aamer, Y.~J. Lee and E.~Vartiainen, \emph{Journal of
  Raman Spectroscopy}, 2012, \textbf{43}, 637--643\relax
\mciteBstWouldAddEndPuncttrue
\mciteSetBstMidEndSepPunct{\mcitedefaultmidpunct}
{\mcitedefaultendpunct}{\mcitedefaultseppunct}\relax
\EndOfBibitem
\bibitem[Camp \emph{et~al.}(2016)Camp, Lee, and Cicerone]{Camp2015}
C.~H. Camp, Y.~J. Lee and M.~T. Cicerone, \emph{Journal of Raman Spectroscopy},
  2016, \textbf{47}, 408--415\relax
\mciteBstWouldAddEndPuncttrue
\mciteSetBstMidEndSepPunct{\mcitedefaultmidpunct}
{\mcitedefaultendpunct}{\mcitedefaultseppunct}\relax
\EndOfBibitem
\bibitem[Rajwade \emph{et~al.}(2013)Rajwade, Rangarajan, and
  Banerjee]{Rajwade:IEEE:2013}
A.~Rajwade, A.~Rangarajan and A.~Banerjee, \emph{IEEE Transactions on Pattern
  Analysis and Machine Intelligence}, 2013, \textbf{35}, 849--862\relax
\mciteBstWouldAddEndPuncttrue
\mciteSetBstMidEndSepPunct{\mcitedefaultmidpunct}
{\mcitedefaultendpunct}{\mcitedefaultseppunct}\relax
\EndOfBibitem
\bibitem[Pegoraro \emph{et~al.}(2012)Pegoraro, Slepkov, Ridsdale, Moffatt, and
  Stolow]{Pegoraro:JBP:2011}
A.~F. Pegoraro, A.~D. Slepkov, A.~Ridsdale, D.~J. Moffatt and A.~Stolow,
  \emph{Journal of Biophotonics}, 2012,  n/a--n/a\relax
\mciteBstWouldAddEndPuncttrue
\mciteSetBstMidEndSepPunct{\mcitedefaultmidpunct}
{\mcitedefaultendpunct}{\mcitedefaultseppunct}\relax
\EndOfBibitem
\bibitem[Parekh \emph{et~al.}(2010)Parekh, Lee, Aamer, and
  Cicerone]{Parekh:BJ:2010}
S.~H. Parekh, Y.~J. Lee, K.~A. Aamer and M.~T. Cicerone, \emph{Biophysical
  Journal}, 2010, \textbf{99}, 2695--2704\relax
\mciteBstWouldAddEndPuncttrue
\mciteSetBstMidEndSepPunct{\mcitedefaultmidpunct}
{\mcitedefaultendpunct}{\mcitedefaultseppunct}\relax
\EndOfBibitem
\bibitem[Okuno \emph{et~al.}(2010)Okuno, Kano, Leproux, Couderc, Day, Bonn, and
  Hamaguchi]{Okuno:ACIE:2010}
M.~Okuno, H.~Kano, P.~Leproux, V.~Couderc, J.~P.~R. Day, M.~Bonn and H.-o.~O.
  Hamaguchi, \emph{Angew Chem Int Ed Engl}, 2010, \textbf{49}, 6773--7\relax
\mciteBstWouldAddEndPuncttrue
\mciteSetBstMidEndSepPunct{\mcitedefaultmidpunct}
{\mcitedefaultendpunct}{\mcitedefaultseppunct}\relax
\EndOfBibitem
\bibitem[Camp~Jr \emph{et~al.}(2014)Camp~Jr, Lee, Heddleston, Hartshorn, Hight,
  Rich, Lathia, and Cicerone]{CampJr2014}
C.~H. Camp~Jr, Y.~J. Lee, J.~M. Heddleston, C.~M. Hartshorn, W.~R. Hight, J.~N.
  Rich, J.~D. Lathia and M.~T. Cicerone, \emph{Nat Photon}, 2014, \textbf{8},
  627--634\relax
\mciteBstWouldAddEndPuncttrue
\mciteSetBstMidEndSepPunct{\mcitedefaultmidpunct}
{\mcitedefaultendpunct}{\mcitedefaultseppunct}\relax
\EndOfBibitem
\bibitem[Pohling \emph{et~al.}(2011)Pohling, Buckup, Pagenstecher, and
  Motzkus]{Pohling:BMOE:2011}
C.~Pohling, T.~Buckup, A.~Pagenstecher and M.~Motzkus, \emph{Biomedical optics
  express}, 2011, \textbf{2}, 2110--2116\relax
\mciteBstWouldAddEndPuncttrue
\mciteSetBstMidEndSepPunct{\mcitedefaultmidpunct}
{\mcitedefaultendpunct}{\mcitedefaultseppunct}\relax
\EndOfBibitem
\bibitem[Lee \emph{et~al.}(2007)Lee, Liu, and Cicerone]{Lee2007}
Y.~J. Lee, Y.~Liu and M.~T. Cicerone, \emph{Optics Letters}, 2007, \textbf{32},
  3370--3372\relax
\mciteBstWouldAddEndPuncttrue
\mciteSetBstMidEndSepPunct{\mcitedefaultmidpunct}
{\mcitedefaultendpunct}{\mcitedefaultseppunct}\relax
\EndOfBibitem
\bibitem[Wang \emph{et~al.}(2016)Wang, Osseiran, Igras, Nichols, Roider,
  Pruessner, Tsao, Fisher, and Evans]{Wang:SciRep:2016}
H.~Wang, S.~Osseiran, V.~Igras, A.~J. Nichols, E.~M. Roider, J.~Pruessner,
  H.~Tsao, D.~E. Fisher and C.~L. Evans, \emph{Scientific Reports}, 2016,
  \textbf{6}, 37986--\relax
\mciteBstWouldAddEndPuncttrue
\mciteSetBstMidEndSepPunct{\mcitedefaultmidpunct}
{\mcitedefaultendpunct}{\mcitedefaultseppunct}\relax
\EndOfBibitem
\bibitem[Orringer \emph{et~al.}(2017)Orringer, Pandian, Niknafs, Hollon, Boyle,
  Lewis, Garrard, Hervey-Jumper, Garton, Maher, Heth, Sagher, Wilkinson,
  Snuderl, Venneti, Ramkissoon, McFadden, Fisher-Hubbard, Lieberman, Johnson,
  Xie, Trautman, Freudiger, and Camelo-Piragua]{Orringer:NatBME:2017}
D.~A. Orringer, B.~Pandian, Y.~S. Niknafs, T.~C. Hollon, J.~Boyle, S.~Lewis,
  M.~Garrard, S.~L. Hervey-Jumper, H.~J.~L. Garton, C.~O. Maher, J.~A. Heth,
  O.~Sagher, D.~A. Wilkinson, M.~Snuderl, S.~Venneti, S.~H. Ramkissoon, K.~A.
  McFadden, A.~Fisher-Hubbard, A.~P. Lieberman, T.~D. Johnson, X.~S. Xie, J.~K.
  Trautman, C.~W. Freudiger and S.~Camelo-Piragua, \emph{Nature Biomedical
  Engineering}, 2017, \textbf{1}, 0027\relax
\mciteBstWouldAddEndPuncttrue
\mciteSetBstMidEndSepPunct{\mcitedefaultmidpunct}
{\mcitedefaultendpunct}{\mcitedefaultseppunct}\relax
\EndOfBibitem
\bibitem[Otsuka \emph{et~al.}(2014)Otsuka, Satoh, Kyogaku, Hashimoto, Itoh, and
  Ozeki]{Otsuka:BiOS:2014}
Y.~Otsuka, S.~Satoh, M.~Kyogaku, H.~Hashimoto, K.~Itoh and Y.~Ozeki, SPIE BiOS,
  2014, pp. 89470C--89470C\relax
\mciteBstWouldAddEndPuncttrue
\mciteSetBstMidEndSepPunct{\mcitedefaultmidpunct}
{\mcitedefaultendpunct}{\mcitedefaultseppunct}\relax
\EndOfBibitem
\bibitem[Ozeki \emph{et~al.}(2009)Ozeki, Dake, Kajiyama, Fukui, and
  Itoh]{Ozeki:OptEx:2009}
Y.~Ozeki, F.~Dake, S.~Kajiyama, K.~Fukui and K.~Itoh, \emph{Optics Express},
  2009, \textbf{17}, 3651--3658\relax
\mciteBstWouldAddEndPuncttrue
\mciteSetBstMidEndSepPunct{\mcitedefaultmidpunct}
{\mcitedefaultendpunct}{\mcitedefaultseppunct}\relax
\EndOfBibitem
\bibitem[Saar \emph{et~al.}(2010)Saar, Freudiger, Reichman, Stanley, Holtom,
  and Xie]{Saar:Science:2011}
B.~G. Saar, C.~W. Freudiger, J.~Reichman, C.~M. Stanley, G.~R. Holtom and X.~S.
  Xie, \emph{Science}, 2010, \textbf{330}, 1368--70\relax
\mciteBstWouldAddEndPuncttrue
\mciteSetBstMidEndSepPunct{\mcitedefaultmidpunct}
{\mcitedefaultendpunct}{\mcitedefaultseppunct}\relax
\EndOfBibitem
\bibitem[Freudiger \emph{et~al.}(2012)Freudiger, Pfannl, Orringer, Saar, Ji,
  Zeng, Ottoboni, Ying, Waeber, Sims, De~Jager, Sagher, Philbert, Xu, Kesari,
  Xie, and Young]{Freudiger2012}
C.~W. Freudiger, R.~Pfannl, D.~A. Orringer, B.~G. Saar, M.~Ji, Q.~Zeng,
  L.~Ottoboni, W.~Ying, C.~Waeber, J.~R. Sims, P.~L. De~Jager, O.~Sagher, M.~A.
  Philbert, X.~Xu, S.~Kesari, X.~S. Xie and G.~S. Young, \emph{Lab Invest},
  2012, \textbf{92}, 1492--1502\relax
\mciteBstWouldAddEndPuncttrue
\mciteSetBstMidEndSepPunct{\mcitedefaultmidpunct}
{\mcitedefaultendpunct}{\mcitedefaultseppunct}\relax
\EndOfBibitem
\bibitem[Ozeki \emph{et~al.}(2012)Ozeki, Umemura, Otsuka, Satoh, Hashimoto,
  Sumimura, Nishizawa, Fukui, and Itoh]{Ozeki:NatPhot:2012}
Y.~Ozeki, W.~Umemura, Y.~Otsuka, S.~Satoh, H.~Hashimoto, K.~Sumimura,
  N.~Nishizawa, K.~Fukui and K.~Itoh, \emph{Nat Photon}, 2012, \textbf{6},
  845--851\relax
\mciteBstWouldAddEndPuncttrue
\mciteSetBstMidEndSepPunct{\mcitedefaultmidpunct}
{\mcitedefaultendpunct}{\mcitedefaultseppunct}\relax
\EndOfBibitem
\bibitem[Berto \emph{et~al.}(2014)Berto, Andresen, and
  Rigneault]{Berto:PRL:2014}
P.~Berto, E.~R. Andresen and H.~Rigneault, \emph{Phys. Rev. Lett.}, 2014,
  \textbf{112}, year\relax
\mciteBstWouldAddEndPuncttrue
\mciteSetBstMidEndSepPunct{\mcitedefaultmidpunct}
{\mcitedefaultendpunct}{\mcitedefaultseppunct}\relax
\EndOfBibitem
\bibitem[Fu \emph{et~al.}(2017)Fu, Yang, and Xie]{Fu:JACS:2016}
D.~Fu, W.~Yang and X.~S. Xie, \emph{J Am Chem Soc}, 2017, \textbf{139},
  583--586\relax
\mciteBstWouldAddEndPuncttrue
\mciteSetBstMidEndSepPunct{\mcitedefaultmidpunct}
{\mcitedefaultendpunct}{\mcitedefaultseppunct}\relax
\EndOfBibitem
\bibitem[Freudiger \emph{et~al.}(2014)Freudiger, Yang, Holtom, Peyghambarian,
  Xie, and Kieu]{Freudiger:NatPhot:2014}
C.~W. Freudiger, W.~Yang, G.~R. Holtom, N.~Peyghambarian, X.~S. Xie and K.~Q.
  Kieu, \emph{Nature Photonics}, 2014, \textbf{8}, 153--159\relax
\mciteBstWouldAddEndPuncttrue
\mciteSetBstMidEndSepPunct{\mcitedefaultmidpunct}
{\mcitedefaultendpunct}{\mcitedefaultseppunct}\relax
\EndOfBibitem
\bibitem[Nandakumar \emph{et~al.}(2009)Nandakumar, Kovalev, and
  Volkmer]{Nandakumar:NJP:2009}
P.~Nandakumar, A.~Kovalev and A.~Volkmer, \emph{New Journal of Physics}, 2009,
  \textbf{11}, 033026--\relax
\mciteBstWouldAddEndPuncttrue
\mciteSetBstMidEndSepPunct{\mcitedefaultmidpunct}
{\mcitedefaultendpunct}{\mcitedefaultseppunct}\relax
\EndOfBibitem
\bibitem[Freudiger \emph{et~al.}(2008)Freudiger, Min, Saar, Lu, Holtom, He,
  Tsai, Kang, and Xie]{Freudiger2008}
C.~W. Freudiger, W.~Min, B.~G. Saar, S.~Lu, G.~R. Holtom, C.~W. He, J.~C. Tsai,
  J.~X. Kang and X.~S. Xie, \emph{Science}, 2008, \textbf{322},
  1857--1861\relax
\mciteBstWouldAddEndPuncttrue
\mciteSetBstMidEndSepPunct{\mcitedefaultmidpunct}
{\mcitedefaultendpunct}{\mcitedefaultseppunct}\relax
\EndOfBibitem
\bibitem[Fu \emph{et~al.}(2013)Fu, Holtom, Freudiger, Zhang, and
  Xie]{Fu:JPCB:2013}
D.~Fu, G.~Holtom, C.~Freudiger, X.~Zhang and X.~S. Xie, \emph{The Journal of
  Physical Chemistry B}, 2013, \textbf{117}, 4634--4640\relax
\mciteBstWouldAddEndPuncttrue
\mciteSetBstMidEndSepPunct{\mcitedefaultmidpunct}
{\mcitedefaultendpunct}{\mcitedefaultseppunct}\relax
\EndOfBibitem
\bibitem[Alshaykh \emph{et~al.}(2017)Alshaykh, Liao, Sandoval, Gitzinger,
  Forget, Leaird, Cheng, and Weiner]{Alshaykh:OptLett:2017}
M.~S. Alshaykh, C.-S. .~S. Liao, O.~E. Sandoval, G.~Gitzinger, N.~Forget, D.~E.
  Leaird, J.-X. .~X. Cheng and A.~M. Weiner, \emph{Opt Lett}, 2017,
  \textbf{42}, 1548--1551\relax
\mciteBstWouldAddEndPuncttrue
\mciteSetBstMidEndSepPunct{\mcitedefaultmidpunct}
{\mcitedefaultendpunct}{\mcitedefaultseppunct}\relax
\EndOfBibitem
\bibitem[Liao \emph{et~al.}(2015)Liao, Slipchenko, Wang, Li, Lee, Oglesbee, and
  Cheng]{Liao:Light:2015}
C.-S. .~S. Liao, M.~N. Slipchenko, P.~Wang, J.~Li, S.-Y. .~Y. Lee, R.~A.
  Oglesbee and J.-X. .~X. Cheng, \emph{Light: Science \& Applications}, 2015,
  \textbf{4}, e265\relax
\mciteBstWouldAddEndPuncttrue
\mciteSetBstMidEndSepPunct{\mcitedefaultmidpunct}
{\mcitedefaultendpunct}{\mcitedefaultseppunct}\relax
\EndOfBibitem
\bibitem[Mittal \emph{et~al.}(2013)Mittal, Balu, Krasieva, Potma, Elkeeb,
  Zachary, and WilderSmith]{Mittal:LasersInSurgeryAndMedicine:2013}
R.~Mittal, M.~Balu, T.~Krasieva, E.~O. Potma, L.~Elkeeb, C.~B. Zachary and
  P.~WilderSmith, \emph{Lasers in Surgery and Medicine}, 2013, \textbf{45},
  496--502\relax
\mciteBstWouldAddEndPuncttrue
\mciteSetBstMidEndSepPunct{\mcitedefaultmidpunct}
{\mcitedefaultendpunct}{\mcitedefaultseppunct}\relax
\EndOfBibitem
\bibitem[Lu \emph{et~al.}(2016)Lu, Calligaris, Olubiyi, Norton, Yang,
  Santagata, Xie, Golby, and Agar]{Lu:CancerRes:2016}
F.-K. .~K. Lu, D.~Calligaris, O.~I. Olubiyi, I.~Norton, W.~Yang, S.~Santagata,
  X.~S. Xie, A.~J. Golby and N.~Y.~R. Agar, \emph{Cancer Research}, 2016\relax
\mciteBstWouldAddEndPuncttrue
\mciteSetBstMidEndSepPunct{\mcitedefaultmidpunct}
{\mcitedefaultendpunct}{\mcitedefaultseppunct}\relax
\EndOfBibitem
\bibitem[Ji \emph{et~al.}(2015)Ji, Lewis, Camelo-Piragua, Ramkissoon, Snuderl,
  Venneti, Fisher-Hubbard, Garrard, Fu, Wang, Heth, Maher, Sanai, Johnson,
  Freudiger, Sagher, Xie, and Orringer]{Ji:SciTransMed:2015}
M.~Ji, S.~Lewis, S.~Camelo-Piragua, S.~H. Ramkissoon, M.~Snuderl, S.~Venneti,
  A.~Fisher-Hubbard, M.~Garrard, D.~Fu, A.~C. Wang, J.~A. Heth, C.~O. Maher,
  N.~Sanai, T.~D. Johnson, C.~W. Freudiger, O.~Sagher, X.~S. Xie and D.~A.
  Orringer, \emph{Sci Transl Med}, 2015, \textbf{7}, 309ra163\relax
\mciteBstWouldAddEndPuncttrue
\mciteSetBstMidEndSepPunct{\mcitedefaultmidpunct}
{\mcitedefaultendpunct}{\mcitedefaultseppunct}\relax
\EndOfBibitem
\bibitem[Cui \emph{et~al.}(2017)Cui, Wang, and Yue]{Cui:SPIE:2017}
S.~Cui, P.~Wang and S.~Yue, Proc. of SPIE, 2017, pp. 100690M--1\relax
\mciteBstWouldAddEndPuncttrue
\mciteSetBstMidEndSepPunct{\mcitedefaultmidpunct}
{\mcitedefaultendpunct}{\mcitedefaultseppunct}\relax
\EndOfBibitem
\bibitem[Egawa \emph{et~al.}(2016)Egawa, Tokunaga, Hosoi, Iwanaga, and
  Ozeki]{Egawa:JBO:2016}
M.~Egawa, K.~Tokunaga, J.~Hosoi, S.~Iwanaga and Y.~Ozeki, \emph{J Biomed Opt},
  2016, \textbf{21}, 86017\relax
\mciteBstWouldAddEndPuncttrue
\mciteSetBstMidEndSepPunct{\mcitedefaultmidpunct}
{\mcitedefaultendpunct}{\mcitedefaultseppunct}\relax
\EndOfBibitem
\bibitem[Galli \emph{et~al.}(2016)Galli, Uckermann, Temme, Leipnitz, Meinhardt,
  Koch, Schackert, Steiner, and Kirsch]{Galli:JBPhot:2016}
R.~Galli, O.~Uckermann, A.~Temme, E.~Leipnitz, M.~Meinhardt, E.~Koch,
  G.~Schackert, G.~Steiner and M.~Kirsch, \emph{J Biophotonics}, 2016\relax
\mciteBstWouldAddEndPuncttrue
\mciteSetBstMidEndSepPunct{\mcitedefaultmidpunct}
{\mcitedefaultendpunct}{\mcitedefaultseppunct}\relax
\EndOfBibitem
\bibitem[Yue \emph{et~al.}(2014)Yue, Li, Lee, Lee, Shao, Song, Cheng,
  Masterson, Liu, Ratliff, and Cheng]{Yue:CellMet:2014}
S.~Yue, J.~Li, S.-Y. .~Y. Lee, H.~Lee, T.~Shao, B.~Song, L.~Cheng,
  T.~Masterson, X.~Liu, T.~Ratliff and J.-X. .~X. Cheng, \emph{Cell
  Metabolism}, 2014, \textbf{19}, 393--406\relax
\mciteBstWouldAddEndPuncttrue
\mciteSetBstMidEndSepPunct{\mcitedefaultmidpunct}
{\mcitedefaultendpunct}{\mcitedefaultseppunct}\relax
\EndOfBibitem
\bibitem[Zhang \emph{et~al.}(2012)Zhang, Roeffaers, Basu, Daniele, Fu,
  Freudiger, Holtom, and Xie]{Zhang:PCP:2012}
X.~Zhang, M.~B.~J. Roeffaers, S.~Basu, J.~R. Daniele, D.~Fu, C.~W. Freudiger,
  G.~R. Holtom and X.~S. Xie, \emph{Chemphyschem}, 2012, \textbf{13},
  1054--9\relax
\mciteBstWouldAddEndPuncttrue
\mciteSetBstMidEndSepPunct{\mcitedefaultmidpunct}
{\mcitedefaultendpunct}{\mcitedefaultseppunct}\relax
\EndOfBibitem
\bibitem[Wang \emph{et~al.}(2016)Wang, Zheng, Lin, and
  Huang]{Wang:ProcSPIE:2016}
Z.~Wang, W.~Zheng, J.~Lin and Z.~Huang, \emph{Proc. SPIE}, 2016, \textbf{9712},
  97120G--97120G--6\relax
\mciteBstWouldAddEndPuncttrue
\mciteSetBstMidEndSepPunct{\mcitedefaultmidpunct}
{\mcitedefaultendpunct}{\mcitedefaultseppunct}\relax
\EndOfBibitem
\bibitem[Tam \emph{et~al.}(2015)Tam, Lim, Wistuba, Tamrazi, Kuo, Ziv, Wong,
  Shih, Webster, Fischer, Nagrath, Davis, White, and Ahrar]{Tam:JVIR:2016}
A.~L. Tam, H.~J. Lim, I.~I. Wistuba, A.~Tamrazi, M.~D. Kuo, E.~Ziv, S.~Wong,
  A.~J. Shih, R.~J. Webster, G.~S. Fischer, S.~Nagrath, S.~E. Davis, S.~B.
  White and K.~Ahrar, \emph{Journal of vascular and interventional radiology :
  JVIR}, 2015, \textbf{27}, 8--19\relax
\mciteBstWouldAddEndPuncttrue
\mciteSetBstMidEndSepPunct{\mcitedefaultmidpunct}
{\mcitedefaultendpunct}{\mcitedefaultseppunct}\relax
\EndOfBibitem
\bibitem[Frank \emph{et~al.}(1995)Frank, McCreery, and
  Redd]{Frank:AnalChem:1995}
C.~J. Frank, R.~L. McCreery and D.~C. Redd, \emph{Analytical chemistry}, 1995,
  \textbf{67}, 777--783\relax
\mciteBstWouldAddEndPuncttrue
\mciteSetBstMidEndSepPunct{\mcitedefaultmidpunct}
{\mcitedefaultendpunct}{\mcitedefaultseppunct}\relax
\EndOfBibitem
\bibitem[Balu \emph{et~al.}(2010)Balu, Liu, Chen, Tromberg, and
  Potma]{Balu:OptEx:2010}
M.~Balu, G.~Liu, Z.~Chen, B.~J. Tromberg and E.~O. Potma, \emph{Opt. Express},
  2010, \textbf{18}, 2380--2388\relax
\mciteBstWouldAddEndPuncttrue
\mciteSetBstMidEndSepPunct{\mcitedefaultmidpunct}
{\mcitedefaultendpunct}{\mcitedefaultseppunct}\relax
\EndOfBibitem
\bibitem[Saar \emph{et~al.}(2011)Saar, Johnston, Freudiger, Xie, and
  Seibel]{Saar:OptLett:2011}
B.~G. Saar, R.~S. Johnston, C.~W. Freudiger, X.~S. Xie and E.~J. Seibel,
  \emph{Optics letters}, 2011, \textbf{36}, 2396--2398\relax
\mciteBstWouldAddEndPuncttrue
\mciteSetBstMidEndSepPunct{\mcitedefaultmidpunct}
{\mcitedefaultendpunct}{\mcitedefaultseppunct}\relax
\EndOfBibitem
\bibitem[Hammoudi \emph{et~al.}(2011)Hammoudi, Wong, Gao, Luo, Wong, Yang, and
  Wang]{Hammoudi:OptEx:2011}
A.~A. Hammoudi, K.~K. Wong, L.~Gao, P.~Luo, S.~T.~C. Wong, Y.~Yang and Z.~Wang,
  \emph{Optics Express}, 2011, \textbf{19}, 7960--7970\relax
\mciteBstWouldAddEndPuncttrue
\mciteSetBstMidEndSepPunct{\mcitedefaultmidpunct}
{\mcitedefaultendpunct}{\mcitedefaultseppunct}\relax
\EndOfBibitem
\bibitem[Heuke \emph{et~al.}(2013)Heuke, Vogler, Meyer, Akimov, Kluschke,
  R\"{o}wert-Huber, Lademann, Dietzek, and Popp]{Heuke:Healthcare:2013}
S.~Heuke, N.~Vogler, T.~Meyer, D.~Akimov, F.~Kluschke, H.-J. .~J.
  R\"{o}wert-Huber, J.~Lademann, B.~Dietzek and J.~Popp, \emph{Healthcare},
  2013, \textbf{1}, 64--83\relax
\mciteBstWouldAddEndPuncttrue
\mciteSetBstMidEndSepPunct{\mcitedefaultmidpunct}
{\mcitedefaultendpunct}{\mcitedefaultseppunct}\relax
\EndOfBibitem
\bibitem[Liao and Cheng(2015)]{Liao:SciAdv:2015}
C.-S. Liao and J.-X. Cheng, \emph{Science Advances}, 2015\relax
\mciteBstWouldAddEndPuncttrue
\mciteSetBstMidEndSepPunct{\mcitedefaultmidpunct}
{\mcitedefaultendpunct}{\mcitedefaultseppunct}\relax
\EndOfBibitem
\bibitem[Spivack \emph{et~al.}(1994)Spivack, Khanna, Tafra, Juillard, and
  Giuliano]{Spivack:AOS:1994}
B.~Spivack, M.~M. Khanna, L.~Tafra, G.~Juillard and A.~E. Giuliano,
  \emph{Archives of Surgery}, 1994, \textbf{129}, 952--957\relax
\mciteBstWouldAddEndPuncttrue
\mciteSetBstMidEndSepPunct{\mcitedefaultmidpunct}
{\mcitedefaultendpunct}{\mcitedefaultseppunct}\relax
\EndOfBibitem
\bibitem[Lacroix \emph{et~al.}(2001)Lacroix, Abi-Said, Fourney, Gokaslan, Shi,
  DeMonte, Lang, McCutcheon, Hassenbusch, Holland, Hess, Michael, Miller, and
  Sawaya]{Lacroix:JNeuro:2001}
M.~Lacroix, D.~Abi-Said, D.~R. Fourney, Z.~L. Gokaslan, W.~Shi, F.~DeMonte,
  F.~F. Lang, I.~E. McCutcheon, S.~J. Hassenbusch, E.~Holland, K.~Hess,
  C.~Michael, D.~Miller and R.~Sawaya, \emph{Journal of Neurosurgery}, 2001,
  \textbf{95}, 190--198\relax
\mciteBstWouldAddEndPuncttrue
\mciteSetBstMidEndSepPunct{\mcitedefaultmidpunct}
{\mcitedefaultendpunct}{\mcitedefaultseppunct}\relax
\EndOfBibitem
\bibitem[Jermyn \emph{et~al.}(2015)Jermyn, Mok, Mercier, Desroches, Pichette,
  Saint-Arnaud, Bernstein, Guiot, Petrecca, and
  Leblond]{Jermyn:SciTransMed:2015}
M.~Jermyn, K.~Mok, J.~Mercier, J.~Desroches, J.~Pichette, K.~Saint-Arnaud,
  L.~Bernstein, M.-C.~C. Guiot, K.~Petrecca and F.~Leblond, \emph{Sci Transl
  Med}, 2015, \textbf{7}, 274ra19\relax
\mciteBstWouldAddEndPuncttrue
\mciteSetBstMidEndSepPunct{\mcitedefaultmidpunct}
{\mcitedefaultendpunct}{\mcitedefaultseppunct}\relax
\EndOfBibitem
\bibitem[Kong \emph{et~al.}(2014)Kong, Zaabar, Rakha, Ellis, Koloydenko, and
  Notingher]{Kong:PMB:2014}
K.~Kong, F.~Zaabar, E.~Rakha, I.~Ellis, A.~Koloydenko and I.~Notingher,
  \emph{Physics in Medicine and Biology}, 2014, \textbf{59}, 6141\relax
\mciteBstWouldAddEndPuncttrue
\mciteSetBstMidEndSepPunct{\mcitedefaultmidpunct}
{\mcitedefaultendpunct}{\mcitedefaultseppunct}\relax
\EndOfBibitem
\bibitem[Shim and Wilson(1996)]{Shim:PCPB:1996}
M.~G. Shim and B.~C. Wilson, \emph{Photochemistry and Photobiology}, 1996,
  \textbf{63}, 662--671\relax
\mciteBstWouldAddEndPuncttrue
\mciteSetBstMidEndSepPunct{\mcitedefaultmidpunct}
{\mcitedefaultendpunct}{\mcitedefaultseppunct}\relax
\EndOfBibitem
\bibitem[Galli \emph{et~al.}(2013)Galli, Uckermann, Koch, Schackert, Kirsch,
  and Steiner]{Galli:JBO:2013}
R.~Galli, O.~Uckermann, E.~Koch, G.~Schackert, M.~Kirsch and G.~Steiner,
  \emph{Journal of Biomedical Optics}, 2013, \textbf{19}, 071402\relax
\mciteBstWouldAddEndPuncttrue
\mciteSetBstMidEndSepPunct{\mcitedefaultmidpunct}
{\mcitedefaultendpunct}{\mcitedefaultseppunct}\relax
\EndOfBibitem
\bibitem[Klein \emph{et~al.}(2012)Klein, Gigler, Aschenbrenner, Monetti, Bunk,
  Jamitzky, Morfill, Stark, and Schlegel]{Klein:BJ:2012}
K.~Klein, A.~Gigler, T.~Aschenbrenner, R.~Monetti, W.~Bunk, F.~Jamitzky,
  G.~Morfill, R.~Stark and J.~Schlegel, \emph{Biophysical Journal}, 2012,
  \textbf{102}, 360--368\relax
\mciteBstWouldAddEndPuncttrue
\mciteSetBstMidEndSepPunct{\mcitedefaultmidpunct}
{\mcitedefaultendpunct}{\mcitedefaultseppunct}\relax
\EndOfBibitem
\bibitem[Gr\"{o}schel \emph{et~al.}(2016)Gr\"{o}schel, Bommer, Hutter,
  Budczies, Bonekamp, Heining, Horak, Fr\"{o}hlich, Uhrig, H\"{u}bschmann,
  Ge\"{o}rg, Richter, Pfarr, Pf\"{u}tze, Wolf, Schirmacher, J\"{a}ger, von
  Kalle, Brors, Glimm, Weichert, Stenzinger, and
  Fr\"{o}hling]{Groschel:CSH:2016}
S.~Gr\"{o}schel, M.~Bommer, B.~Hutter, J.~Budczies, D.~Bonekamp, C.~Heining,
  P.~Horak, M.~Fr\"{o}hlich, S.~Uhrig, D.~H\"{u}bschmann, C.~Ge\"{o}rg,
  D.~Richter, N.~Pfarr, K.~Pf\"{u}tze, S.~Wolf, P.~Schirmacher, D.~J\"{a}ger,
  C.~von Kalle, B.~Brors, H.~Glimm, W.~Weichert, A.~Stenzinger and
  S.~Fr\"{o}hling, \emph{Cold Spring Harb Mol Case Stud}, 2016, \textbf{2},
  a001180\relax
\mciteBstWouldAddEndPuncttrue
\mciteSetBstMidEndSepPunct{\mcitedefaultmidpunct}
{\mcitedefaultendpunct}{\mcitedefaultseppunct}\relax
\EndOfBibitem
\bibitem[Mountzios \emph{et~al.}(2010)Mountzios, Dimopoulos, Soria, Sanoudou,
  and Papadimitriou]{Mountzios:CROH:2010}
G.~Mountzios, M.-A. .~A. Dimopoulos, J.-C. .~C. Soria, D.~Sanoudou and C.~A.
  Papadimitriou, \emph{Critical Reviews in Oncology/Hematology}, 2010,
  \textbf{75}, 94--109\relax
\mciteBstWouldAddEndPuncttrue
\mciteSetBstMidEndSepPunct{\mcitedefaultmidpunct}
{\mcitedefaultendpunct}{\mcitedefaultseppunct}\relax
\EndOfBibitem
\bibitem[Navin \emph{et~al.}(2010)Navin, Krasnitz, Rodgers, Cook, Meth,
  Kendall, Riggs, Eberling, Troge, and Grubor]{Navin:CSH:2010}
N.~Navin, A.~Krasnitz, L.~Rodgers, K.~Cook, J.~Meth, J.~Kendall, M.~Riggs,
  Y.~Eberling, J.~Troge and V.~Grubor, \emph{Genome research}, 2010,
  \textbf{20}, 68--80\relax
\mciteBstWouldAddEndPuncttrue
\mciteSetBstMidEndSepPunct{\mcitedefaultmidpunct}
{\mcitedefaultendpunct}{\mcitedefaultseppunct}\relax
\EndOfBibitem
\bibitem[Curtis \emph{et~al.}(2012)Curtis, Shah, Chin, Turashvili, Rueda,
  Dunning, Speed, Lynch, Samarajiwa, Yuan, Gr{\"a}f, Ha, Haffari, Bashashati,
  Russell, McKinney, Group, Langer{\o}d, Green, Provenzano, Wishart, Pinder,
  Watson, Markowetz, Murphy, Ellis, Purushotham, B{\o}rresen-Dale, Brenton,
  Tavar{\'e}, Caldas, and Aparicio]{Curtis:Nature:2012}
C.~Curtis, S.~P. Shah, S.-F. .~F. Chin, G.~Turashvili, O.~M. Rueda, M.~J.
  Dunning, D.~Speed, A.~G. Lynch, S.~Samarajiwa, Y.~Yuan, S.~Gr{\"a}f, G.~Ha,
  G.~Haffari, A.~Bashashati, R.~Russell, S.~McKinney, M.~Group, A.~Langer{\o}d,
  A.~Green, E.~Provenzano, G.~Wishart, S.~Pinder, P.~Watson, F.~Markowetz,
  L.~Murphy, I.~Ellis, A.~Purushotham, A.-L. .~L. B{\o}rresen-Dale, J.~D.
  Brenton, S.~Tavar{\'e}, C.~Caldas and S.~Aparicio, \emph{Nature}, 2012,
  \textbf{486}, 346\relax
\mciteBstWouldAddEndPuncttrue
\mciteSetBstMidEndSepPunct{\mcitedefaultmidpunct}
{\mcitedefaultendpunct}{\mcitedefaultseppunct}\relax
\EndOfBibitem
\bibitem[Natrajan \emph{et~al.}(2016)Natrajan, Sailem, Mardakheh, Garcia, Tape,
  Dowsett, Bakal, and Yuan]{Natrajan:PLOS:2016}
R.~Natrajan, H.~Sailem, F.~K. Mardakheh, M.~A. Garcia, C.~J. Tape, M.~Dowsett,
  C.~Bakal and Y.~Yuan, \emph{PLoS Med}, 2016, \textbf{13}, e1001961\relax
\mciteBstWouldAddEndPuncttrue
\mciteSetBstMidEndSepPunct{\mcitedefaultmidpunct}
{\mcitedefaultendpunct}{\mcitedefaultseppunct}\relax
\EndOfBibitem
\bibitem[Ideguchi \emph{et~al.}(2013)Ideguchi, Holzner, Bernhardt, Guelachvili,
  Picque, and Hansch]{Ideguchi:Nature:2013}
T.~Ideguchi, S.~Holzner, B.~Bernhardt, G.~Guelachvili, N.~Picque and T.~W.
  Hansch, \emph{Nature}, 2013, \textbf{502}, 355--358\relax
\mciteBstWouldAddEndPuncttrue
\mciteSetBstMidEndSepPunct{\mcitedefaultmidpunct}
{\mcitedefaultendpunct}{\mcitedefaultseppunct}\relax
\EndOfBibitem
\bibitem[Hashimoto \emph{et~al.}(2016)Hashimoto, Takahashi, Ideguchi, and
  Goda]{Hashimoto:SciRep:2016}
K.~Hashimoto, M.~Takahashi, T.~Ideguchi and K.~Goda, \emph{Sci Rep}, 2016,
  \textbf{6}, 21036\relax
\mciteBstWouldAddEndPuncttrue
\mciteSetBstMidEndSepPunct{\mcitedefaultmidpunct}
{\mcitedefaultendpunct}{\mcitedefaultseppunct}\relax
\EndOfBibitem
\bibitem[Ploetz \emph{et~al.}(2007)Ploetz, Laimgruber, Berner, Zinth, and
  Gilch]{Ploetz:APB:2007}
E.~Ploetz, S.~Laimgruber, S.~Berner, W.~Zinth and P.~Gilch, \emph{Applied
  Physics B-Lasers and Optics}, 2007, \textbf{87}, 389--393\relax
\mciteBstWouldAddEndPuncttrue
\mciteSetBstMidEndSepPunct{\mcitedefaultmidpunct}
{\mcitedefaultendpunct}{\mcitedefaultseppunct}\relax
\EndOfBibitem
\bibitem[Rock \emph{et~al.}(2013)Rock, Bonn, and Parekh]{Rock:OE:2013}
W.~Rock, M.~Bonn and S.~H. Parekh, \emph{Optics Express}, 2013, \textbf{21},
  15113\relax
\mciteBstWouldAddEndPuncttrue
\mciteSetBstMidEndSepPunct{\mcitedefaultmidpunct}
{\mcitedefaultendpunct}{\mcitedefaultseppunct}\relax
\EndOfBibitem
\bibitem[Shi \emph{et~al.}(2010)Shi, Li, Xu, Psaltis, and Liu]{Shi:PRL:2010}
K.~Shi, H.~Li, Q.~Xu, D.~Psaltis and Z.~Liu, \emph{Phys. Rev. Lett.}, 2010,
  \textbf{104}, 093902\relax
\mciteBstWouldAddEndPuncttrue
\mciteSetBstMidEndSepPunct{\mcitedefaultmidpunct}
{\mcitedefaultendpunct}{\mcitedefaultseppunct}\relax
\EndOfBibitem
\bibitem[Heinrich \emph{et~al.}(2008)Heinrich, Hofer, Ritsch, Ciardi, Bernet,
  and Ritsch-Marte]{Heinrich:OpEx:2008}
C.~Heinrich, A.~Hofer, A.~Ritsch, C.~Ciardi, S.~Bernet and M.~Ritsch-Marte,
  \emph{Opt Express}, 2008, \textbf{16}, 2699--2708\relax
\mciteBstWouldAddEndPuncttrue
\mciteSetBstMidEndSepPunct{\mcitedefaultmidpunct}
{\mcitedefaultendpunct}{\mcitedefaultseppunct}\relax
\EndOfBibitem
\bibitem[Toytman \emph{et~al.}(2007)Toytman, Cohn, Smith, Simanovskii, and
  Palanker]{Toytman:OptLett:2007}
I.~Toytman, K.~Cohn, T.~Smith, D.~Simanovskii and D.~Palanker, \emph{Opt Lett},
  2007, \textbf{32}, 1941--1943\relax
\mciteBstWouldAddEndPuncttrue
\mciteSetBstMidEndSepPunct{\mcitedefaultmidpunct}
{\mcitedefaultendpunct}{\mcitedefaultseppunct}\relax
\EndOfBibitem
\bibitem[Lei \emph{et~al.}(2011)Lei, Winterhalder, Selm, and
  Zumbusch]{Lei:JBMO:2011}
M.~Lei, M.~Winterhalder, R.~Selm and A.~Zumbusch, \emph{JOURNAL OF BIOMEDICAL
  OPTICS}, 2011, \textbf{16}, year\relax
\mciteBstWouldAddEndPuncttrue
\mciteSetBstMidEndSepPunct{\mcitedefaultmidpunct}
{\mcitedefaultendpunct}{\mcitedefaultseppunct}\relax
\EndOfBibitem
\bibitem[Berto \emph{et~al.}(2012)Berto, Gachet, Bon, Monneret, and
  Rigneault]{Berto:PRL:2012}
P.~Berto, D.~Gachet, P.~Bon, S.~Monneret and H.~Rigneault, \emph{Physical
  Review Letters}, 2012, \textbf{109}, 093902--\relax
\mciteBstWouldAddEndPuncttrue
\mciteSetBstMidEndSepPunct{\mcitedefaultmidpunct}
{\mcitedefaultendpunct}{\mcitedefaultseppunct}\relax
\EndOfBibitem
\bibitem[Slipchenko \emph{et~al.}(2012)Slipchenko, Oglesbee, Zhang, Wu, and
  Cheng]{Slipchenko:JB:2012}
M.~N. Slipchenko, R.~A. Oglesbee, D.~Zhang, W.~Wu and J.-X.~X. Cheng, \emph{J
  Biophotonics}, 2012\relax
\mciteBstWouldAddEndPuncttrue
\mciteSetBstMidEndSepPunct{\mcitedefaultmidpunct}
{\mcitedefaultendpunct}{\mcitedefaultseppunct}\relax
\EndOfBibitem
\end{mcitethebibliography}

\end{document}